\documentclass[pra,twocolumn,aps,tightenlines,superscriptaddress,floatfix]{revtex4}

\usepackage{graphicx}
\usepackage{dcolumn}
\usepackage{bm}
\usepackage{hyperref}

\usepackage{calc}

\usepackage{amsmath}
\usepackage{amssymb}

\usepackage{subfigure}

\usepackage{rgb_labref}
\usepackage{rgb_misc}
\usepackage{rgb_math}

\newcommand{\epsf}[2]{
\begin{figure}
\includegraphics[width=\columnwidth]{#1}
\caption{#2\labf{#1}}
\end{figure}}

\newcommand{\ket}[1]{\left| #1 \right\rangle}
\newcommand{\bra}[1]{\left\langle #1 \right|}

\newcommand{\ho}[1]{\hbar\, \omega_{#1}}
\newcommand{\rhx}[2]{\rho_{\{#1\}\{#2\}}}

\begin{document}

\bibliographystyle{apsrev}

\title{Applications of Coherent Population Transfer

to Quantum Information Processing}

\author{R.\ G.\ Beausoleil}
 \email{ray.beausoleil@hp.com}
 \affiliation{Hewlett-Packard Laboratories,
13837 175$^\textrm{th}$ Pl.\ NE, Redmond, WA 98052--2180, USA}
 \author{W.\ J.\ Munro}
 \affiliation{Hewlett-Packard Laboratories, Filton Road, Stoke Gifford,
 Bristol BS34 8QZ, United Kingdom}
 \author{T.\ P.\ Spiller}
 \affiliation{Hewlett-Packard Laboratories, Filton Road, Stoke Gifford,
 Bristol BS34 8QZ, United Kingdom}

\date{\today}

\begin{abstract}
We develop a theoretical framework for the exploration of quantum
mechanical coherent population transfer phenomena, with the
ultimate goal of constructing faithful models of devices for
classical and quantum information processing applications. We
begin by outlining a general formalism for weak-field quantum
optics in the Schr\"{o}dinger picture, and we include a general
phenomenological representation of Lindblad decoherence
mechanisms. We use this formalism to describe the interaction of a
single stationary multilevel atom with one or more propagating
classical or quantum laser fields, and we describe in detail
several manifestations and applications of electromagnetically
induced transparency. In addition to providing a clear description
of the nonlinear optical characteristics of electromagnetically
transparent systems that lead to ``ultraslow light,'' we verify
that --- in principle --- a multi-particle atomic or molecular
system could be used as either a low power optical switch or a
quantum phase shifter. However, we demonstrate that the presence
of significant dephasing effects on the metastable levels destroys
the induced transparency. Finally, a detailed calculation of the
relative quantum phase induced by a system of atoms on a
superposition of spatially distinct Fock states predicts that a
significant quasi-Kerr nonlinearity and a low entropy cannot be
simultaneously achieved in the presence of arbitrary spontaneous
emission rates. Within our model, we identify the constraints that
need to be met for this system to act as a one-qubit and a
two-qubit conditional phase gate.
\end{abstract}

 \maketitle

 \section{INTRODUCTION}

 \subsection{Quantum Information Processing}

With developments in theory and experiment over the last few
years, there has been a dramatic growth of interest in the area of
quantum information processing and
communication.\cite{lo98,bouw00,niel00} There are now strong
indicators that this fundamental research field could lead to a
whole new quantum information technology in the future. Working
practical quantum cryptosystems exist already.\cite{stuc02,gisi02}
It is known that large (many qubit) quantum computers, if they can
be built, would be capable of performing certain tasks (such as
factoring large composite integers or searching) much more
efficiently than conventional classical computers. Rather smaller
(tens or hundreds of qubits) quantum processors would be able to
perform quantum simulations unreachable with any classical machine
and such processors also have the potential to extend the working
distances and applicability of quantum communications.

There are currently numerous possible  routes for quantum
computing hardware.\cite{fort00,clar01} Many of these are based on
coherent condensed matter systems. While at present such systems
generally exhibit less coherence than qubits based on fundamental
entities (such as ions or atoms), the goals are to increase this
sufficiently for error correction techniques to be applicable and
to utilize the potential fabrication advantage to make condensed
matter many-qubit processors. Even if this proves to be the way
forward, it seems certain that there will be a need for separate
quantum processors to communicate with each other in a quantum
coherent manner. Coherent electromagnetic fields are likely the
best candidates for realizing this goal. In addition, some quantum
information processing may be performed directly on photon qubits,
using non-linear \cite{turc95} or linear quantum optical
processes.\cite{knil01} So, photon qubits (or other quantum
coherent states of the electromagnetic field) certainly have a
number of important uses for quantum computing. However, they play
center stage when it comes to communication, because the best way
to send quantum information over large distances is certainly
using light, either down optical fibers or even through free
space.\cite{stuc02}

Given all this, the study of coherent interactions between light
and matter is an extremely important topic.\cite{luki01a,monr02}
It is hard to see how large scale quantum information technology
can emerge without the ability to easily interconvert traveling
photon qubits and stationary matter qubits.\cite{divi97,divi00}
Such interconversion would enable the construction of quantum
networks \cite{cira97}, where communication between matter qubit
nodes is mediated through photons. The ability to perform quantum
gates (one- and two-qubit) directly on qubits encoded into photons
is highly desirable for communication and computing. In addition,
the coherent storage of photon qubits would open up new
possibilities for quantum information processing.

Another very important application of quantum phenomena is to
achieve tasks in classical (optical) information processing and
communication. These applications will likely be the first to
impact on future information technology. In this work we study
aspects of coherent population transfer phenomena and discuss how
the emergent effects may be applicable for producing large optical
phase shifts and for switching. These effects in themselves are
useful for conventional optical data processing. However, if in
addition they can be demonstrated to work on non-classical input
states of light, then they have the potential to form the basis of
one- and two-qubit gates. We present a detailed calculation of the
relative quantum phase induced by a system of atoms on a
superposition of spatially distinct Fock states (a ``dual rail''
photon qubit) and identify the conditions under which this
performs as a useful one-qubit phase gate. We also consider the
qubit limit of one of the control fields applied to the system,
and thus how a conditional two-qubit phase gate can be realized.

 \subsection{Electromagnetically Induced Transparency}

The basic interaction between an atom and a photon is weak, which
is why photons are so good for communication purposes. In order to
enhance the interaction, one can either seek to use many actual
atoms---an ensemble---or many ``images'' of a single atom, in
effect created by the mirrors of a very high-$Q$ cavity. The
latter cavity QED approach has produced numerous impressive
experimental and theoretical results over the last few years,
\cite{kimb98} but from the perspective of using atoms to
manipulate light for information processing, we focus on the
former ensemble approach.\cite{luki03}

Photons, or light in general, can interact strongly with an
ensemble of atoms, leading to a number of very interesting
effects. The basis of many of these is the idea of
electromagnetically induced transparency (EIT), a specific example
of coherent population transfer.\cite{harr97} Here, quantum
interference can be used to effectively cancel the would-be
absorption in an atomic medium, rendering it transparent.
Ordinarily, a light signal on resonance with an atomic transition
$\ket{1} \rightarrow \ket{2}$ ($\ket{1}$ being the lower and
long-lived level) interacts strongly with the atoms. However, if a
third long-lived level $\ket{3}$ comes into play and a control
field resonant with $\ket{1} \rightarrow \ket{3}$ is applied, the
atoms are stimulated into states which cannot absorb,\cite{arim96,
boll91} and so in effect the atomic medium is transparent to the
signal. These so-called ``dark states'' are coherent
superpositions of $\ket{1}$ and $\ket{3}$. We discuss this effect
in detail for such three-level atoms. As will be seen, the
transparency occurs only exactly on resonance. A transparency
window, necessary for practical applications, can be opened by use
of four-level atoms. Our focus in this work is the use of such EIT
systems for the optical information processing applications.
However, other important and interesting directions and
applications exist, which can be followed up in detail elsewhere.

The intimate link between absorption and dispersion also means
that EIT media can be used to manipulate the group velocity of
light pulses.\cite{harr92} This can be reduced dramatically below
the speed of light in vacuum $c$ through reducing the power in the
control field and increasing the atomic density of the EIT medium.
Various experiments demonstrated the potential of this effect, but
the significant breakthrough came in 1999 when Hau {\it et al.}
\cite{hau99} produced a group velocity of 17 m/s for light pulses
in sodium vapor. Further developments continue to emerge. It is
something of an oversimplification to regard a light pulse as
simply slowing down as it propagates through an EIT medium. The
strong interaction of the light with the atoms can effectively be
diagonalized through the concept of quasiparticles (well known in
condensed matter physics). In this case the propagation through
the EIT system is described by quasiparticles called dark-state
polaritons \cite{flei00}, which are a coherent superposition of
photons and spins. These polaritons move at a velocity given by
the group velocity of their photonic component. An EIT system can
therefore act as a "delay line" for light pulses, an effect which
in itself has significant application potential for communication.
However, the ultimate limit of this effect is a complete slowing,
leading to quantum memory for
photons.\cite{flei02,juze02,mewe02,flei01} As the (externally
controllable) polariton group velocity is reduced towards zero,
the photonic component of the polariton also reduces. In effect,
the quantum state of the light field is stored in long-lived spin
states of atoms within the EIT medium. This clearly has
significant potential for quantum communication and information
processing applications.

In a sense the whole scenario can also be reversed, with atomic
ensembles (as opposed to the photons) being the principal
information carriers, with interactions being controlled through
electromagnetic fields. A number of novel quantum phenomena, also
with potential for communication and processing applications, can
arise in this case. It is possible to make quantum nondemolition
measurements on the collective spin degree of freedom of atomic
ensembles \cite{kuzm00} using light. The interaction between such
ensembles and light has also been employed to create a level of
entanglement between the collective spin degrees of freedom of two
atomic ensembles \cite{juls01}, a first step towards QIP with
collective spins. Small ensembles of atoms can also exhibit
analogous effects to the well known solid state mesoscopic
phenomena such as Coulomb blockade. In very small capacitance
sub-micron devices the strong Coulomb energy makes just a few
energy levels relevant, and enables the manipulation of individual
electronic charges. Likewise in small atomic ensembles the strong
dipole-dipole interactions make just a few energy levels relevant,
and it is possible to manipulate individual excitations of these
spin systems and exhibit "dipole blockade"
phenomena.\cite{luki01b,bren99,jaks00,luki00a}

In this work we develop a theoretical framework for the
exploration of quantum mechanical coherent population transfer
phenomena, with a particular emphasis on the aspects of
nonrelativistic quantum electrodynamics that play a strong role in
potential classical and quantum information processing
applications. We begin by outlining a general formalism for
quantum optics in the Schr\"{o}dinger picture, and we include a
general phenomenological representation of Lindblad decoherence
mechanisms. We use this formalism to describe the interaction of a
single stationary multilevel atom with one or more propagating
classical or quantum laser fields, assuming that this interaction
is sufficiently weak that the population of the ground state is
never significantly depleted. In addition to providing a clear
description of the nonlinear optical characteristics of
electromagnetically transparent systems that lead to ``ultraslow
light,'' we verify that --- in principle --- a multi-particle
atomic or molecular system could be used as either a low power
optical switch or a quantum phase shifter. However, we demonstrate
that the presence of significant dephasing effects destroys the
induced transparency, and that increasing the number of particles
weakly interacting with the probe field only reduces the
nonlinearity further.

 \section{\labs{efield}REPRESENTATION OF THE ELECTROMAGNETIC FIELD}

We wish to simplify our treatment of the quantum electromagnetic
field as much as possible without introducing approximations which
will result in significant disagreements with experimental
results. Therefore, we begin our discussion of field quantization
with the model of a unidirectionally propagating traveling wave
shown in \fig{resonator}.\cite{cohe92} As a result of the finite
round-trip length $L$ of the cavity, the electromagnetic field is
comprised of a superposition of discrete longitudinal modes with
different wavevectors $\mathbf{k}$, angular frequencies
$\omega_\mathbf{k} \equiv |\mathbf{k}| c$, and polarization unit
vectors $\hat{\bm{\epsilon}}_{\mathbf{k} \lambda}$, where $\lambda
\in \{1, 2\}$, that satisfy the orthogonality condition
$\hat{\bm{\epsilon}}^*_{\mathbf{k}^\prime \lambda^\prime}
\mathbf{\cdot} \hat{\bm{\epsilon}}_{\mathbf{k} \lambda} =
\delta_{\mathbf{k}^\prime, \mathbf{k}}\, \delta_{\lambda^\prime,
\lambda}$. The field interacts with an ensemble of atoms located
within the cavity; since the resonator is lossless, the only
mechanism able to alter the number of photons in a given mode is
the stimulated emission and absorption of photons in that mode.

\epsf{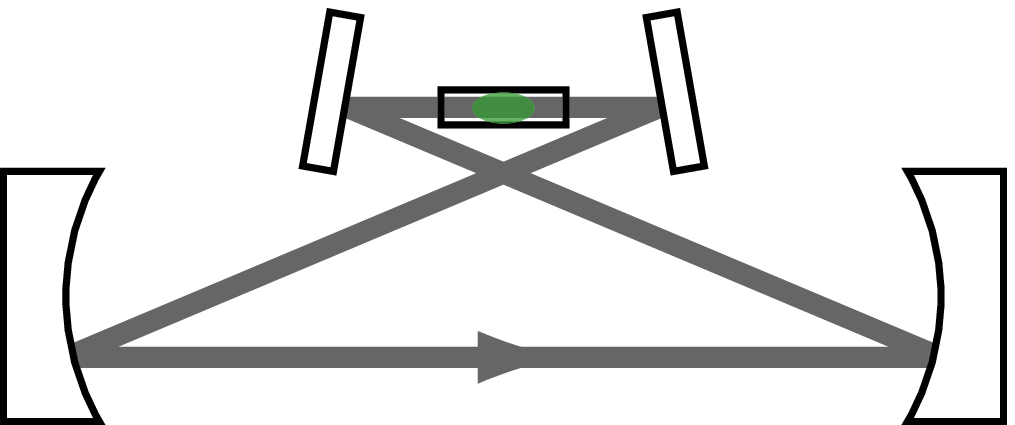}{Discrete model of a traveling-wave quantum
electromagnetic field interacting with an ensemble of atoms within
a lossless resonator.}

In the laboratory, a typical experiment will most likely employ a
continuously tunable multi-frequency traveling-wave field
interacting with an atom in free space. Our model can represent
such an experiment if two conditions are satisfied:\cite{cohe92}
 \begin{enumerate}
 \item The resonator enforces an amplitude and
 local transverse spatial distribution on
 the model field that is identical to that of the experiment.
 Therefore, the exact number of photons in a given mode $\mathbf{k}$ is not
 important when modeling a classical field, provided that
 the mean field intensity (approximately given by
 $\left<n_V\right> \ho{\mathbf{k}}\, c/V$, where $\left<n_V\right>$ is the
 mean number of modes occupying the mode volume $V$) and mode shape are
 preserved.
 \item The cavity length $L$ is sufficiently long that the
 spontaneous emission spectrum of the atom is not altered.
 \end{enumerate}
In principle, we can model the tunability of the laser field by
assuming that the positions of the resonator mirrors can be
microscopically adjusted.

In the Schr\"{o}dinger picture,\footnote{In the Heisenberg
picture, $\mathbf{k} \cdot \mathbf{r} \longrightarrow \mathbf{k}
\cdot \mathbf{r} - \omega_\mathbf{k} t$.} we can represent the
quantized transverse electric field operator as a discrete sum
over the allowed wavevectors $\mathbf{k}$ and possible
polarizations $\hat{\bm{\epsilon}}_{\mathbf{k} \lambda}$:
 \begin{equation}
 \begin{split}
\mathbf{E}(\mathbf{r}) &\equiv \sum_{\mathbf{k} \lambda}
\hat{\bm{\epsilon}}_{\mathbf{k} \lambda}\,
\mathcal{E}_\mathbf{k}\, U_\mathbf{k}(\mathbf{r})\, e^{i
\mathbf{k} \cdot \mathbf{r}}\, a_{\mathbf{k} \lambda} \\
 &+ \sum_{\mathbf{k} \lambda}
\hat{\bm{\epsilon}}^*_{\mathbf{k} \lambda}\,
\mathcal{E}_\mathbf{k}\, U^*_\mathbf{k}(\mathbf{r})\, e^{-i
\mathbf{k} \cdot \mathbf{r}}\, a^\dag_{\mathbf{k} \lambda}
 \end{split}
 \end{equation}
where the dimensionless resonator eigenfunction
$U_\mathbf{k}(\mathbf{r})$ describes the transverse spatial
dependence of a field with a characteristic spot size
$w_\mathbf{k}$, and satisfies both $U_\mathbf{k}(0) = 1$ and the
volume normalization condition
 \begin{equation}\labe{Unorm}
 \mathcal{V} \equiv
 \int_V d^3r\, \left| U_\mathbf{k}(\mathbf{r}) \right|^2
= \frac{\pi}{2} w^2_\mathbf{k}\, L.
 \end{equation}
Therefore, the coefficient $\mathcal{E}_\mathbf{k} \equiv
\sqrt{\ho{\mathbf{k}}/\epsilon_0 \pi w^2_\mathbf{k} L}$ has the
dimensions of an electric field.

We will apply the creation and annihilation operators
$a^\dag_{\mathbf{k} \lambda}$ and $a_{\mathbf{k} \lambda}$ to the
corresponding Fock (number) states through the ladder operator
equations\cite{loud00}
 \begin{subequations}
 \begin{eqnarray}
a^\dag_{\mathbf{k} \lambda} \left| n_{\mathbf{k} \lambda} \right>
&=& \sqrt{n_{\mathbf{k} \lambda} + 1}\left| n_{\mathbf{k} \lambda}
+ 1 \right> \\ a_{\mathbf{k} \lambda} \left| n_{\mathbf{k}
\lambda} \right> &=& \sqrt{n_{\mathbf{k} \lambda}}\left|
n_{\mathbf{k} \lambda} - 1 \right>
 \end{eqnarray}
 \end{subequations}
 The Fock state $\left| n_{\mathbf{k} \lambda} \right>$ is an
 eigenstate of the unperturbed Hamiltonian\footnote{Note that we have explicitly
 neglected the quantum electrodynamic contribution to the Hamiltonian from the vacuum.}
 \begin{equation}
 H = \sum_{\mathbf{k} \lambda} \ho{\mathbf{k}}\, a^\dag_{\mathbf{k}
 \lambda} a_{\mathbf{k} \lambda},
 \end{equation}
 with the eigenvalue $n_{\mathbf{k} \lambda}\, \ho{\mathbf{k}}$,
 giving the total energy stored in photons of mode
 $\mathbf{k} \lambda$. In fact, we can ensure that we have chosen the correct
 value for the normalization constant $\mathcal{E}_\mathbf{k}$ by integrating
 the corresponding expectation value of the electromagnetic energy density
 $\epsilon_0 \mathbf{E}(\mathbf{r}) \cdot \mathbf{E}(\mathbf{r})$
 over the
 mode volume. Applying \eqr{Unorm}, we obtain
 \begin{equation}
\epsilon_0 \int_V d^3r\, \left<n_{\mathbf{k}
\lambda}\right|\mathbf{E}(\mathbf{r}) \cdot \mathbf{E}(\mathbf{r})
\left|n_{\mathbf{k} \lambda}\right> = n_{\mathbf{k} \lambda}\,
\ho{\mathbf{k}},
\end{equation}
as expected.

In this work, we will pass from the quantum regime to the
classical limit through the coherent state $\left| \alpha(t)
\right>$,\cite{cohe92} defined in terms of the single-mode (i.e.,
single frequency $\omega$ and single polarization $\lambda$) Fock
states $\left| n \right>$ as
 \begin{equation} \labe{csdef}
\left| \alpha(t) \right> = e^{-\half |\alpha(t)|^2}\,
\sum^\infty_{n = 0} \frac{\alpha^n(t)}{\sqrt{n!}} \left| n \right>
,
 \end{equation}
 where $\alpha(t)$ is defined in terms of the mean occupation
 number $\left< n_V \right>$ as
 \begin{equation} \labe{alphadef}
 \alpha(t) = \sqrt{\left< n_V \right>}\, e^{-i \omega t} .
 \end{equation}
 Since $\left| \alpha(t) \right>$ is an eigenstate of the
 single-mode destruction operator $a$ with eigenvalue $\alpha(t)$,
 we have
 \begin{subequations}
 \begin{eqnarray}
 a \ket{\alpha(t)} &=& \alpha(t) \ket{\alpha(t)}, \nd \\
 \bra{\alpha(t)} a^\dag &=& \bra{\alpha(t)} \alpha^*(t) ,
 \end{eqnarray}
 \end{subequations}
giving
 \begin{equation}
 \begin{split}
 \bra{\alpha(t)} \mathbf{E}(\mathbf{r}) \ket{\alpha(t)} &=
 \mathcal{E}\, U(\mathbf{r})\, e^{i
\mathbf{k} \cdot \mathbf{r}} \alpha(t) + c.c.\\
 &\equiv \half E(\mathbf{r})\, e^{i (
\mathbf{k} \cdot \mathbf{r} - \omega t)} + c.c. ,
 \end{split}
 \end{equation}
where the associated classical field amplitude is
 \begin{equation}
 E(\mathbf{r}) = 2\, \sqrt{\left< n_V \right>
 \frac{\hbar \omega}{\epsilon_0 \pi w^2 L}}\, U(\mathbf{r}) .
 \end{equation}
Therefore, for a given resonator mode containing exactly $n$
photons, we can associate the quantity $2\, \mathcal{E} \sqrt{n}$
with a classical field amplitude $E$ at $\mathbf{r} = 0$. In
practice, we will allow this classical amplitude --- and therefore
the mean photon number $\left< n_V \right>$ --- to vary slowly in
time when we study the adiabatic interaction of pump-probe pulses
in our discussions of semiclassical coherent population transfer.
However, we must be careful when passing from the quantum to the
classical regime when the number of photons in a pulse is
relatively small. In \sct{eit4}, we show that weak coherent fields
that are used to manipulate a quantum optical nonlinearity can be
disrupted, in the sense that the field can no longer be
represented as a coherent state after the interaction has ceased.
In all cases, the atom + field system dynamics can be correctly
described by tracking the evolution of individual energy (i.e.,
Fock) eigenstates, and then evaluating the sum given by
\eqr{csdef} either analytically or numerically.

 \section{ELECTROMAGNETICALLY INDUCED TRANSPARENCY}
One of the most striking examples of coherent population transfer
is electromagnetically induced transparency, where the absorption
of a weak probe field by an atomic or molecular medium is
mitigated by a second control field.\cite{mara98,mara00} In
general, we will be concerned with systems that are weakly pumped,
in the sense that we will assume that the perturbed atom(s) will
remain primarily in the ground state.

 \begin{figure}
   \centering
   \subfigure[Semiclassical energy levels]{\labf{lambda_2}
 \includegraphics[width=2.6in]{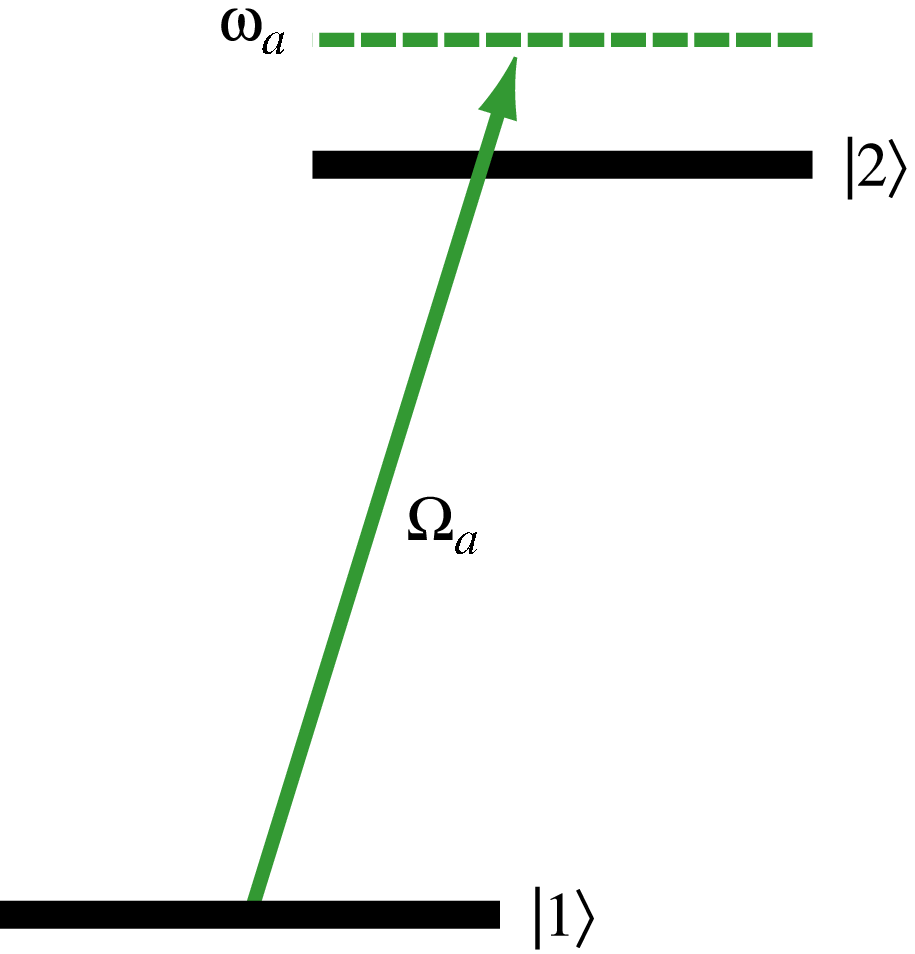}}\qquad
   \subfigure[Quantum energy manifolds]{\labf{quantum_2}
 \includegraphics[width=2.6in]{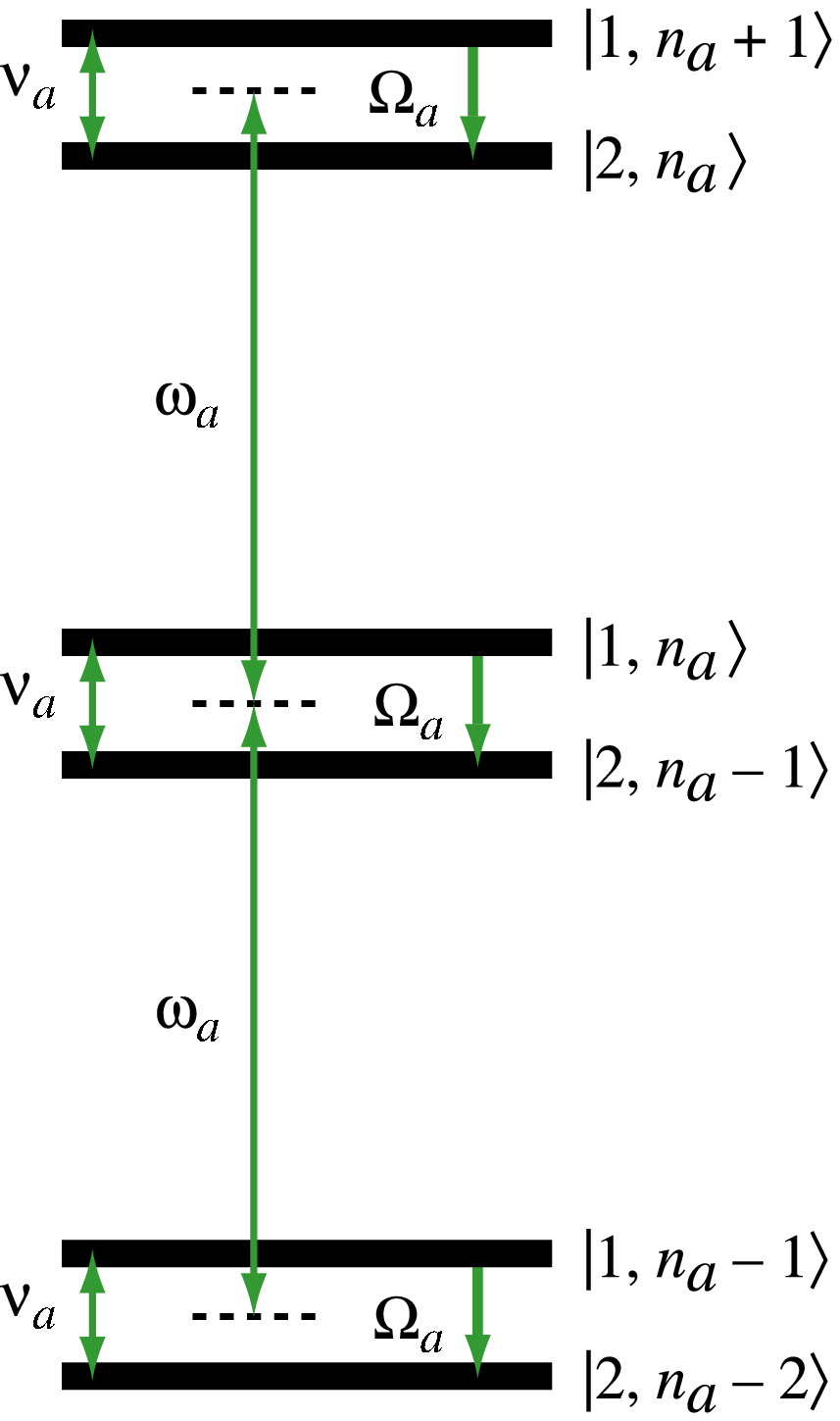}}
   \caption{\labf{two_level_atom} Electric dipole interaction
   between a two-level atom and a nearly resonant
   electromagnetic field.  In the semiclassical view, the atomic energy levels
   are separated by the energy $\ho{21}$,
   and coupled by a field oscillating at the frequency $\omega_a = \omega_{21} + \nu_a$.
   In the quantum view, the states of the atom + photons system
   separate into manifolds coupled internally by resonant
   transitions; the energy difference
   $\hbar \nu_a$ between the two levels of a
   given resonant manifold is far smaller than the difference
   $\ho{a}$ between two adjacent manifolds. }
 \end{figure}

 \subsection{Quantum Optics of the Two-Level Atom\labs{eit2}}

The electric dipole interaction between a two-level atom and a
nearly resonant electromagnetic field is shown schematically in
\fig{two_level_atom}. In the semiclassical view of \fig{lambda_2},
the two atomic energy levels are separated by the energy $\ho{21}
\equiv \ho{2} - \ho{1}$, and coupled by a field oscillating at the
angular frequency $\omega_a \approx \omega_{21}$. However, in the
quantum view of \fig{quantum_2}, the outer-product states of the
atom + photons system separate into manifolds coupled internally
by resonant transitions;\cite{cohe92} if we define the detuning
parameter
 \begin{equation}
 \labe{detuna}
 \nu_a \equiv \omega_a - \left(\omega_2 - \omega_1\right) ,
 \end{equation}
then the energy difference $\hbar \nu_a$ between the two levels of
a given resonant manifold is far smaller than the difference
$\ho{a}$ between two adjacent manifolds.

Based on \fig{quantum_2}, we write the unperturbed Hamiltonian of
the system in the Schr\"{o}dinger picture\cite{cohe77} as
\begin{equation}
H_0 = \ho{1}\, \sigma_{11} + \ho{2}\, \sigma_{22} + \ho{a}\,
a^\dag a,
\end{equation}
where $a^\dag$ and $a$ are respectively the creation and
annihilation operators for photons with energy $\ho{a}$ and
polarization vector $\hat{\bm{\epsilon}}_a$, and $\sigma_{ij}
\equiv \ket{i} \bra{j} = \sigma^\dag_{ji}$ are the atomic raising
and lowering operators. Now, given an atom initially in the ground
state and $n_a$ photons stored in the (lossless) resonator, the
two eigenstates of the unperturbed Hamiltonian $H_0$ are $\ket{1,
n_a}$ and $\ket{2, n_a - 1}$, with the eigenvalues
 \begin{equation}
  \bra{1, n_a} H_0 \ket{1, n_a} =
  \ho{1} + n_a\, \ho{a} ,
 \end{equation}
and
 \begin{equation}
 \bra{2, n_a - 1} H_0 \ket{2, n_a - 1} = \ho{2} + \left(n_a - 1\right) \ho{a} .
 \end{equation}

 In the long-wavelength approximation, the electric dipole/field
 coupling interaction Hamiltonian for a single atom at $\mathbf{r} = 0$ is given explicitly
 by\cite{cohe89,cohe92,loud00}
 \begin{equation}
 \begin{split}
 V &= -\mathbf{d} \cdot \mathbf{E}(\mathbf{0}) \\
 &= -\left( \mathbf{d}_{12}\, \sigma_{12} + \mathbf{d}_{21}\, \sigma_{21}
 \right)\\
 &\qquad \cdot \left( \hat{\bm{\epsilon}}_a\, \mathcal{E}_a\, a +
 \hat{\bm{\epsilon}}^*_a\, \mathcal{E}_a\, a^\dag\right)
 \end{split}
 \end{equation}
In the basis of the unperturbed Schr\"{o}dinger eigenstates, the
matrix elements of the interaction Hamiltonian are
 \begin{equation}
 \bra{2, n_a - 1} V \ket{1, n_a} = \hat{\bm{\epsilon}}_a
 \cdot \mathbf{d}_{21}\, \mathcal{E}_a
 \sqrt{n_a} ,
 \end{equation}
Therefore, if we define the effective coupling constant
 \begin{equation}
 \labe{gadef}
  g_a \equiv \hat{\bm{\epsilon}}_a \cdot \mathbf{d}_{21}\, \mathcal{E}_a
 \end{equation}
 and the effective Rabi frequency
 \begin{equation}
 \labe{rabiadef}
 \Omega_a = \frac{g_a}{\hbar}  \sqrt{n_a} ,
 \end{equation}
then in the unperturbed Schr\"{o}dinger basis
 \begin{equation} \labe{fock_2}
 \begin{split}
 \left\{ \right. & \ket{1, n_a} , \\
 & \ket{2, n_a - 1} \left. \right\}
 \end{split}
 \end{equation}
 we obtain the total Hamiltonian
 \begin{equation} \labe{H2m}
 H = -\hbar\, \begin{bmatrix}
 0 & \Omega_a^* \\
 \Omega_a & \nu_a
 \end{bmatrix} ,
 \end{equation}
where we have subtracted $E_0 \equiv \bra{1, n_a} H_0 \ket{1,
n_a}$ from both diagonal terms. Note that we are using the bare
eigenstates (rather than dressed states\cite{cohe92}) of the
unperturbed Hamiltonian in the Schr\"{o}dinger picture; the
resulting perturbed Hamiltonian agrees with that of the
corresponding semiclassical system in the interaction picture.

The evolution of the wavefunction
 \begin{equation} \labe{ket2}
 \ket{\psi(t)} \equiv c_1(t) \ket{1, n_a} + c_2(t) \ket{2,
 n_a - 1}
 \end{equation}
is governed by the Schr\"{o}dinger equation\cite{cohe77}
 \begin{equation}
 i \hbar \frac{d}{d t} \ket{\psi(t)} = H \ket{\psi(t)} ,
 \end{equation}
which has the formal solution
 \begin{equation}
 \ket{\psi(t)} = U(t) \ket{\psi(0)} ,
 \end{equation}
where the evolution operator $U(t)$ is given by
 \begin{widetext}
 \begin{equation}
 U(t) = e^{-\frac{i}{\hbar}\, H\, t} = e^{\frac{i}{2}\, \nu_a t}\,
 \begin{bmatrix}
 \cos\left(\Omega_R\, t\right) - i \frac{\nu_a}{2\, \Omega_R}
 \sin\left(\Omega_R\,
 t\right) & i \frac{\Omega_a}{\Omega_R} \sin\left(\Omega_R\,
 t\right) \\
 i \frac{\Omega_a}{\Omega_R} \sin\left(\Omega_R\,
 t\right) & \cos\left(\Omega_R\, t\right) + i \frac{\nu_a}{2\, \Omega_R}
 \sin\left(\Omega_R\,
 t\right)
 \end{bmatrix} ,
 \end{equation}
 \end{widetext}
and
 \begin{equation} \labe{rabi2def}
 \Omega_R \equiv \half \sqrt{\nu_a^2 + 4 \left|\Omega_a\right|^2} .
 \end{equation}

Decoherence phenomena significantly complicate the evolution of
the two-level manifold of \fig{quantum_2}. For example,
spontaneous emission by the excited atom can scatter a photon into
the free-space boundary volume (the ``environment'') enclosing the
idealized resonator of \fig{resonator} at a rate\cite{cohe92}
 \begin{equation} \labe{einsteina21}
 A_{21} = \frac{\omega_{21}^3 \left|\mathbf{d}_{21}\right|^2}
 {3\, \pi\, \epsilon_0\, \hbar\, c^3} ,
 \end{equation}
causing a transition from the the central manifold of
\fig{quantum_2} to the lower manifold. Strictly speaking, then, a
completely general model describing incoherent population transfer
phenomena must incorporate multiple manifolds. However, in this
work, we are primarily concerned with applications of
nonrelativistic quantum electrodynamics to quantum information
processing, particularly high-fidelity quantum gate operations. In
this case, the resonantly-coupled manifold of interest is
$\left\{\ket{1,1}, \ket{2,0}\right\}$; therefore, spontaneous
emission by the excited atom causes a transition from the product
state $\ket{2,0}$ to $\ket{1,0}$, a state lying in the lowest
manifold. (In \sct{efield}, we explicitly assumed that the
resonator of \fig{resonator} does not modify the spontaneous
emission spectrum, so we do not consider spontaneous emission into
the cavity mode.) This transition effectively destroys the
information carried by the quantum state of the electromagnetic
field, and therefore a detailed analytical representation of the
subsequent evolution of the state of the system is uninteresting.
In practice, we can recover our resonant-coupling approximation by
extending our product states to append an entry indicating whether
a photon with frequency $\omega_a$ has been emitted into the
environment, giving us a new single-photon basis, extended from
\eqr{fock_2} to
 \begin{equation} \labe{fock_2e}
 \begin{split}
 \left\{ \right. & \ket{1, n_a, 0} , \\
 & \ket{2, n_a - 1, 0} , \\
 & \ket{1, n_a - 1, 1} \left. \right\}.
 \end{split}
 \end{equation}

The introduction of decoherence into our model will prevent us
from describing the two-level system state using the pure vector
given by \eqr{ket2}. Therefore, in the extended basis of
\eqr{fock_2e}, we introduce the corresponding density
matrix\cite{cohe77} of the atom-photon system as
 \begin{equation} \labe{rho2}
 \rho = \begin{bmatrix}
 \rho_{1 1} & \rho_{1 2} & \rho_{1 e} \\
 \rho_{2 1} & \rho_{2 2} & \rho_{2 e} \\
 \rho_{e 1} & \rho_{e 2} & \rho_{e e}
 \end{bmatrix} .
 \end{equation}
We can then use the corresponding total Hamiltonian
 \begin{equation} \labe{H2e}
 H = -\hbar\, \begin{bmatrix}
 0 & \Omega^*_a & 0 \\
 \Omega_a & \nu_a & 0 \\
 0 & 0 & 0
 \end{bmatrix} ,
 \end{equation}
and an appropriate set of initial conditions to solve the density
matrix equations of motion\cite{cohe77}
 \begin{equation}
 \labe{rhodot}
 \dot{\rho}(t) = -\frac{i}{\hbar}\,\left[H, \rho(t)\right] -
 \Gamma[ \rho(t) ] ,
 \end{equation}
where we have incorporated damping through the decoherence
operator $\Gamma[ \rho(t) ]$. We adopt the Lindblad
form\cite{lind76,niel00} of $\Gamma[ \rho(t) ]$, given by
 \begin{equation} \labe{Lindblad}
\Gamma(\rho) = \half \sum_m \gamma_m \left( \left[ \rho\,
L^\dag_m, L_m \right] + \left[ L^\dag_m, L_m\, \rho \right]
\right) ,
 \end{equation}
to preserve both positive probabilities and a positive
semidefinite density operator. The Lindblad operator $L_m$
represents a general dissipative process occurring at the rate
$\gamma_m$. For example, we can describe the depopulation of the
product state $\ket{j, \dots}$ (due either to spontaneous emission
from the atomic state $\ket{j}$ to the environment $\ket{e}$, or
to photon transmission or scattering losses in the cavity) at the
rate $\gamma_{j}^\prime$ using the lowering operator
 \begin{equation}
 L^\prime_{j} \equiv \sigma_{ej} = \ket{e} \bra{j}
 \end{equation}
and pure dephasing of the state $\ket{j, \dots}$ at the rate
$\gamma^{\prime \prime}_{j}$ using the operator
 \begin{equation}
L^{\prime \prime}_{j} \equiv \frac{1}{\sqrt{2}}\, \left(I - 2
\sigma_{jj}\right) ,
 \end{equation}
where $I$ is the identity matrix. For example, in the case $j =
2$, we have
 \begin{equation}
 L^\prime_{2} = \begin{bmatrix}
 0 & 0 & 0 \\
 0 & 0 & 0 \\
 0 & 1 & 0
 \end{bmatrix}
 \end{equation}
and
 \begin{equation}
L^{\prime \prime}_{2} = \frac{1}{\sqrt{2}}\, \left(\sigma_{11} -
\sigma_{22} + \sigma_{33}\right) = \frac{1}{\sqrt{2}}
\begin{bmatrix}
 1 & 0 & 0 \\
 0 & -1 & 0 \\
 0 & 0 & 1
 \end{bmatrix} .
 \end{equation}
Since we have assumed in \sct{efield} that the resonator of
\fig{resonator} does not modify the spontaneous emission spectrum
of the atom, we have $\gamma_{2}^\prime = A_{21}$. Therefore, if
we assume that atomic level $\ket{1}$ is metastable and that the
cavity is lossless, then we obtain the decoherence operator
 \begin{equation} \labe{Gamma2}
 \Gamma(\rho) = \begin{bmatrix}
 0 & \gamma_{21}\, \rho_{12} & \gamma_{e1}\, \rho_{1e} \\
 \gamma_{21}\, \rho_{21} & \gamma_{22}\, \rho_{22} & \gamma_{e2}\, \rho_{2e} \\
 \gamma_{e1}\, \rho_{e1} & \gamma_{e2}\, \rho_{e2} & -\gamma_{22}\, \rho_{22}
 \end{bmatrix} ,
 \end{equation}
where
 \begin{subequations}
 \begin{eqnarray}
 \gamma_{21} = \gamma_{12} &\equiv& \half \gamma^\prime_{2} + \gamma^{\prime \prime}_{1} + \gamma^{\prime \prime}_{2} , \labe{gamma21def}\\
 \gamma_{22} &\equiv& \gamma^\prime_{2} , \\
 \gamma_{e1} = \gamma_{1e} &\equiv& \gamma^{\prime \prime}_{1} , \nd \\
 \gamma_{e2} = \gamma_{2e} &\equiv& \half \gamma^\prime_{2} + \gamma^{\prime \prime}_{2} .
 \end{eqnarray}
 \end{subequations}

We substitute \eqr{rho2}, \eqr{H2e}, and \eqr{Gamma2} into
\eqr{rhodot} to obtain
 \begin{subequations}  \labe{rho2dot}
 \begin{eqnarray}
 \dot{\rho}_{11}(t) &=& -2 \Im\left[\rho_{21}(t)\, \Omega_a^*\right]
 , \labe{rho2dot_11} \\
 \dot{\rho}_{22}(t) &=& -\gamma_{22} \rho_{22}(t) + 2 \Im\left[\rho_{21}(t)\, \Omega_a^*\right]
 , \nd \labe{rho2dot_22} \\
 \dot{\rho}_{21}(t) &=& i \left(\nu_a + i \gamma_{21}\right) \rho_{21}(t) \nonumber \\
 &&\quad + i\, \Omega_a
 \left[ \rho_{11}(t) - \rho_{22}(t) \right] \labe{rho2dot_21}
 \end{eqnarray}
 \end{subequations}
and then solve for the elements $\rho_{11}(t)$, $\rho_{22}(t)$,
and $\rho_{21}(t)$ using a ``bootstrapping'' method for the
initial condition $\rho_{ij}(0) = \delta_{i 1}\, \delta_{j 1}$. If
we assume that the interaction is unsaturated (i.e.,
$\left|\Omega_a\right|/\gamma_{21} \ll 1$) so that $\rho_{11}(t)
\gg \rho_{22}(t)$ for all $t$, and that
--- to zeroth order in $\left|\Omega_a\right|$ --- $\rho_{11}(t)$
varies slowly compared to $\sqrt{\nu_a^2 + \gamma_{21}^2}$, then
$\rho_{21}(t)$ will adiabatically follow $\rho_{11}(t)$ and assume
some quasi-steady-state value $\tilde{\rho}_{21}$. Therefore, if
we ignore short-term transients we must obtain
\begin{equation}\labe{rho11apprx}
\rho_{11}(t) \cong \exp\left[-2 \Im\left(\tilde{\rho}_{21}\,
\Omega_a^*\right) t\right] ,
\end{equation}
and subsequently from \eqr{rho2dot_21}
 \begin{equation}
\rho_{21}(t) \cong \left[1 - e^{i \left(\nu_a + i
\gamma_{21}\right) t}\right]\, \tilde{\rho}_{21}\, \rho_{11}(t) ,
 \end{equation}
where
 \begin{equation}\labe{rho21tilde}
\tilde{\rho}_{21} \equiv -\frac{\Omega_a}{\nu_a + i \gamma_{21}} =
-\frac{\nu_a - i \gamma_{21}}{\nu_a^2 + \gamma_{21}^2} \Omega_a.
 \end{equation}
Collecting results and solving for $\rho_{22}(t)$, we quickly find
 \begin{subequations}  \labe{rho2_t}
 \begin{eqnarray}
 \rho_{11}(t) &=& e^{-t/\tau_a} ,\\
 \rho_{22}(t) &=& \frac{1 - e^{-\gamma_{22} t}}
 {\gamma_{22}\, \tau_a }\, e^{-t/\tau_a} , \\
 \rho_{21}(t) &=& - \frac{1 - e^{i \left(\nu_a + i \gamma_{21}\right) t}}
 {\nu_a + i \gamma_{21}}\, \Omega_a\, e^{-t/\tau_a} ,
 \end{eqnarray}
 \end{subequations}
 where
 \begin{equation} \labe{taua2}
 \frac{1}{\tau_a} \equiv 2 \Im\left(\tilde{\rho}_{21}\,
\Omega_a^*\right) = \frac{2\, \gamma_{21}
\left|\Omega_a\right|^2}{\nu^2_a +
 \gamma^2_{21}} ,
 \end{equation}
 and $\rho_{ee}(t) = 1 - \rho_{11}(t) - \rho_{22}(t)$.
Therefore, $\trace\left(\rho^2\right) = 1 -
O(|\Omega_a/\gamma_{21}|^2)$, and the quasi-steady-state density
matrix describes a pure state only in the weak-field limit.

In the limit where the photon number $n_a$ is sufficiently small
that $\left|\Omega_a/\gamma_{21}\right|^2 \ll 1$,
$\tilde{\rho}_{21}$ is a valid approximation for $\rho_{21}(t)$
only when the time $t$ satisfies $1/\gamma_{21} \ll t \ll \tau_a$.
It is worth noting that this steady state does not result if the
system parameters are chosen so that the constraint $\gamma_{21}
\tau_a \gg 1$ is not strictly satisfied. In \fig{rho_21_t}, we
have plotted the real and imaginary parts of the off-diagonal
density matrix element $\rho_{21}(t)$ as a function of time for
the case where $\nu_a/\gamma_{21} = 3$. Even though $\gamma_{21}
\tau_a = 55.6$, the magnitude of $\rho_{21}(t)$ decays
significantly on a time scale that is only a few times the
transient lifetime. The characteristic time $\tau_a$ can be
lengthened by increasing the detuning $\nu_a$ or by decreasing the
Rabi frequency $\Omega_a$.

 \begin{figure}
   \centering
   \subfigure[Real part]{\labf{re_rho_21}
 \includegraphics[width=3.25in]{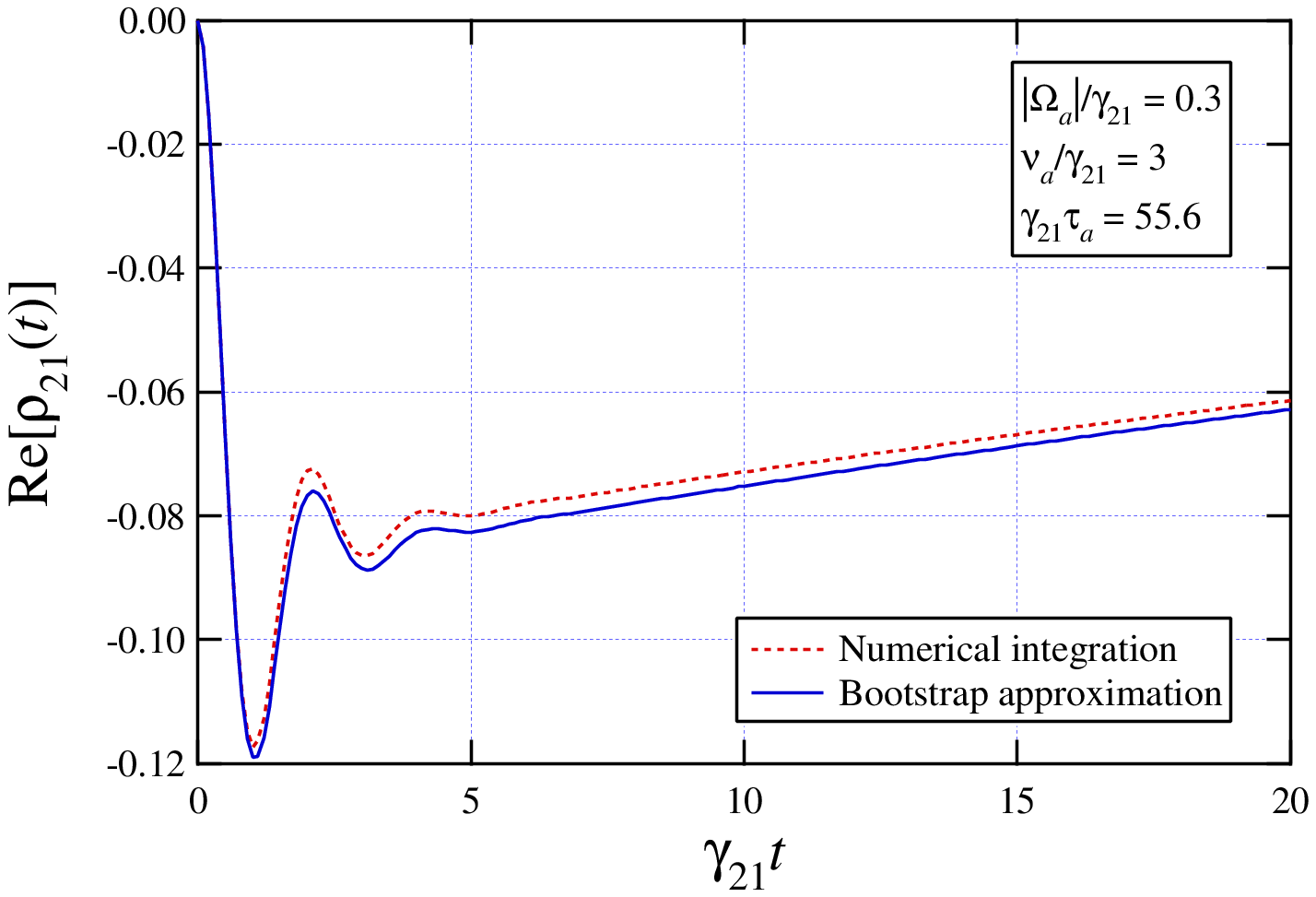}}\quad
   \subfigure[Imaginary part]{\labf{im_rho_21}
 \includegraphics[width=3.25in]{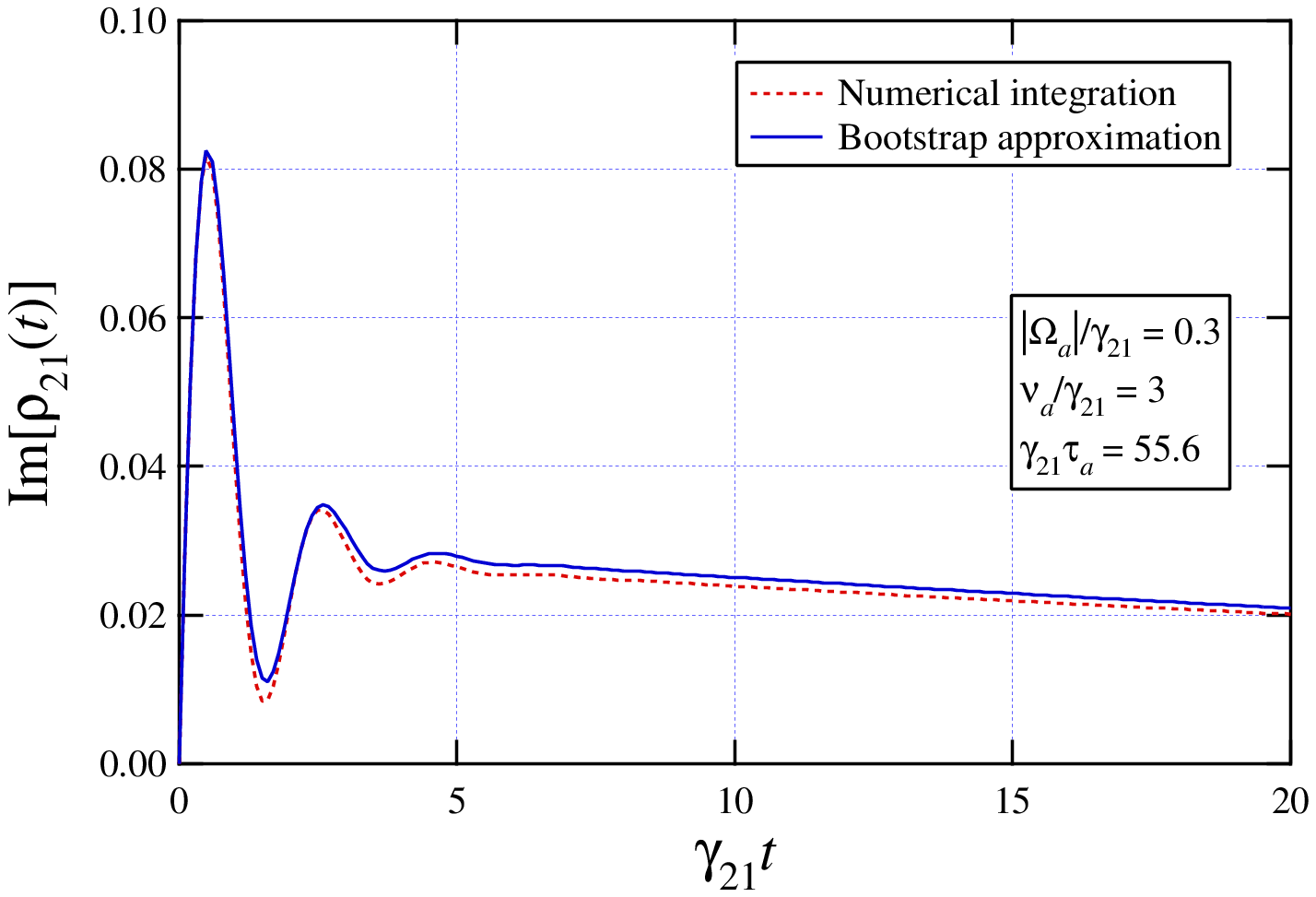}}
   \caption{\labf{rho_21_t} Real and imaginary parts of the off-diagonal
   density matrix element $\rho_{21}(t)$. After transient contributions with frequency
   $\nu_a$ and lifetime $1/\gamma_{21}$, both terms adiabatically
   follow $\rho_{11}(t)$ and decay with the characteristic
   time $1/\tau_a$.}
 \end{figure}

The expectation value of the microscopic polarization of the atom
at $\mathbf{r} = 0$ is simply
\begin{equation} \labe{Poldef}
\left\langle\mathbf{d}(0)\right\rangle = \text{Tr}\left(\rho\,
\mathbf{d}\right) = \tilde{\rho}_{21} \mathbf{d}_{12} +
\tilde{\rho}_{12} \mathbf{d}_{21} ,
\end{equation}
where the first term arises from the annihilation process and the
second from the creation process. We define the corresponding
complex linear susceptibility $\chi^{(1)}\left(-\omega_a,
\omega_a\right)$ in terms of the macroscopic polarization
$\mathbf{P}(0)$ and the associated classical field amplitude $E_a
\equiv 2\, \mathcal{E}_a \sqrt{n_a}$ using the expression
\begin{equation} \labe{Pchidef2}
\mathbf{P}(0) \equiv
\frac{\left\langle\mathbf{d}(0)\right\rangle}{\mathcal{V}} \equiv
\hat{\bm{\epsilon}}_a \frac{\epsilon_0}{2}
\chi^{(1)}\left(-\omega_a, \omega_a\right) E_a + c.c. ,
\end{equation}
where $\mathcal{V}$ is the effective mode volume given by
\eqr{Unorm}. Therefore, we can calculate the complex
susceptibility using the definition
\begin{equation} \labe{chidef2}
\chi^{(1)}\left(-\omega_a, \omega_a\right) \equiv \frac{
\mathcal{V}\, \hat{\bm{\epsilon}}^*_a \cdot
\mathbf{d}_{12}}{\epsilon_0\, \mathcal{E}_a \sqrt{n_a}}\,
\tilde{\rho}_{21} = \frac{2}{\omega_a} \,
\frac{\tilde{\rho}_{21}\, \Omega_a^*}{n_a},
\end{equation}
which, after applying \eqr{gadef} and \eqr{rabiadef} to
\eqr{rho21tilde}, gives
\begin{equation} \labe{chi2}
\chi^{(1)}\left(-\omega_a, \omega_a\right) = - \frac{\nu_a - i
 \gamma_{21}}{\nu^2_a + \gamma^2_{21}}\, \frac{2 \left|\Omega_a\right|^2}{\omega_a n_a}.
 \end{equation}

By convention, the real refractive index $\eta(\omega_a)$ and the
linear absorption coefficient $\kappa(\omega_a)$ are defined in
terms of the real and imaginary parts of the susceptibility as
 \begin{subequations} \labe{chi_2}
 \begin{eqnarray}
 \eta^2\left(\omega_a\right) &\equiv& 1 +
 \Re\left[\chi^{(1)}\left(-\omega_a, \omega_a\right)\right], \nd \\
 \kappa\left(\omega_a\right) &\equiv&
 \frac{\omega_a}{\eta\left(\omega_a\right) c}\,\Im\left[\chi^{(1)}
 \left(-\omega_a, \omega_a\right)\right].
 \end{eqnarray}
 \end{subequations}
 We have plotted the real and imaginary parts of \eqr{chi2} in
 \fig{eta_kappa_2}, with units chosen so that $\kappa(\nu_a = 0) = 1$. Note that the frequency with the largest dispersion (at
 $\nu_a = 0$) corresponds to the frequency with the greatest
 absorption.

 \begin{figure}
   \centering
   \subfigure[Refractive index: $\Re \chi(-\omega_a, \omega_a)$]{\labf{eta_2}
 \includegraphics[width=3.25in]{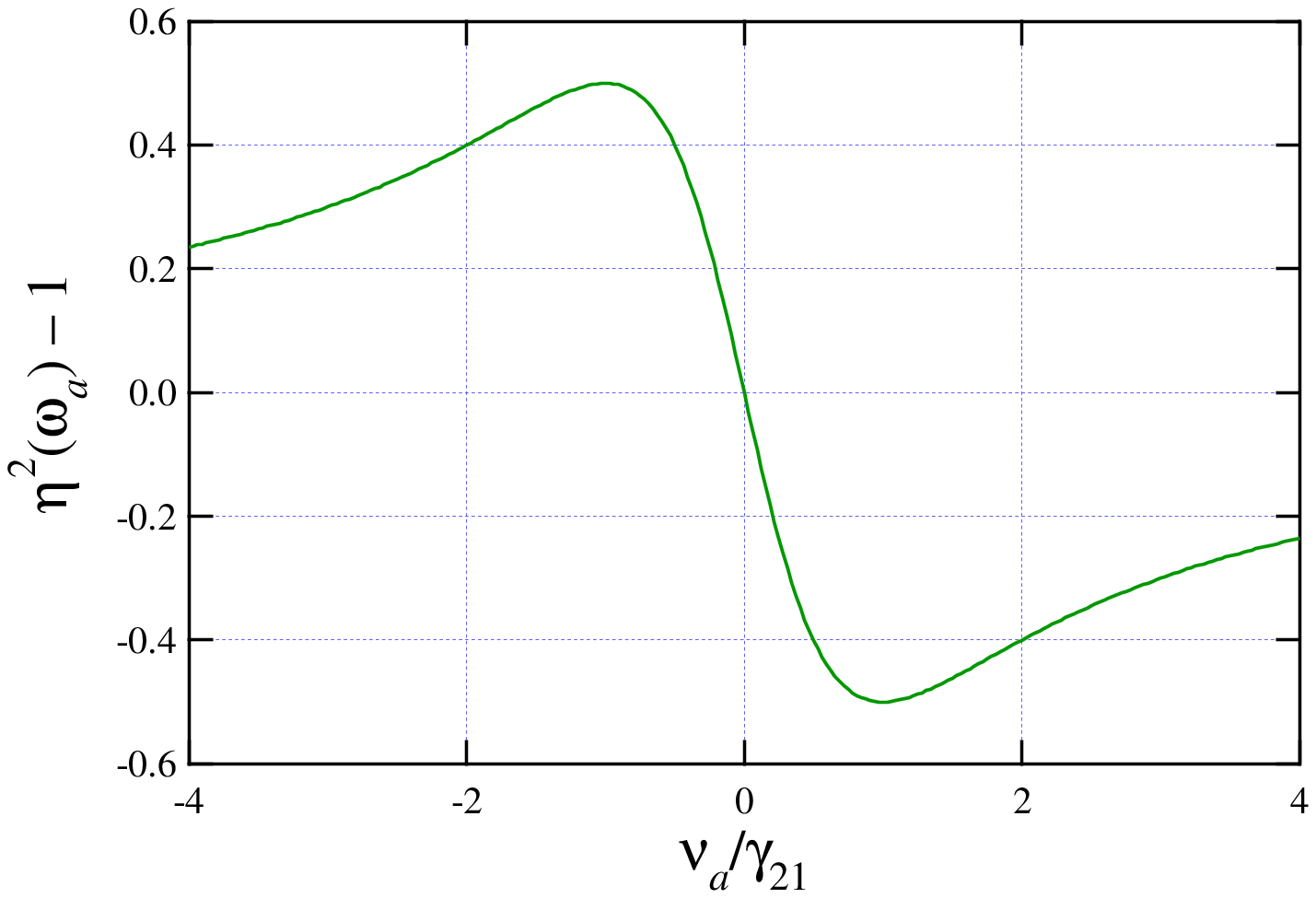}}\quad
   \subfigure[Absorption coefficient: $\Im \chi(-\omega_a, \omega_a)$]{\labf{kappa_2}
 \includegraphics[width=3.25in]{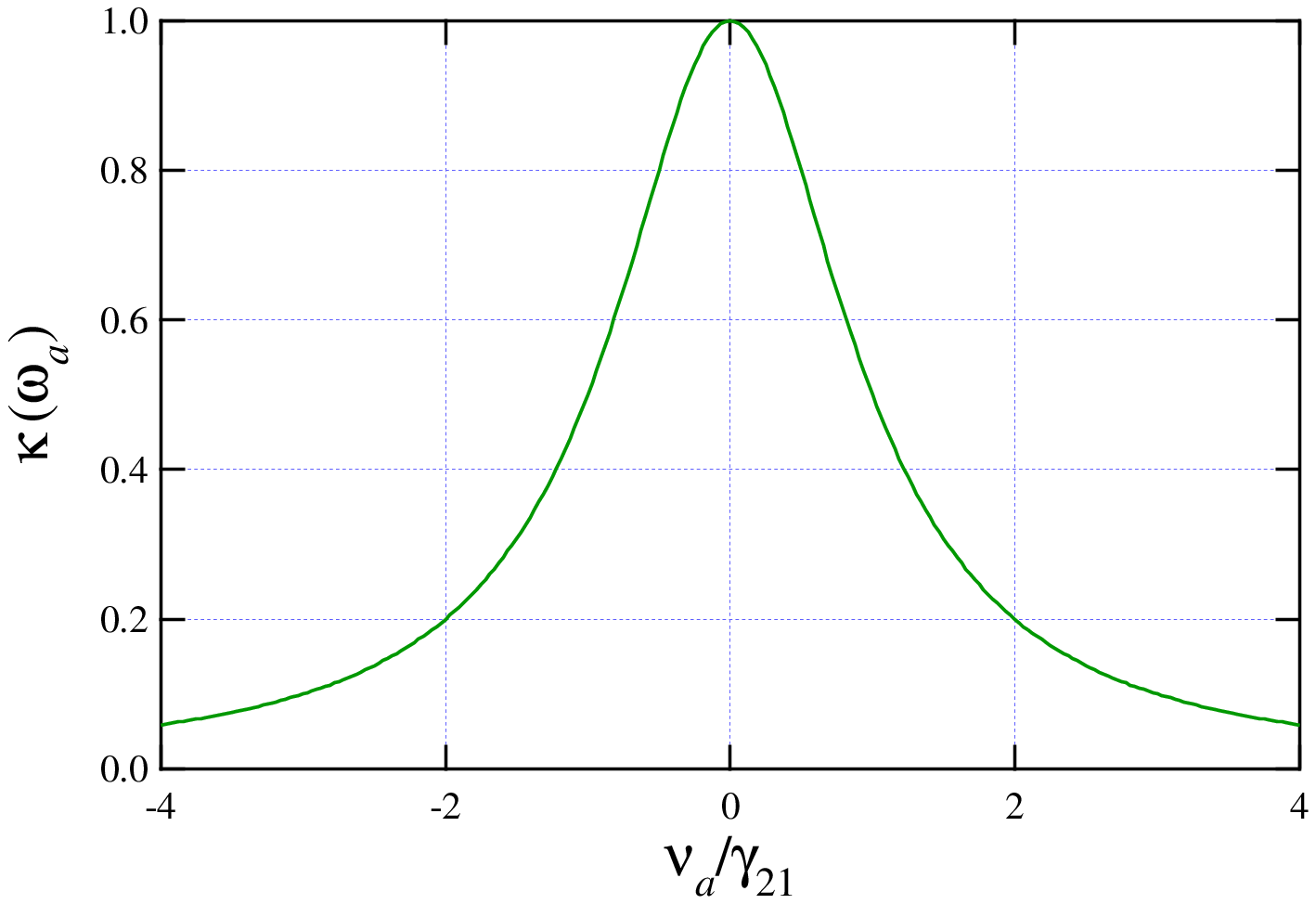}}
   \caption{\labf{eta_kappa_2} Refractive index and linear absorption
   coefficient for the two-level atom shown in \fig{two_level_atom}.
   Both the dispersion and absorption are largest at $\nu_a = 0$.}
 \end{figure}

We can increase the susceptibility defined by \eqr{chidef2} by
increasing the number of atoms placed in the interaction region
shown in \fig{resonator}. We assume that $N$ atoms are at rest
near $\mathbf{r} = 0$ in a volume that is small compared to
$\mathcal{A}\, z_R$, where $\mathcal{A} \equiv \half \pi w_0^2$ is
the effective area of the fundamental laser mode at the beam
waist, and $z_R \equiv \pi w_0^2/\lambda$ is the Rayleigh length
of the mode. Then, in a traveling-wave cavity, the Rabi frequency
of each atom has a negligible spatial dependence, and is given by
\eqr{rabiadef}. In the unsaturated, weak-field case, the lower
level of the quantum manifold shown in \fig{quantum_2} has become
$N$-fold degenerate, since any one of the $N$ atoms can be excited
to the upper atomic level via resonant excitation. We therefore
define the unperturbed $N$-atom basis
 \begin{equation}
 \begin{split}
 \left\{ \right. & \ket{\{1\}, n_a, 0} , \\
 & \ket{\left\{2^{(k)}\right\}, n_a - 1, 0} , \\
 & \ket{\{1\}, n_a - 1, 1} \left. \right\}
 \end{split}
 \end{equation}
where the entry $\{1\}$ represents the $N$-element string $1, 1,
\dots , 1$, describing all atoms in the ground state $\ket{1}$,
and $\left\{2^{(k)}\right\}$ represents the same string, with the
element at position $k$ replaced by a `2', indicating that atom
$k$ has been excited to the upper level $\ket{2}$. In this basis,
we write the $(N + 2) \times (N + 2)$ density matrix as
 \begin{equation} \labe{rho2en}
 \rho = \begin{bmatrix}
 \rhx{1}{1} & \rhx{1}{2^{(1)}} & \rhx{1}{2^{(2)}} & \cdots & \rhx{1}{e} \\
 \rhx{2^{(1)}}{1} & \rhx{2^{(1)}}{2^{(1)}} & \rhx{2^{(1)}}{2^{(2)}} & \cdots & \rhx{2^{(1)}}{e} \\
 \rhx{2^{(2)}}{1} & \rhx{2^{(2)}}{2^{(1)}} & \rhx{2^{(2)}}{2^{(2)}} & \cdots & \rhx{2^{(2)}}{e} \\
 \vdots   &   \vdots   &  \vdots    & \ddots & \vdots  \\
 \rhx{e}{1} & \rhx{e}{2^{(1)}} & \rhx{e}{2^{(2)}} & \cdots & \rhx{e}{e} \\
 \end{bmatrix}
 \end{equation}
and, if we neglect any interactions between the atoms, the
Hamiltonian as
 \begin{equation} \labe{H2en}
 H = -\hbar\, \begin{bmatrix}
 0        & \Omega_a^* & \Omega_a^* & \cdots & 0 \\
 \Omega_a & \nu_a      & 0          & \cdots & 0 \\
 \Omega_a & 0          & \nu_a      & \cdots & 0 \\
 \vdots   & \vdots     & \vdots     & \ddots & \vdots  \\
 0        & 0          & 0          & \cdots & 0
 \end{bmatrix} .
 \end{equation}

We can construct the $N$-atom Lindblad decoherence operator
$\Gamma(\rho)$ by allowing each atom to scatter an absorbed photon
to the environment at the rate $\gamma_{2}^\prime$ (independent of
position), and we assume that the pure dephasing of the state
$\ket{j, \dots}$ occurs at the rate $\gamma^{\prime \prime}_{j}$.
If we repeat the single-atom quasi-steady-state approach that led
to \eqs{rho2_t}, then in the limit
$\left|\Omega_a\right|/\gamma_{21} \ll 1$ in the noninteracting
$N$-atom case, we obtain
 \begin{subequations}  \labe{rho2_ss_n}
 \begin{eqnarray}
\rho_{11}(t) &\longrightarrow& \rhx{1}{1}(t) , \\ \rho_{22}(t)
&\longrightarrow& \rhx{2^{(k)}}{2^{(k)}}(t) , \nd
\\ \rho_{21}(t) &\longrightarrow& \rhx{2^{(k)}}{1}(t) ,
 \end{eqnarray}
 \end{subequations}
where \eqs{rho2_ss_n} have the same form as \eqs{rho2_t} ---
independent of $k$ --- but with the new time constant
 \begin{equation} \labe{tauan2}
 \frac{1}{\tau_a(N)} \equiv \frac{2\, N\, \gamma_{21} \left|\Omega_a\right|^2}{\nu^2_a +
 \gamma^2_{21}} .
 \end{equation}
Note that we have implicitly assumed that the population of the
system ground state has not been significantly depleted (i.e., $N
\rho_{22}(t) \ll \rho_{11}(t)$).

Extending \eqr{Poldef} to the $N$-atom ensemble near $\mathbf{r} =
0$, we obtain an expectation value of the microscopic polarization
given by
 \begin{equation} \labe{PoldefN}
 \begin{split}
 \left\langle\mathbf{d}_N (0)\right\rangle = \text{Tr}\left(\rho\, \mathbf{d}\right) &=
 \sum_{k = 1}^N \left(\tilde{\rho}_{\{2^{(k)}\}\{1\}} \mathbf{d}_{12} + \tilde{\rho}_{\{1\}\{2^{(k)}\}}
 \mathbf{d}_{21}\right) \\
 &= \tilde{\rho}_{21}^{\{N\}} \mathbf{d}_{12} + \tilde{\rho}_{12}^{\{N\}}
 \mathbf{d}_{21} ,
 \end{split}
 \end{equation}
where, in the quasi-steady-state regime where the time $t$
satisfies $1/\gamma_{21} \ll t \ll \tau_a(N)$,
 \begin{equation} \labe{rho2_21_n}
 \tilde{\rho}_{21}^{\{N\}} \cong -\frac{N\, \Omega_a}{\nu_a + i
 \gamma_{21}}  = - \frac{\nu_a - i
 \gamma_{21}}{\nu^2_a + \gamma^2_{21}}\, N\, \Omega_a.
 \end{equation}
Note that $\rho_{21}^{\{N\}}$ scales linearly with $N$ only in the
weak-field limit where the probability that more than one atom has
been excited to the upper level is negligible.

In \sct{qip2}, we will calculate both the quantum phase shift and
the scattering rate encountered by a single photon optically
coupled with one or more two-level atoms in the interaction region
of the model resonator shown in \fig{resonator}. We can anticipate
those results now using the definition of the complex
susceptibility given by \eqr{chidef2} in a semiclassical
calculation. Suppose that our resonator encloses a coherent state
with $\alpha(t) = \sqrt{n_a} e^{-i \omega t}$ and contains $N$
atoms in an interaction region of length $l$ having the effective
volume $\half \pi w_0^2 l$. Then we can approximate the total
phase shift per round trip of the corresponding electromagnetic
field as
 \begin{equation}
 \Delta \varphi \approx \frac{\Delta \eta\left(\omega_a\right) \omega_a l}{c} ,
 \end{equation}
where $ \Delta \eta\left(\omega_a\right) \equiv
\eta\left(\omega_a\right) - 1 \ll 1$, and the fraction of the
circulating electromagnetic power absorbed per round trip as
 \begin{equation}
 \frac{\Delta P_a}{P_a} \approx \kappa\left(\omega_a\right) l .
 \end{equation}
If we substitute \eqr{rho2_21_n} into \eqr{chidef2}, then we
obtain the $N$-atom susceptibility
$\chi^{(1)\{N\}}\left(-\omega_a, \omega_a\right)$, and the
corresponding refractive index change $\Delta
\eta\left(\omega_a\right) \cong
 \half \Re\left[\chi^{(1)\{N\}}\left(-\omega_a, \omega_a\right)\right]$.
If $l \ll L$, then the round-trip time is given by $\Delta t =
L/c$, and we obtain for the corresponding classical
electromagnetic field amplitude at $\mathbf{r} = 0$
 \begin{equation} \labe{Easc}
 E(t) = E(0)\, e^{-i \left(\omega_a - W_a\right) t}
 \end{equation}
where the complex frequency shift $W_a$ is given by
 \begin{equation}
 W_a \equiv \frac{\Delta \varphi}{\Delta t} + i \frac{\Delta P_a}{2 P_a \Delta t}
 = \frac{\omega_a}{2}\, \chi^{(1)\{N\}}\left(-\omega_a,
 \omega_a\right) ,
 \end{equation}
or, after applying \eqr{chidef2},
 \begin{equation} \labe{Wadef}
 W_a = \tilde{\rho}^{\{N\}}_{21}\, \frac{\Omega_a^*}{n_a} .
 \end{equation}
Therefore, in the classical linear-optical limit, the mean phase
shift accumulated by the $n_a$-photon coherent state at frequency
$\omega_a$ after an elapsed time $t$ is
 \begin{equation}
 \varphi(t) = \Re\left(W_a\right) t = -\frac{\nu_a}{\nu_a^2 +
 \gamma_{21}^2}\, \frac{N \left|\Omega_a\right|^2}{n_a}\, t ,
 \end{equation}
and the mean rate at which photons are absorbed and scattered by
the atoms is, as expected,
 \begin{equation}
 2 \Im\left(W_a\right) n_a = 2 \Im\left(\tilde{\rho}^{\{N\}}_{21} \Omega_a^*\right) = \frac{1}{\tau_a(N)}
 ,
 \end{equation}
where $\tau_a(N)$ is given by \eqr{tauan2}. We see that --- in the
limit $\gamma_{22}\, \tau_a(N) \gg 1$ --- this scattering rate is
equivalent to $\gamma_{22}\, \tilde{\rho}^{\{N\}}_{22}$, as
expected from \eqr{rho2dot_22}. This semiclassical result is
entirely consistent with that of the $N$-atom quantum calculation
found in \eqr{rho2_ss_n}, as predicted by extending
\eqr{rho2dot_11} to the $N$-atom case.

 \begin{figure}
   \centering
   \subfigure[Semiclassical energy levels]{\labf{lambda_3}
 \includegraphics[width=3.25in]{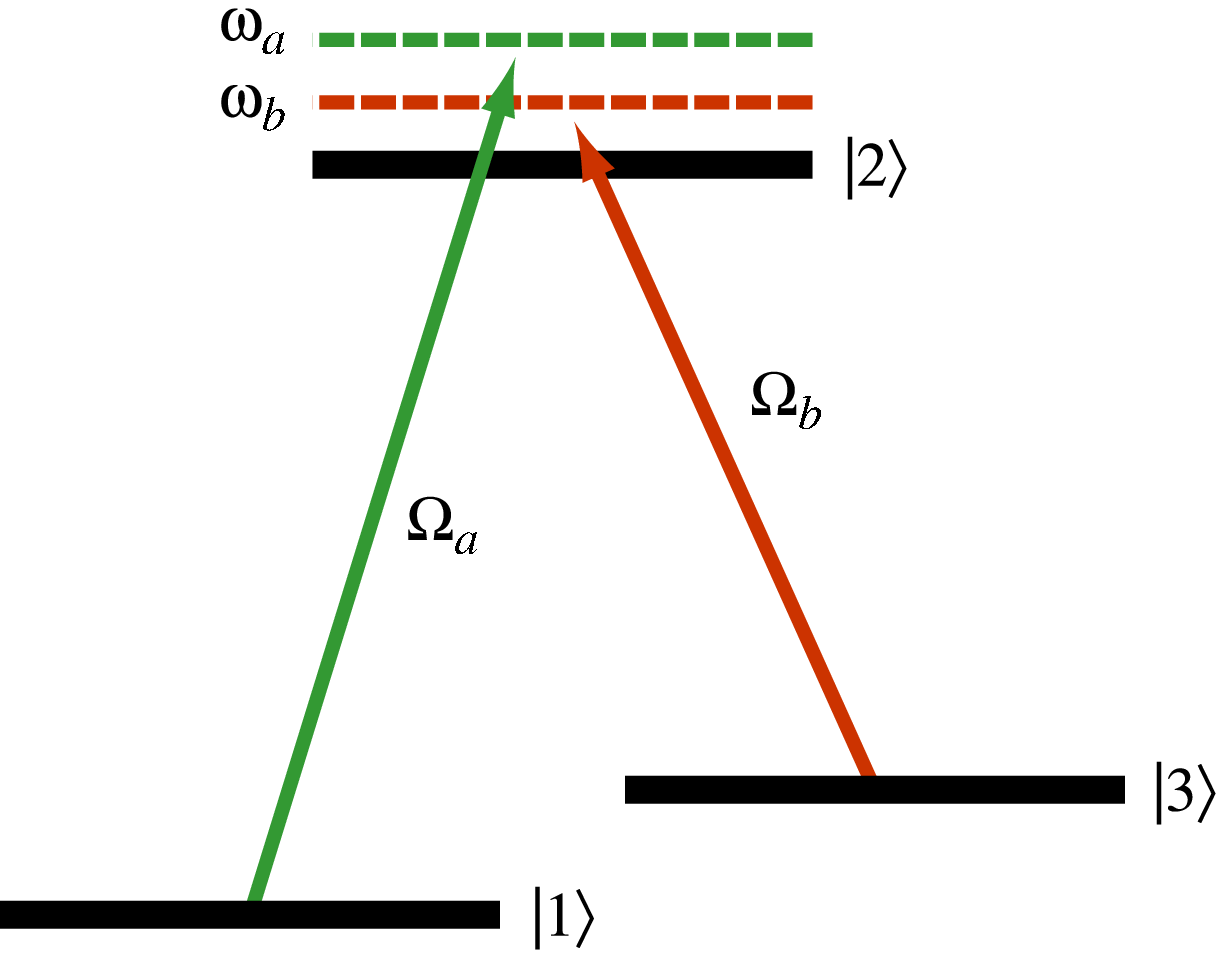}}\quad
   \subfigure[Quantum energy manifold]{\labf{quantum_3}
 \includegraphics[width=3.25in]{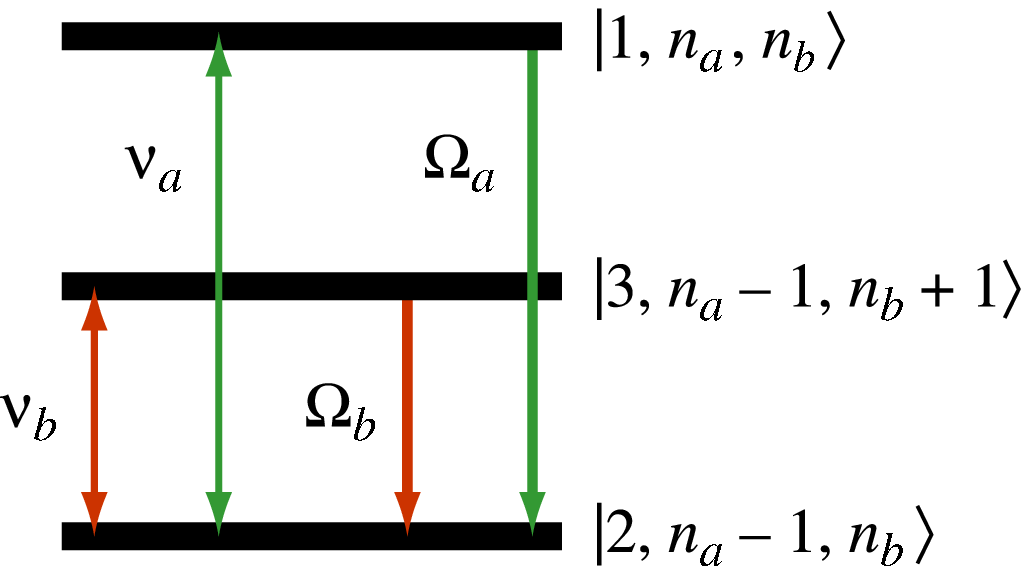}}
   \caption{\labf{three_level_atom} Interaction between a three-level atom and a
    nearly resonant two-frequency
   electromagnetic field. Note that the annihilation of a
   photon of frequency $\omega_k$ is represented by the complex number $\Omega_k$.}
 \end{figure}

 \subsection{Transparency of the Three-Level $\Lambda$ Atom\labs{eit3}}
In \sct{eit2}, we developed a formalism to describe the quantum
optics of a two-level atom. In this formalism, the effective Rabi
frequency $\Omega_j$ represents the annihilation of a photon in
mode $j$, and $\nu_j$ is the diagonal element of $H$ corresponding
to the (positively) detuned interaction between mode $j$ and the
adjacent atomic transition in a given manifold of the energy-level
diagram of the total system. In this section, we will follow the
same formalism to analyze the quantum optical properties of the
three-level atom shown in \fig{three_level_atom}. Note that the
upper atomic energy level $\ket{2}$ and the new metastable level
$\ket{3}$ are coupled by a control field with angular frequency
$\omega_b \approx \omega_2 - \omega_3$. It is the destructive
quantum interference established by this control field that
results in transparency (i.e., vanishing absorption) for the probe
field at $\omega_a$.

As in the case of the two-level atom, we reduce the semiclassical
atomic system depicted in \fig{lambda_3} to the quantum manifold
given by \fig{quantum_3}. We will work in a manifold corresponding
to an atom in state $\left|1\right>$, with $n_a$ photons in mode
$a$ and $n_b$ photons in mode $b$. We again extend our basis to
include energy dissipation to the environment by appending an
entry to each product state, indicating the occurrence of
scattering of a photon of frequency $\omega_a$ or $\omega_b$.
Therefore, the environment can be represented by the nonresonant
submanifold $\left\{ \ket{1, n_a - 1, n_b, 1}, \ket{1, n_a, n_b -
1, 1}\right\}$ that captures dissipated energy and preserves the
trace of the density matrix. Referencing \sct{eit2}, by inspection
in the extended unperturbed Schr\"{o}dinger basis
 \begin{equation} \labe{fock_3e}
 \begin{split}
 \left\{ \right. & \ket{1, n_a, n_b, 0} , \\
 & \ket{2, n_a - 1, n_b, 0} , \\
 & \ket{3, n_a - 1, n_b + 1, 0} , \\
 & \left\{ \ket{1, n_a - 1, n_b, 1}, \ket{1, n_a, n_b - 1, 1}\right\} \left. \right\}
 \end{split}
 \end{equation}
we obtain the total Hamiltonian
 \begin{equation} \labe{H3}
 H = -\hbar\, \begin{bmatrix}
 0 & \Omega_a^* & 0 & 0 \\
 \Omega_a & \nu_a & \Omega_b & 0 \\
 0 & \Omega_b^* & \nu_a - \nu_b & 0 \\
 0 & 0 & 0 & 0
 \end{bmatrix} ,
 \end{equation}
where we have defined the detuning parameter
 \begin{equation}
 \labe{detunb}
 \nu_b \equiv \omega_b - \left(\omega_2 - \omega_3\right) ,
 \end{equation}
 the effective coupling constant
 \begin{equation}
 \labe{gbdef}
  g_b \equiv \hat{\bm{\epsilon}}_b \cdot \mathbf{d}_{23}\,
  \mathcal{E}_b,
 \end{equation}
 and the effective Rabi frequency
 \begin{equation}
 \labe{rabibdef}
 \Omega_b = \frac{g_b}{\hbar}  \sqrt{n_b + 1} ,
 \end{equation}
 and we have subtracted the energy
 \begin{equation*}
 E_0 \equiv \bra{1, n_a, n_b, 0} H_0 \ket{1, n_a, n_b, 0}
 \end{equation*}
  from all diagonal terms.

 The dynamics of a strongly-coupled system (e.g., a system having intracavity fields
 that are sufficiently intense that dephasing can be ignored) is often described
 using dressed states,\cite{cohe92} where the submatrix of the
 Hamiltonian corresponding to the vectors $\ket{2, n_a - 1, n_b, 0}$ and
 $\ket{3, n_a - 1, n_b + 1, 0}$ is diagonalized in the case of perfectly resonant tuning.
 For our purposes, weak fields and linearized
 (Lindblad) decoherence models generally allow the unperturbed
 eigenvectors to serve as a reasonably accurate basis set.
 For example, if we diagonalize the Hamiltonian given by \eqr{H3}
 with $\nu_a = \nu_b = 0$, we find the nontrivial eigenvalues
 \begin{subequations}
 \begin{eqnarray}
 \Omega_0 &=& 0 , \\
 \Omega_- &=& -\Omega_R , \nd \\
 \Omega_+ &=& +\Omega_R ,
 \end{eqnarray}
 \end{subequations}
where $\Omega_R^2 \equiv \left|\Omega_a\right|^2 +
\left|\Omega_b\right|^2$, and the nonzero eigenvectors
 \begin{subequations}
 \begin{eqnarray}
 \ket{0} &=& \left\{-\frac{\Omega_b}{\Omega_R}, 0, \frac{\Omega_a}{\Omega_R}, 0\right\}, \\
 \ket{-} &=& \frac{1}{\sqrt{2}} \left\{\frac{\Omega_a^*}{\Omega_R}, 1,  \frac{\Omega_b^*}{\Omega_R}, 0\right\}, \nd \\
 \ket{+} &=& \frac{1}{\sqrt{2}} \left\{\frac{\Omega_a^*}{\Omega_R}, -1,  \frac{\Omega_b^*}{\Omega_R}, 0\right\},
 \end{eqnarray}
 \end{subequations}
in the basis of \eqr{fock_3e}. Therefore, if we assume that the
system is entirely in the ground state at $t = 0$, we find
 \begin{equation}
 \begin{split}
 \ket{\psi(t)} &= -\frac{\Omega_b^*}{\Omega_R} \ket{0} \\ &+
 \frac{\Omega_a}{\sqrt{2} \Omega_R} \left(\ket{-} e^{i\, \Omega_R t} + \ket{+} e^{-i\, \Omega_R
 t}\right) .
 \end{split}
 \end{equation}
Therefore, in the strongly-coupled case, the population of the
state $\ket{2, n_a - 1, n_b, 0}$ is given by
$\left|\Omega_a/\Omega_R\right|^2 \sin^2\left(\Omega_R t\right)$.
We anticipate, then, that in a weakly coupled system with
appreciable decoherence this population will remain small for all
$t$ if $\left|\Omega_a/\Omega_b\right|^2 \ll 1$ and/or
$\left|\Omega_R\right| \gg \gamma_{21}$.

We now generalize the phenomenological discussion of decoherence
presented in \sct{eit2} to describe more complex atomic
energy-level schemes. We define $\gamma^\prime_{j}$ as the total
free-space depopulation rate to the environment of product state
$\ket{j, \dots}$ arising both from spontaneous emission from
$\ket{j}$ to all lower atomic levels and from transmission and
scattering losses in the cavity, and $\gamma^{\prime \prime}_{j}$
as the pure dephasing rate for the state $\ket{j, \dots}$. In
general, the decoherence coefficient $\gamma_{i j}$ of the term
$\gamma_{i j}\, \rho_{i j}$ appearing in the Lindblad decoherence
operator given by \eqr{Lindblad} can be quickly written down using
a straightforward set of rules:
 \begin{enumerate}
 \item In all cases, $\gamma_{i j} = \gamma_{j i}$.
 \item If $i = e$ and $j \ne e$, then
 \begin{equation}
 \gamma_{ej} = \half \gamma_{j}^\prime + \gamma_{j}^{\prime \prime}
 .
 \end{equation}
 \item If $i \ne e$ and $i = j$, then
 \begin{equation}
 \gamma_{jj} = \gamma_{j}^\prime .
 \end{equation}
 \item If $i, j \ne e$ and $i \ne j$, then
 \begin{equation}
 \gamma_{ij} = \half \left(\gamma_{i}^\prime + \gamma_{j}^\prime\right)
  + \gamma_{i}^{\prime \prime} + \gamma_{j}^{\prime \prime} .
 \end{equation}
 \item The value of the term corresponding to $i = j = e$ in
 $\Gamma(\rho)$ is chosen to ensure that $\trace[\Gamma(\rho)] = 0$.
 \end{enumerate}

In the following analysis, we assume that atomic level $\ket{3}$
is metastable so that $\gamma^\prime_{3} = 0$. We substitute
\eqr{H3} and \eqr{Lindblad} into \eqr{rhodot} and seek the
quasi-steady-state solution in the unsaturated weak-field limit
$|\Omega_a/\gamma_{21}|^2 \ll 1$, assuming that $\rho_{22}(t) \ll
\rho_{11}(t)$ and $\Omega_a \rho_{32}(t) \ll \Omega_b^*
\rho_{31}(t)$ for all $t > 0$. Then we obtain $\tilde{\rho}_{11}
\cong 1$, $\tilde{\rho}_{k 2} = \tilde{\rho}^*_{2 k} \cong 0$,
$\tilde{\rho}_{k 3} = \tilde{\rho}^*_{3 k} \cong 0$ (where $k \in
\{2, 3\}$), and
 \begin{subequations} \labe{rho_3_ss}
 \begin{eqnarray}
 \tilde{\rho}_{21} &\cong& -\frac{(\nu_a - \nu_b + i \gamma_{31})\,
 \Omega_a}{(\nu_a + i \gamma_{21})(\nu_a - \nu_b + i \gamma_{31}) -
 \left|\Omega_b\right|^2} , \labe{rho_21_3} \\
 \tilde{\rho}_{31} &\cong& \frac{\Omega_a\, \Omega_b^*}{(\nu_a + i
 \gamma_{21})(\nu_a - \nu_b + i \gamma_{31}) -
 \left|\Omega_b\right|^2} ,
 \end{eqnarray}
 \end{subequations}
with $\tilde{\rho}_{12} = \tilde{\rho}^*_{21}$ and
$\tilde{\rho}_{13} = \tilde{\rho}^*_{31}$, where $\gamma_{21}$ is
given by \eqr{gamma21def} and
 \begin{equation} \labe{gamma31def}
 \gamma_{31} \equiv \gamma^{\prime \prime}_{1}  + \gamma^{\prime \prime}_{3} .
 \end{equation}
Since \eqr{rho2dot_11} remains valid for the three-level atom +
photons system, we can follow the same bootstrap procedure to
obtain the approximate solution for $\rho_{11}(t)$ given by
\eqr{rho11apprx}, and
 \begin{subequations}
 \begin{eqnarray}
\rho_{21}(t) &\cong& \tilde{\rho}_{21} \left(1 - e^{-\gamma_{21}
t}\right) \rho_{11}(t) , \nd \\
\rho_{31}(t) &\cong& \tilde{\rho}_{31} \left(1 - e^{-\gamma_{31}
t}\right) \rho_{11}(t) .
 \end{eqnarray}
 \end{subequations}
Therefore, the steady-state solutions given by \eqs{rho_3_ss} are
valid at any time $t$ where the laser parameters have been chosen
to allow the inequality
 \begin{equation} \labe{sstineq3}
1/\gamma_{21}, 1/\gamma_{31} \ll t \ll \tau_a \equiv \left[2
\Im(\tilde{\rho}_{21} \Omega_a^*) \right]^{-1}
 \end{equation}
to be satisfied.

 \begin{figure}
   \centering
   \subfigure[Refractive index: $\Re \chi(-\omega_a, \omega_a)$]{\labf{eta_3}
 \includegraphics[width=3.25in]{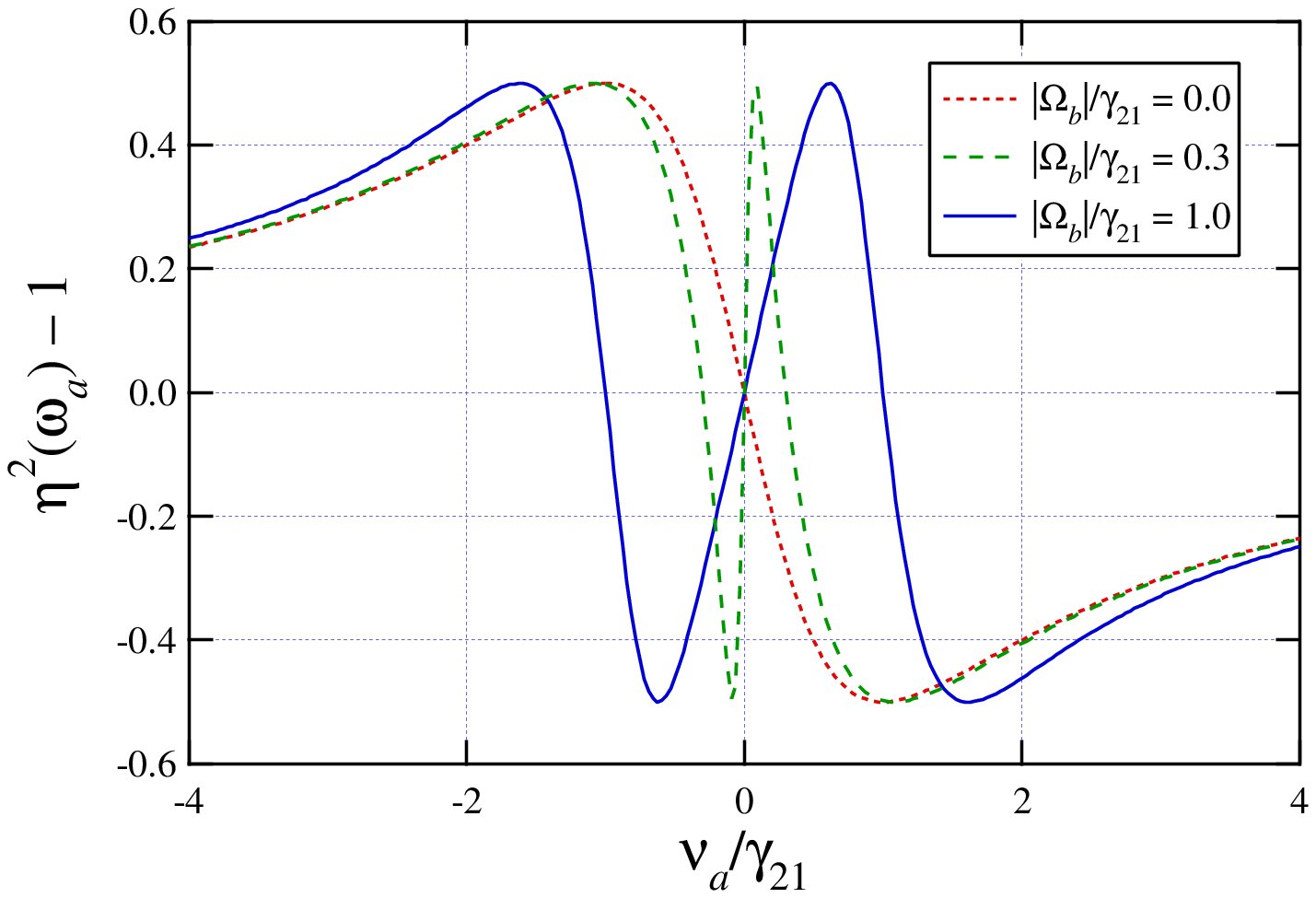}}\quad
   \subfigure[Absorption coefficient: $\Im \chi(-\omega_a, \omega_a)$]{\labf{kappa_3}
 \includegraphics[width=3.25in]{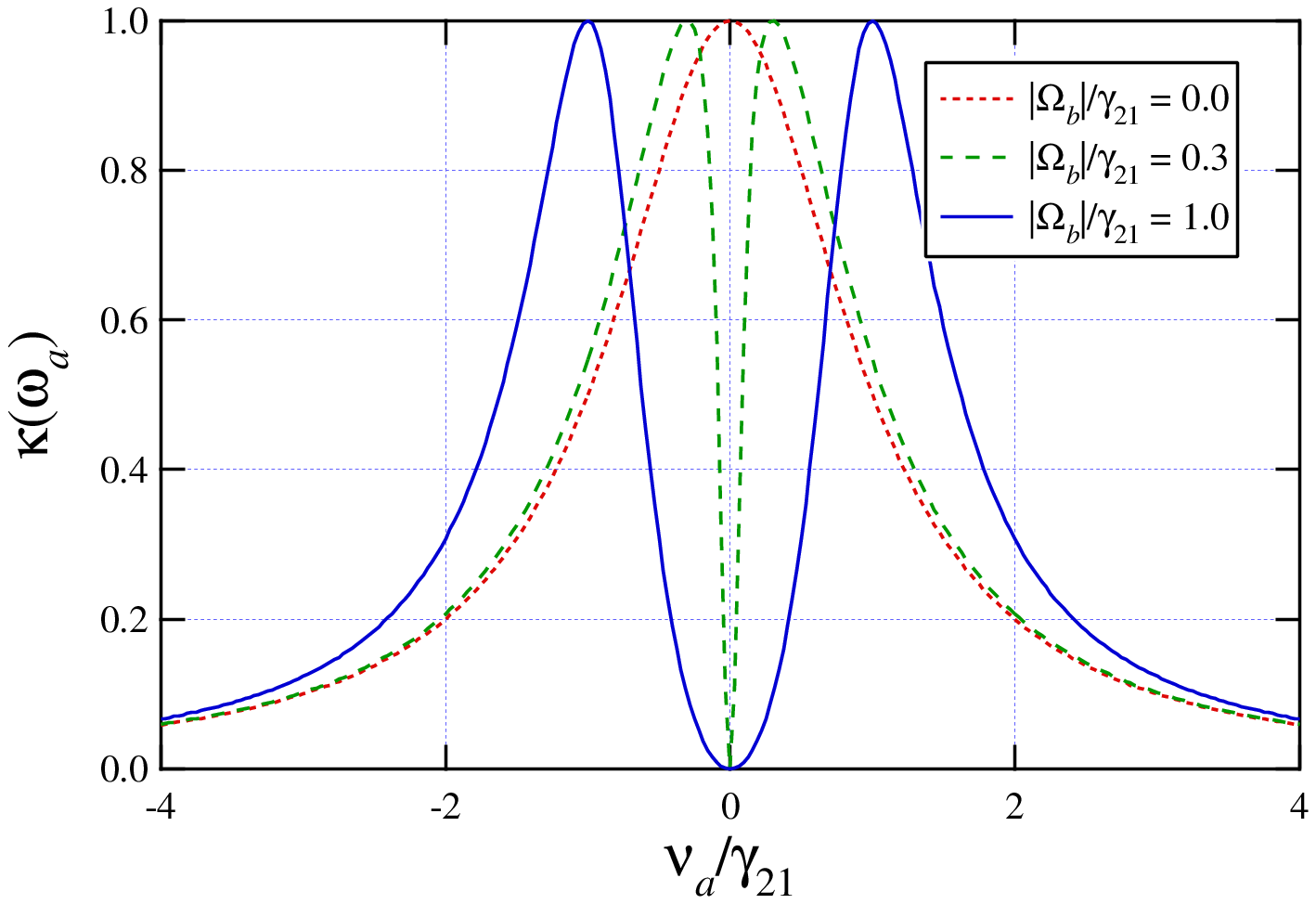}}
   \caption{\labf{eta_kappa_3} Refractive index and linear absorption
   coefficient for the three-level atom shown in \fig{three_level_atom},
   with $\nu_b = \gamma_{31} = 0$. Both the dispersion and transmission
   window are sharpest when $\left|\Omega_b/\gamma_{21}\right|^2 \ll 1$.}
 \end{figure}

Substituting \eqr{rho_21_3} into \eqr{chidef2}, we obtain the
susceptibility (linear in $E_a$)
 \begin{equation} \labe{chi3}
 \begin{split}
\chi^{(1)}&\left(-\omega_a, \omega_a\right) = -\frac{2
\left|\Omega_a\right|^2}{\omega_a n_a} \\ &\times \frac{(\nu_a -
\nu_b + i \gamma_{31})}{(\nu_a + i \gamma_{21})(\nu_a - \nu_b + i
\gamma_{31}) -
 \left|\Omega_b\right|^2} .
 \end{split}
 \end{equation}
 When $\gamma_{31} \rightarrow 0$, the new detuning terms in
\eqr{chi3} have remarkable implications for both the refractive
index and the absorption coefficient given by \eqr{chi_2}. In
\fig{eta_kappa_3}, we assume that the control field has been tuned
to the resonance frequency of the $\ket{3} \rightarrow \ket{2}$
atomic transition so that $\nu_b = 0$, and we plot the real and
imaginary parts of the complex susceptibility as a function of the
normalized detuning $\nu_a/\gamma_{21}$ for several values of the
normalized control Rabi frequency $|\Omega_b|/\gamma_{21}$. When
$|\Omega_b| = 0$, the absorption and dispersion curves reduce to
those of the two-level atom shown in \fig{eta_kappa_2}. However,
when $|\Omega_b| > 0$, near $\nu_a = 0$ the absorption vanishes
completely over a frequency range with a full width at
half-maximum (FWHM) of
 \begin{equation} \labe{fwhm}
\Delta \nu_a = \sqrt{4 \left|\Omega_b\right|^2 + \gamma_{21}^2} -
\gamma_{21} .
 \end{equation}
 Over the same frequency range, the dispersion of the refractive
 index $d \eta(\omega)/d\omega$ is proportional to
$\gamma_{21}/|\Omega_b|^2$, resulting in a significant increase in
the group refractive index $\eta_g(\omega) \cong \eta(\omega) +
\omega\, d \eta(\omega)/d\omega$ and a corresponding reduction in
the group velocity $c/\eta_g(\omega_a)$ at frequency $\omega_a$.
However, the dispersion cannot be made arbitrarily large by
reducing the amplitude of the control field, because in the limit
$|\Omega_b/\gamma_{21}|^2 \ll 1$ the width of the transparency
window given by \eqr{fwhm} becomes $\Delta \nu_a \cong 2
|\Omega_b|^2/\gamma_{21}$. Instead, the magnitude of the control
field must be chosen to allow all significant spectral components
of the probe pulse with carrier frequency $\omega_a$ to be
transmitted with the maximum possible dispersion. (In an
inhomogeneously broadened medium such as a dilute gas, only those
atoms with Doppler-shifted resonance frequencies that are
coincident with $\omega_a$ influence the group velocity of the
probe beam.\cite{java01,lee02})

Note that the transparency predicted by \eqr{rho_21_3} arises
whenever $\nu_a = \nu_b$, generating a pathway for a ``virtual
transition'' between the unperturbed atomic energy levels
$\ket{1}$ and $\ket{3}$. In the steady-state case, the absorption
by the atom of a photon of frequency $\omega_a$ \emph{never}
occurs, in the sense that \emph{a measurement of the state of the
system will never find the atom in the level} $\ket{2}$. Instead,
the control field creates a coherent superposition of the $\ket{1}
\rightarrow \ket{2}$ and $\ket{3} \rightarrow \ket{2}$ paths in
Hilbert space such that destructive interference effectively
reduces the $\ket{1} \rightarrow \ket{2}$ transition rate to zero.

We can estimate the Rabi frequency resulting from a particular
choice of experimental parameters by noting that both the
spontaneous emission rate $A_{21}$ given by \eqr{einsteina21} and
the Rabi frequency given by \eqr{rabiadef} depend on the electric
dipole matrix element $\mathbf{d}_{21}$. If we assume that
$\hat{\bm{\epsilon}}_a \cdot \mathbf{d}_{21} =
\left|\mathbf{d}_{21}\right|$, then we find that
 \begin{equation} \labe{Rabirule}
 \left|\Omega_a\right|^2 = \frac{1}{8 \pi}\,
 \frac{\sigma_a}{\mathcal{A}}\, A_{21}\, \Delta \omega_r\, n_a
 ,
 \end{equation}
where $\sigma_a \equiv 3 \lambda_a^2/2 \pi$ is the resonant atomic
absorption cross section at wavelength $\lambda_a \cong 2 \pi
c/\omega_{21}$,\cite{cohe92} $\mathcal{A} \equiv \half \pi w_0^2$
is the effective laser mode cross-sectional area, and $\Delta
\omega_r \equiv 2 \pi c/L$ is the free spectral range of the ring
resonator. If we were simulating the adiabatic interaction of a
pulsed laser field with a stationary atom, then $\Delta \omega_r$
would represent the bandwidth of the pulse profile
function,\cite{blow90,vane02,loud00,chan02,domo02} although more
complex transients can arise when the interaction is
non-adiabatic.\cite{gree01} In the weak-control-pulse case where
$\left|\Omega_b\right|^2/\gamma_{21}^2 \ll 1$, we would require
that $\Delta \omega_r < \Delta \nu_a \cong 2
\left|\Omega_b\right|^2/\gamma_{21}$ for maximum transmission of
the signal pulse at frequency $\omega_a$. If, as an example, we
choose $\Delta \omega_r = \left|\Omega_b\right|^2/\gamma_{21}$,
the Rabi frequency required to open a sufficiently large
transparency window is
 \begin{equation}
\left|\Omega_b\right|^2 = 8 \pi\, \frac{\mathcal{A}}{\sigma_a}\,
\frac{\gamma_{21}}{A_{21}}\, \frac{\left|\Omega_a\right|^2}{n_a}
 \end{equation}
Assuming realistic optical focusing parameters, in free space the
optimum value of the ratio $\sigma_a/\mathcal{A}$ is about 20\%
for Gaussian beams,\cite{vane00,vane01} but in waveguides it can
approach unity.\cite{domo02} If we neglect dephasing and set
$\gamma_{1}^{\prime \prime} = \gamma_{2}^{\prime \prime} =
\gamma_{3}^{\prime \prime} = 0$, then $\gamma_{21} = \half
\gamma_{2}^\prime = \half (A_{21} + A_{23})$, and
$\gamma_{21}/A_{21} \approx 1$ if the branching ratio
$A_{23}/A_{21}$ is approximately unity. Therefore, in order to
open the transparency window just wide enough to admit every
photon in the probe pulse, we must have $\left|\Omega_b\right|
\approx 10 \left|\Omega_a\right|/\sqrt{n_a}$ in free space, and
about half that value in a waveguide. As an example, for
interactions of single probe photons having
$\left|\Omega_a\right|/\gamma_{21} \approx 0.03$ with rubidium
atoms (with a lifetime of level $\ket{2 \equiv 5S_{3/2}, F = 3}$
of 27~ns), the above parameters require
$|\Omega_b|^2/\gamma_{21}^2 \approx 0.1$, giving a FWHM of the
transparency window of about 1~MHz. Therefore, only pulses with
durations longer than 300~ns can safely propagate through the
window.

The assumption that $\gamma_{31} \rightarrow 0$ is generally valid
in dilute gases, where decoherence arises primarily from
spontaneous emission, although the introduction of Doppler
broadening alters the frequency dependence of the susceptibility
given by \eqr{chi3}.\cite{java01,lee02} The ultraslow group
velocity of light propagating through an appropriately prepared
gas sample\cite{hau99} can be reduced dynamically to zero,
allowing probe pulses at frequency $\omega_a$ to be stored in the
resulting atomic coherence and subsequently recovered after a
controllable time delay.\cite{phil01} Both transparency and
storage of light has been demonstrated in a solid-state system
(Pr:YSO) where cooling to liquid helium temperatures reduces the
dephasing interactions between atomic levels $\ket{1}$ and
$\ket{2}$.\cite{turu02} However, in semiconductor-based systems,
where the possibility that $\gamma_{31} \ne 0$ may be higher given
the dependence of the off-diagonal decoherence rates on a common
set of Lindblad parameters, these conclusions become less
certain.\cite{gree02} For example, in units where the absorption
of the equivalent two-level system (corresponding to $|\Omega_b| =
0$) is 1, the absorption coefficient of the general three-level
system at $\nu_a = 0$ is given by
 \begin{equation}
 \kappa(\nu_a = 0) \propto \frac{\gamma_{21} \gamma_{31}}
 {\left|\Omega_b\right|^2 + \gamma_{21} \gamma_{31}}.
 \end{equation}
Hence, if we assume that $\left|\Omega_b\right|/\gamma_{21} = 0.1$
and  $\gamma_{31}/\gamma_{21} = 0.01$, we obtain an absorption
coefficient that is 50\% of the corresponding two-level value, and
the system is no longer transparent. Similarly, the dispersion of
the general three-level system is
 \begin{equation}
 \frac{d \eta(\nu_a = 0)}{d \nu_a} \propto \frac{\gamma_{21}
 \left(\left|\Omega_b\right|^2 - \gamma_{31}^2\right)}{\left(\left|\Omega_b\right|^2
 + \gamma_{21} \gamma_{31}\right)^2},
 \end{equation}
indicating that we must have $|\Omega_b| > \gamma_{31}$ to
maintain a positive dispersion and a reduced group velocity for
the probe field. For $\gamma_{31} \ne 0$, the optimum value of the
control Rabi frequency yielding the largest dispersion is given by
$\Omega_b = \sqrt{\gamma_{21} \gamma_{31} + 2 \gamma_{31}^2} =
0.1$ when $\gamma_{31}/\gamma_{21} = 0.01$, resulting in a
dispersion that is reduced by a factor of four relative to the
case $\gamma_{31} = 0$.

If we follow the approach of \sct{eit2} and extend our single-atom
three-level analysis to include $N$ atoms, we draw essentially the
same conclusion as in the two-level case: the susceptibility is
enhanced by a factor of $N$ in the unsaturated, weak-probe-field
limit. Adopting the notation of \sct{eit2}, in the unperturbed
basis
 \begin{equation} \labe{fock_3en}
 \begin{split}
 \left\{ \right. & \ket{\{1\}, n_a, n_b, 0} , \\
 & \ket{\left\{2^{(k)}\right\}, n_a - 1, n_b, 0} , \\
 & \ket{\left\{3^{(k)}\right\}, n_a - 1, n_b + 1, 0} , \\
 & \left\{ \ket{\{1\}, n_a - 1, n_b, 1}, \ket{\{1\}, n_a, n_b - 1, 1}\right\} \left.
 \right\}
 \end{split}
 \end{equation}
we have the $N$-atom Hamiltonian
 \begin{equation} \labe{H3en}
 H = -\hbar\, \begin{bmatrix}
 0        & \Omega_a^* & 0             & \Omega_a^* & 0             & \cdots & 0 \\
 \Omega_a & \nu_a      & \Omega_b      & 0          & 0             &        & 0 \\
 0        & \Omega_b^* & \nu_a - \nu_b & 0          & 0             &        & 0 \\
 \Omega_a & 0          & 0             & \nu_a      & \Omega_b      &        & 0 \\
 0        & 0          & 0             & \Omega_b^* & \nu_a - \nu_b &        & 0 \\
 \vdots   &            &               &            &               & \ddots &   \\
 0        & 0          & 0             & 0          & 0             &        & 0
 \end{bmatrix} .
 \end{equation}
As in the case of $N$ noninteracting two-level atoms, the
off-diagonal density-matrix elements for each atom are given by
\eqs{rho_3_ss}, and \eqr{PoldefN} holds in the three-level case,
with $\rho_{21}^{\{N\}} = N \rho_{21}$ as before. The aggregate
susceptibility is enhanced by a factor of $N$ only if the
conditions $\rho_{22}^{\{N\}} \ll \rho_{11}^{\{N\}}$ and
$\gamma_{31} \rightarrow 0$ are satisfied; otherwise, when $N \gg
1$ and $\nu_a = \nu_b = 0$, the transparency window is completely
destroyed and the susceptibility reduces to that of $N$ two-level
atoms. However, when these conditions \emph{are} met, the
absorption coefficient approaches zero as $\nu_a - \nu_b
\rightarrow 0$, and the group refractive index is increased by a
factor of $N$. This property allows us to compensate for the need
to open a sufficiently large window to allow a pulse to propagate
through the interaction region without significant loss by
increasing the number of atoms in that region.

 \subsection{Tunable Transparency of the Four-Level $\mathcal{N}$ Atom\labs{eit4}}

The capability to transmit a particular probe field with very high
fidelity and extraordinarily slow group velocity is enabled by the
inclusion of the second control field in \fig{three_level_atom}.
However, since the probe frequency providing the greatest
transparency results in a differential refractive index $\Delta
\eta(\omega_a)$ that is precisely zero, this system cannot be used
to generate a significant relative phase shift. More precise
\emph{deterministic} control of the phase and intensity of the
probe pulse can be obtained by adding the third field shown in
\fig{four_level_atom}, coupling the metastable atomic energy level
$\ket{3}$ with the upper energy level $\ket{4}$. As in the
previous sections, we reduce the semiclassical atomic system
depicted in \fig{lambda_4} to the quantum manifold given by
\fig{quantum_4}, and we will work in a manifold corresponding to
an atom initially in state $\ket{1}$, with $n_a$ photons in mode
$a$, $n_b$ photons in mode $b$, and $n_c$ photons in mode $c$. We
again extend our basis to include energy dissipation to the
environment by appending an entry to each product state,
indicating the occurrence of scattering of a photon of frequency
$\omega_a$, $\omega_b$, or $\omega_c$. Therefore, the environment
should be represented by a nonresonant sub-manifold that captures
dissipated energy and preserves the trace of the density matrix,
resulting in the unperturbed Schr\"{o}dinger basis
 \begin{equation} \labe{fock_4e}
 \begin{split}
 \left\{ \right. & \ket{1, n_a, n_b, n_c, 0} , \\
 & \ket{2, n_a - 1, n_b, n_c, 0} , \\
 & \ket{3, n_a - 1, n_b + 1, n_c, 0} , \\
 & \ket{4, n_a - 1, n_b + 1, n_c - 1, 0} , \\
 & \left\{ \right. \ket{1,
n_a - 1, n_b, n_c, 1}, \\ & \qquad \ket{1, n_a, n_b - 1, n_c, 1},
\\ & \qquad \qquad \ket{1, n_a, n_b, n_c - 1, 1}\left.\right\}
\left. \right\} .
 \end{split}
 \end{equation}

 \begin{figure}
   \centering
   \subfigure[Semiclassical energy levels]{\labf{lambda_4}
 \includegraphics[width=3.25in]{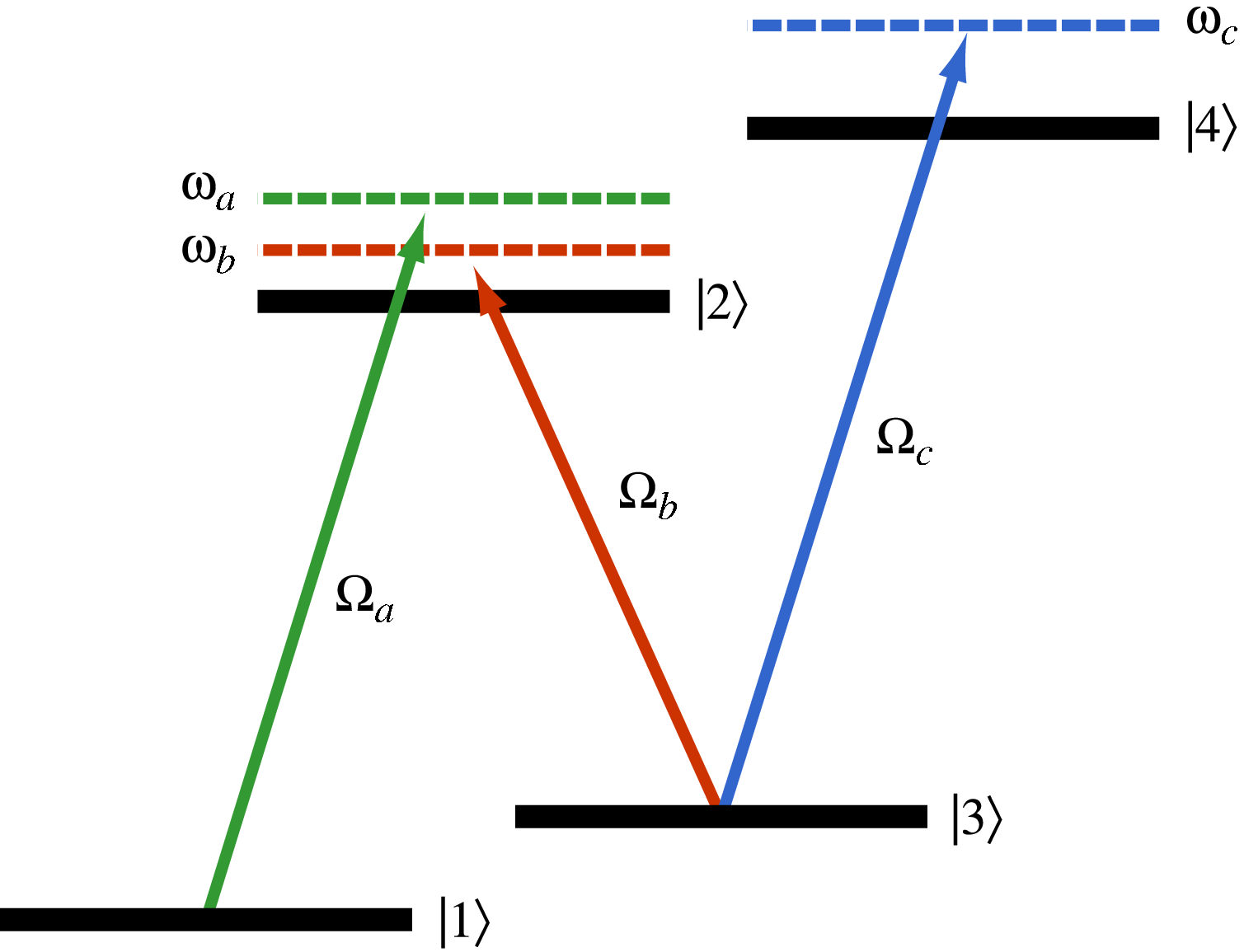}}\quad
   \subfigure[Quantum energy manifold]{\labf{quantum_4}
 \includegraphics[width=3.25in]{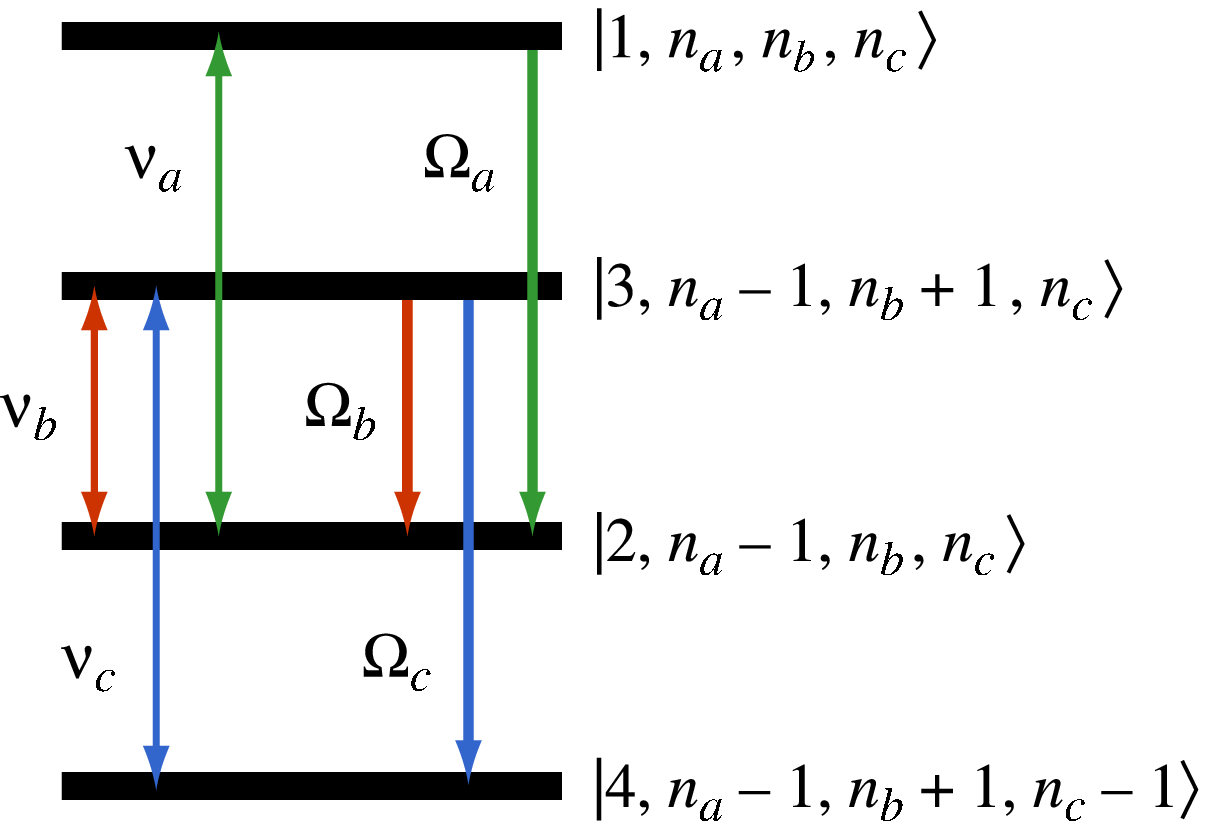}}
   \caption{\labf{four_level_atom} Interaction between a four-level $\mathcal{N}$ atom and a
    nearly resonant three-frequency
   electromagnetic field. Note that the annihilation of a
   photon of frequency $\omega_k$ is represented by the complex number $\Omega_k$.}
 \end{figure}

Referencing \sct{eit3}, by inspection we then obtain the total
Hamiltonian
 \begin{equation} \labe{H4}
 H = -\hbar\, \begin{bmatrix}
 0 & \Omega_a^* & 0 & 0 & 0\\
 \Omega_a & \nu_a & \Omega_b & 0 & 0\\
 0 & \Omega_b^* & \nu_a - \nu_b & \Omega_c^* & 0\\
 0 & 0 & \Omega_c & \nu_a - \nu_b + \nu_c & 0 \\
 0 & 0 & 0 & 0 & 0
 \end{bmatrix} ,
 \end{equation}
where we have defined the detuning parameter
 \begin{equation}
 \labe{detunc}
 \nu_c \equiv \omega_c - \left(\omega_4 - \omega_3\right) ,
 \end{equation}
 the effective coupling constant
 \begin{equation}
 \labe{gcdef}
  g_c \equiv \hat{\bm{\epsilon}}_c \cdot \mathbf{d}_{43}\,
  \mathcal{E}_c,
 \end{equation}
 and the effective Rabi frequency
 \begin{equation}
 \labe{rabicdef}
 \Omega_c = \frac{g_c}{\hbar}  \sqrt{n_c} ,
 \end{equation}
 and subtracted the energy
 \begin{equation*}
 E_0 \equiv \bra{1, n_a, n_b, n_c, 0} H_0 \ket{1, n_a, n_b, n_c, 0}
 \end{equation*}
 from all diagonal terms.

Again, we follow the general decoherence conventions established
in \sct{eit3} to enumerate contributions to the decoherence
operator using \eqr{Lindblad}. We define $\gamma^\prime_{4}$ as
the total free-space depopulation rate to the environment of state
$\ket{4, n_a - 1, n_b + 1, n_c - 1, 0}$ and $\gamma^{\prime
\prime}_{4}$ as the corresponding pure dephasing rate. We
substitute \eqr{H4} and \eqr{Lindblad} into \eqr{rhodot} and seek
the quasi-steady-state solution in the unsaturated weak-field
limit $|\Omega_a/\gamma_{21}|^2 \ll 1$, assuming that
$\rho_{22}(t) \ll \rho_{11}(t)$ , $\Omega_a \rho_{32}(t) \ll
\Omega_b^* \rho_{31}(t) , \Omega_c^* \rho_{41}(t)$, and $\Omega_a
\rho_{42}(t) \ll \Omega_c \rho_{31}(t)$ for all $t > 0$. We
neglect all contributions to density matrix elements of order
$\left|\Omega_a\right|^2$ and higher, and we obtain
$\tilde{\rho}_{11} \cong 1$, $\tilde{\rho}_{k 2} =
\tilde{\rho}^*_{2 k} \cong 0$, $\tilde{\rho}_{k 3} =
\tilde{\rho}^*_{3 k} \cong 0$, $\tilde{\rho}_{k 4} =
\tilde{\rho}^*_{4 k} \cong 0$ (where $k \in \{2, 3, 4\}$), and
 \begin{widetext}
 \begin{subequations} \labe{rho_4_ss}
 \begin{eqnarray}
 \tilde{\rho}_{21} &=& -\frac{\left[(\nu_a - \nu_b + i \gamma_{31})(\nu_a -
 \nu_b + \nu_c + i \gamma_{41}) - \left|\Omega_c\right|^2\right]\,
 \Omega_a}{(\nu_a + i \gamma_{21})\left[(\nu_a - \nu_b + i \gamma_{31})
 (\nu_a - \nu_b + \nu_c + i \gamma_{41}) - \left|\Omega_c\right|^2\right] -
 (\nu_a - \nu_b + \nu_c + i \gamma_{41}) \left|\Omega_b\right|^2}, \labe{rho_21_4} \\
 \tilde{\rho}_{31} &=& \frac{(\nu_a - \nu_b + \nu_c + i \gamma_{41})\, \Omega_a\, \Omega_b^*}
 {(\nu_a + i \gamma_{21})\left[(\nu_a - \nu_b + i \gamma_{31})(\nu_a - \nu_b
 + \nu_c + i \gamma_{41}) - \left|\Omega_c\right|^2\right] -
 (\nu_a - \nu_b + \nu_c + i \gamma_{41}) \left|\Omega_b\right|^2}
 ,  \nd \\
 \tilde{\rho}_{41} &=& -\frac{\Omega_a\, \Omega_b^*\, \Omega_c}{(\nu_a + i \gamma_{21})
 \left[(\nu_a - \nu_b + i \gamma_{31})(\nu_a - \nu_b + \nu_c + i \gamma_{41})
 - \left|\Omega_c\right|^2\right] -
 (\nu_a - \nu_b + \nu_c + i \gamma_{41}) \left|\Omega_b\right|^2},
 \end{eqnarray}
 \end{subequations}
 \end{widetext}
where the remaining off-diagonal density matrix elements are
$\tilde{\rho}_{1 k} = \tilde{\rho}^*_{k 1}$ ($k \in \{2, 3, 4\}$),
$\gamma_{21}$ is given by \eqr{gamma21def}, $\gamma_{31}$ is given
by \eqr{gamma31def}, and $\gamma_{41}$ is defined as
 \begin{equation} \labe{gamma41def}
 \gamma_{41} \equiv \half \gamma^\prime_{4}
 + \gamma^{\prime \prime}_{1} + \gamma^{\prime \prime}_{4}
 \end{equation}
Since \eqr{rho2dot_11} remains valid for the four-level atom +
photons system, we can follow the same bootstrap procedure to
obtain the approximate solution for $\rho_{11}(t)$ given by
\eqr{rho11apprx}, and
 \begin{equation}
\rho_{k1}(t) \cong \tilde{\rho}_{k1} \left(1 - e^{-\gamma_{k1}
t}\right) \rho_{11}(t) ,
 \end{equation}
where $k \in \{2, 3, 4\}$. As in the three-level case, the
steady-state solutions given by \eqs{rho_4_ss} are valid at any
time $t$ where the laser parameters have been chosen to allow the
inequality
 \begin{equation} \labe{sstineq4}
1/\gamma_{21}, 1/\gamma_{31}, 1/\gamma_{41} \ll t \ll \tau_a
\equiv \left[2 \Im(\tilde{\rho}_{21} \Omega_a^*) \right]^{-1}
 \end{equation}
to be satisfied.

    We see immediately that \eqr{rho_21_4} reduces to \eqr{rho_21_3}
if $\sqrt{\nu_c^2 + \gamma_{41}^2} \gg |\Omega_c|^2$. Therefore,
in this limit the four-level system of \fig{four_level_atom}
remains at least approximately transparent. However, in general,
the absorption and dispersion curves shown in \fig{eta_kappa_3}
can be significantly modified by a nonvanishing radiative coupling
between the atomic states $\ket{3}$ and $\ket{4}$. For example,
when all coupling fields are resonantly tuned to the corresponding
transitions (i.e., $\nu_a = \nu_b = \nu_c = 0$), we have
 \begin{equation}
 \rho_{21} = i\, \frac{\left(\gamma_{31} \gamma_{41} + \left|\Omega_c
 \right|^2 \right)\Omega_a}{\gamma_{21} \left(\gamma_{31} \gamma_{41}
 + \left|\Omega_c\right|^2 \right)
 + \gamma_{41} \left|\Omega_b\right|^2}.
 \end{equation}
Since the complex part of $\rho_{21}$ is related to the absorption
coefficient by \eqr{chi_2}, we see that we have lost perfect
transparency when the frequency $\omega_c$ is resonant, even if
$\gamma_{31} = 0$. However, the presence of the fourth atomic
level and the corresponding nearly resonant electromagnetic field
with frequency $\omega_c$ introduces an extremely large Kerr-like
nonlinearity even at relatively low light
intensities.\cite{luki01a,harr90,schm96,imam97,harr99,rebi01,rebi02,braj03}
In particular, we can demonstrate that the four-level system of
\fig{four_level_atom} can behave as either a low-energy quantum
switch\cite{harr98,yan01a,yan01b,braj03} or a quantum
phase-shifter,\cite{schm96} properties which can be harnessed for
both classical and quantum information processing and
distribution.\cite{monr02,vita00,luki00b,luki01b,flei01,flei02}

We assume that the magnitude of the Rabi frequency $\Omega_c$
satisfies the inequality $\left|\Omega_c\right|^2 \ll
\gamma^2_{41}$, so that we can define a complex third-order
susceptibility for the $\ket{1} \longleftrightarrow \ket{2}$
transition by analogy with \eqr{Pchidef2} as
 \begin{equation} \labe{Pchidef4}
 \begin{split}
\mathbf{P}(0) &= \mathcal{V}^{-1} \left(\tilde{\rho}_{21}
\mathbf{d}_{12} + \tilde{\rho}_{12} \mathbf{d}_{21}\right) \\
&\equiv \hat{\bm{\epsilon}}_a \frac{\epsilon_0}{2}
\chi^{(1)}\left(-\omega_a, \omega_a\right) E_a + c.c. \\ & +
\hat{\bm{\epsilon}}_a \frac{\epsilon_0}{8}
\chi^{(3)}\left(-\omega_a, \omega_c, -\omega_c, \omega_a\right)
\left|E_c\right|^2 E_a + c.c. ,
 \end{split}
 \end{equation}
where $\chi^{(1)}\left(-\omega_a, \omega_a\right)$ is the
three-level linear susceptibility given by \eqr{chi3}. If we
expand $\rho_{21}$ in a power series about
$\left|\Omega_c\right|^2 = 0$, we obtain
 \begin{widetext}
 \begin{equation} \labe{chi4_3}
 \chi^{(3)}\left(-\omega_a, \omega_c, -\omega_c, \omega_a\right)
 \propto
 -\frac{\left|\Omega_b\right|^2}{(\nu_a -
\nu_b + \nu_c + i \gamma_{41}) \left[(\nu_a + i \gamma_{21})(\nu_a
- \nu_b + i \gamma_{31}) - \left|\Omega_b\right|^2\right]^2} .
 \end{equation}
 \end{widetext}

As in \sct{eit3}, we set $\nu_b = \gamma_{31} = 0$, we choose the
values of $|\Omega_b|/\gamma_{21} = |\Omega_c|/\gamma_{41} = 0.1$
(consistent with weak fields), and for simplicity we select
$\gamma_{41} = \gamma_{21}$. In \fig{eta_kappa_4} we plot the real
and imaginary parts of the \emph{total} susceptibility as a
function of $\nu_a/\gamma_{21}$ for three values of the relative
detuning $\nu_c/\gamma_{41}$. As we discussed above, when $\nu_c =
0$, the probe field at $\omega_a$ will be strongly absorbed, and
when $\nu_c/\gamma_{41} \gg 1$ transparency is largely restored,
but there is a nonzero contribution to the refractive index at
$\omega_a$. We can see these effects more clearly by setting
$\nu_a = 0$ in \eqr{chi4_3} and then varying $\nu_c$, as is done
in \fig{phase_switch_4}. We note that when $\nu_c = 0$, the
presence of a photon at $\omega_c$ (i.e., $|\Omega_c| \ne 0$)
closes a ``quantum switch'' that causes the absorption of the
probe photon,\cite{harr98,yan01a,yan01b} and when
$\nu_c/\gamma_{41} \approx 30$, the atom acts as a ``quantum phase
shifter'' that is largely transparent but shifts the relative
phase of the probe photon.\cite{schm96} We explore the latter
property of the four-level system in the next section, where we
explicitly calculate the corresponding applied phase shift of a
coherent superposition of single-photon states.

 \begin{figure}
   \centering
   \subfigure[Refractive index: $\Re \chi(\omega_a)$]{\labf{eta_4}
 \includegraphics[width=3.25in]{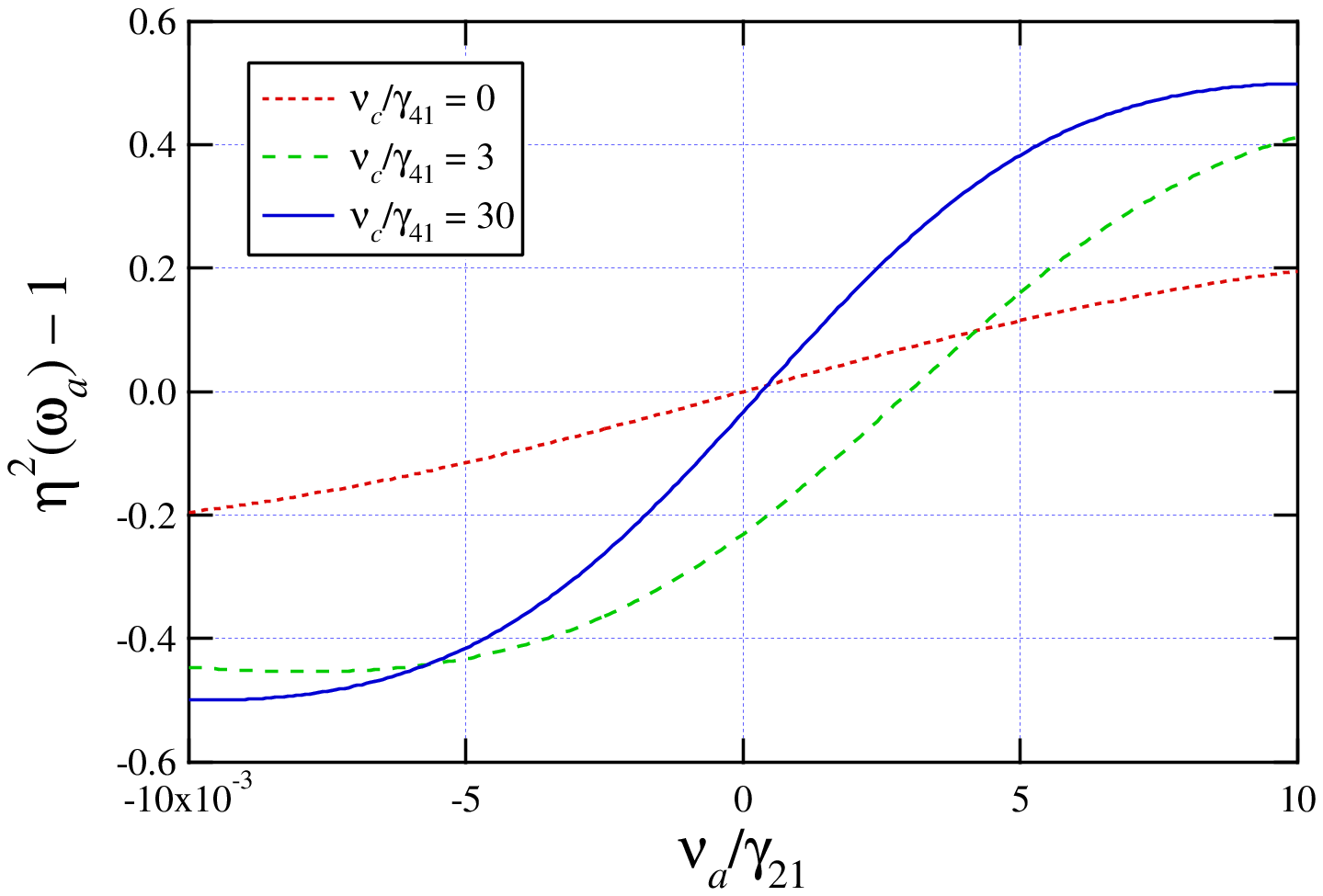}}\quad
   \subfigure[Absorption coefficient: $\Im \chi(\omega_a)$]{\labf{kappa_4}
 \includegraphics[width=3.25in]{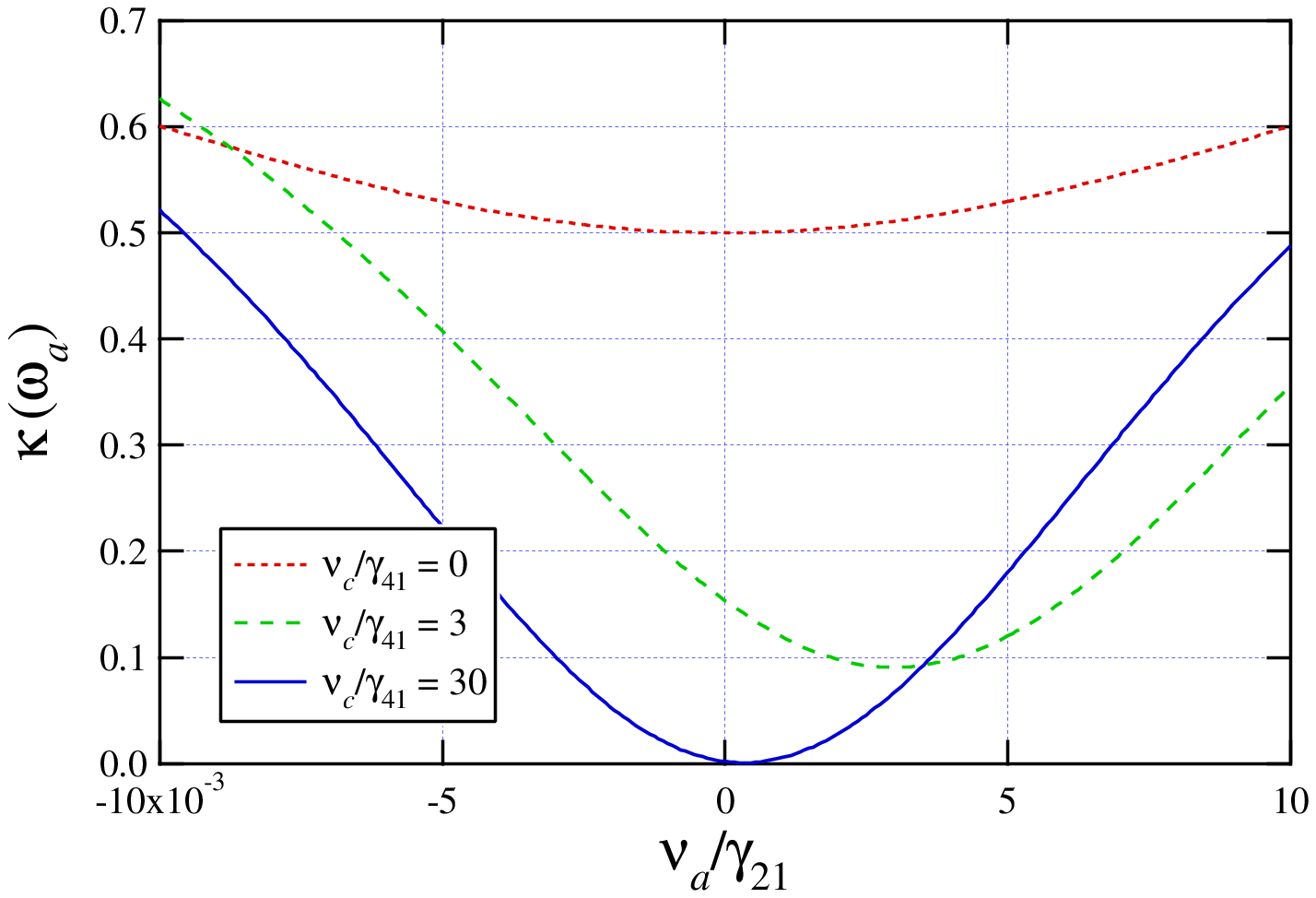}}
   \caption{\labf{eta_kappa_4} Refractive index and linear absorption
   coefficient for the four-level atom shown in \fig{four_level_atom},
   with $\gamma_{31} \rightarrow 0$. Both the dispersion and transmission
   window are sharpest when $\left|\Omega_b/\gamma_{21}\right|^2 \ll 1$.}
 \end{figure}

\epsf{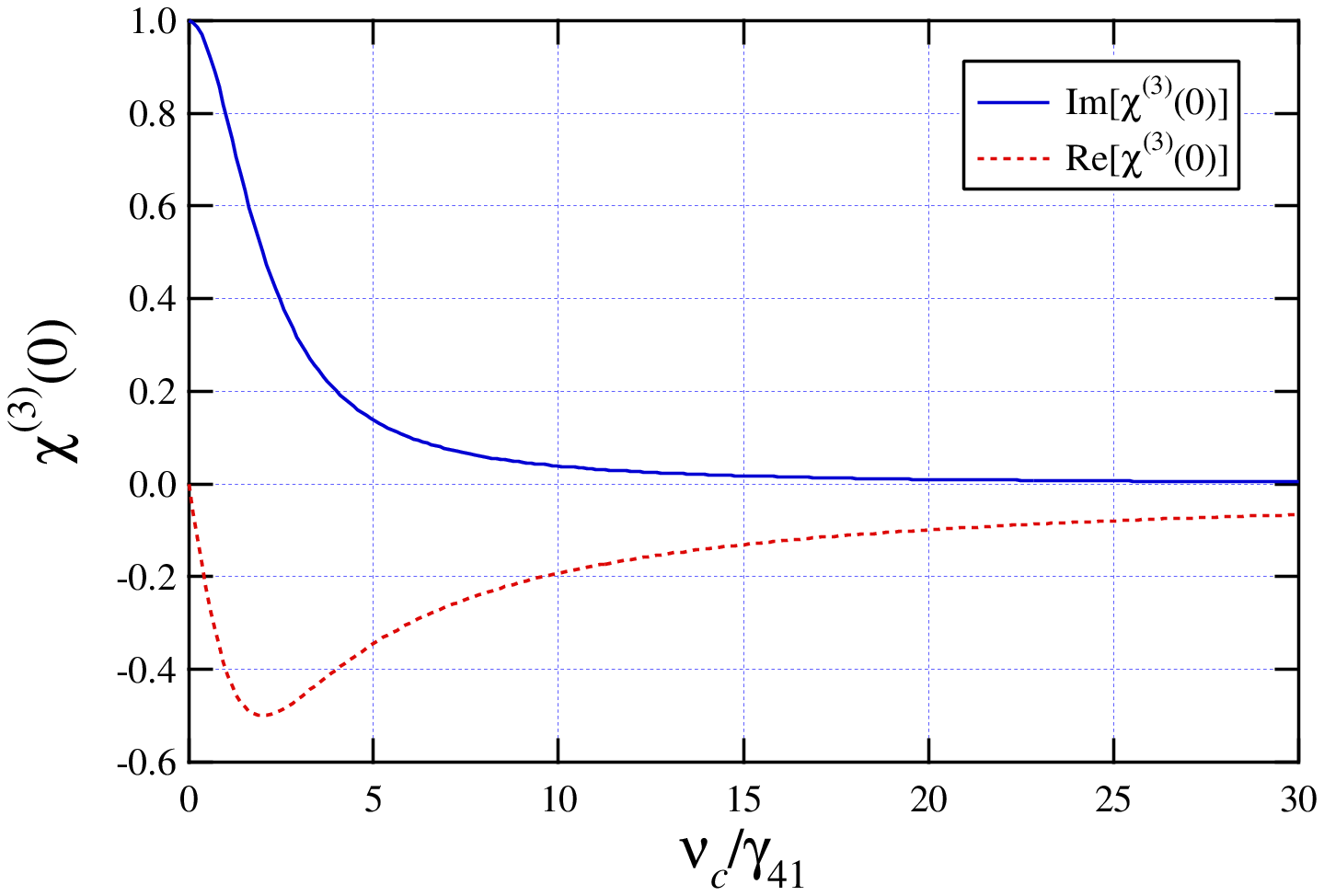}{The behavior of the third-order
susceptibility shown in \fig{eta_kappa_4} at $\nu_a = 0$ as a
function of the normalized detuning $\nu_c/\gamma_{41}$. Note that
as $\nu_c \rightarrow 0$, a large relative absorption arises at
the probe frequency $\omega_a$ (a ``quantum switch''), and that
large values of $\nu_c$ cause a phase shift that is substantial
relative to the small absorption (a ``quantum phase-shifter''). }

Following the approach of \sct{eit3}, an extension of the
single-atom four-level model to include $N$ atoms in the
interaction region reveals that the susceptibility is enhanced by
a factor of $N$ only in the unsaturated, weak-probe-field limit.
Again adopting the notation of \sct{eit2}, in the unperturbed
basis
 \begin{equation} \labe{fock_4en}
 \begin{split}
 \left\{ \right. & \ket{\{1\}, n_a, n_b, n_c, 0} , \\
 & \ket{\left\{2^{(k)}\right\}, n_a - 1, n_b, n_c, 0} , \\
 & \ket{\left\{3^{(k)}\right\}, n_a - 1, n_b + 1, n_c, 0} , \\
 & \ket{\left\{4^{(k)}\right\}, n_a - 1, n_b + 1, n_c - 1, 0} , \\
 & \left\{ \right. \ket{\{1\},
n_a - 1, n_b, n_c, 1}, \\ & \qquad \ket{\{1\}, n_a, n_b - 1, n_c,
1},
\\ & \qquad \qquad \ket{\{1\}, n_a, n_b, n_c - 1, 1}\left.\right\}
\left. \right\}
 \end{split}
 \end{equation}
we have the $N$-atom Hamiltonian
 \begin{widetext}
 \begin{equation} \labe{H4en}
 H = -\hbar\, \begin{bmatrix}
 0        & \Omega_a^* & 0             & 0                     & \Omega_a^* & 0                     & 0                     & \cdots & 0 \\
 \Omega_a & \nu_a      & \Omega_b      & 0                     & 0          & 0                     & 0                     &        & 0 \\
 0        & \Omega_b^* & \nu_a - \nu_b & \Omega_c^*            & 0          & 0                     & 0                     &        & 0 \\
 0        & 0          & \Omega_c      & \nu_a - \nu_b + \nu_c & 0          & 0                     & 0                     &        & 0 \\
 \Omega_a & 0          & 0             & 0                     & \nu_a      & \Omega_b              & 0                     &        & 0 \\
 0        & 0          & 0             & 0                     & \Omega_b^* & \nu_a - \nu_b         & \Omega_c^*            &        & 0 \\
 0        & 0          & 0             & 0                     & 0          & \Omega_c              & \nu_a - \nu_b + \nu_c &        & 0 \\
 \vdots   &            &               &                       &            &                       &                       & \ddots &   \\
 0        & 0          & 0             & 0                     & 0          & 0                     & 0                     &        & 0
 \end{bmatrix} .
 \end{equation}
 \end{widetext}
As in the case of $N$ noninteracting two-level and three-level
atoms, the off- diagonal density-matrix elements for each atom are
given by \eqs{rho_4_ss}, and \eqr{PoldefN} holds in the four-level
case, with $\rho_{21}^{\{N\}} = N \rho_{21}$ as before. The
aggregate susceptibility is enhanced by a factor of $N$ only if
$\rho_{22}^{\{N\}} \ll \rho_{11}^{\{N\}}$ and $\gamma_{31}
\rightarrow 0$.

Our discussions have emphasized that the resonant nonrelativistic
quantum electrodynamic interaction of the four-level system of
\fig{four_level_atom} results in the generation of a giant
third-order optical nonlinearity, commonly described as a Kerr
nonlinearity in the
literature.\cite{luki01a,harr90,schm96,imam97,harr99,rebi01,rebi02,braj03,wang01}
Strictly speaking, an effective Kerr Hamiltonian with the form
 \begin{equation}
H_\text{Kerr} = \hbar \tilde{W}\, a^\dag a\, c^\dag c
 \end{equation}
causes the Fock state $\ket{\psi(0)} \equiv \ket{n_a, n_c}$ to
evolve as
 \begin{equation} \labe{Kerrevol}
 \ket{\psi(t)} = e^{-i \tilde{W} a^\dag a c^\dag c t}
 \ket{\psi(0)} = e^{-i n_a n_c \phi}
 \ket{n_a, n_c} ,
 \end{equation}
where $\phi \equiv \tilde{W} t$. Therefore, if the evolution of
the four-level system shown in \fig{four_level_atom} exhibits this
Kerr behavior, then we can claim that the corresponding
nonlinearity is in fact a Kerr nonlinearity. However, at low light
levels, if one of the optical transitions is driven by a weak
coherent state (rather than a Fock state), we can show that the
structure of this nonlinearity is not strictly of the Kerr type
unless $\left|\Omega_b\right|^2 \gg \left|\Omega_a\right|^2,
\left|\Omega_c\right|^2$. We begin by assuming that all dephasing
rates are zero, and by solving the Schr\"{o}dinger equation for
the basis set given by \eqr{fock_4e} under the influence of
spontaneous emission only. Using our adiabatic bootstrap approach,
we find for the case $\nu_a = \nu_b = 0$ that the $N$-atom ground
state evolves as
\begin{equation} \labe{FockKerr}
\ket{\left\{1\right\}, n_a, n_b, n_c} \longrightarrow e^{-i W t}
\ket{\left\{1\right\}, n_a, n_b, n_c} ,
\end{equation}
where
 \begin{equation} \labe{W41}
W = \frac{N\, \left|\Omega_a\right|^2 \left|\Omega_c\right|^2
}{\nu_c \left|\Omega_b\right|^2 + i \left( \gamma_{41}
\left|\Omega_b\right|^2 + \gamma_{21} \left|\Omega_c\right|^2
\right)} ,
 \end{equation}
and $\gamma_{k 1} = \gamma^\prime_{k}/2$. Note that when the
inequality
\begin{equation}\label{nondem}
\frac{\left|\Omega_b\right|^2}{\gamma_{21}}\,
\frac{\nu_c}{\gamma_{41}} \gg
\frac{\left|\Omega_b\right|^2}{\gamma_{21}} +
 \frac{\left|\Omega_c\right|^2}{\gamma_{41}}
\end{equation}
is satisfied (equivalent to the assumption $\nu_c/\gamma_{41} \gg
1$ in the simple case where $\left|\Omega_b\right|^2/\gamma_{21}
\approx \left|\Omega_c\right|^2/\gamma_{41}$), the probability
that a single photon with frequency $\omega_a$ will be scattered
by the atom becomes vanishingly small. Therefore, the evolution of
the atomic ground state during a prolonged interaction with the
compound Fock state $\ket{n_a, n_b, n_c}$ will be governed
primarily by the real part of \eqr{W41}; since
$\left|\Omega_a\right|^2 \propto n_a$ and $\left|\Omega_c\right|^2
\propto n_c$, the evolution of \eqr{FockKerr} has the Kerr form of
\eqr{Kerrevol} with the nonlinear coefficient
 \begin{equation} \labe{chifock}
 \tilde{W} \equiv \frac{N |\tilde{\Omega}_a|^2 |\tilde{\Omega}_c|^2}{\nu_c |\tilde{\Omega}_b|^2
 n_b}\ ,
 \end{equation}
where the vacuum Rabi frequencies are given by $\tilde{\Omega}_k
\equiv \Omega_k/\sqrt{n_k}$.

Let us now replace the $n_b$-photon Fock state with a coherent
state parameterized by $\alpha_b$ and assess the evolution of the
corresponding unperturbed ground state
 \begin{equation} \labe{psicoh}
 \begin{split}
 \ket{\psi(0)} &\equiv \left|\{1\}, n_a, \alpha_b, n_c\right\rangle\\ &=
 e^{-\half \left|\alpha_b\right|^2} \sum_{n_b =
0}^{\infty}\frac{\alpha_b^{n_b}}{\sqrt{n_b!}} \ket{\{1\}, n_a,
n_b, n_c} .
 \end{split}
 \end{equation}
Since each unperturbed eigenstate evolves according to
\eqr{FockKerr}, after a time $t$ we find
 \begin{equation} \labe{psi_prime}
  \begin{split}
 \ket{\psi(t)} &=
 e^{-\half \left|\alpha_b\right|^2} \sum_{n_b = 0}^{\infty}
\frac{\alpha_b^{n_b}}{\sqrt{n_b!}} \\
&\qquad e^{-i\, n_a n_c \phi(t)\, \left|\alpha_b\right|^2 /n_b}
\ket{\{1\}, n_a, n_b, n_c} ,
 \end{split}
 \end{equation}
where $\phi(t) \equiv \tilde{W} t$ and
 \begin{equation} \labe{chicoh}
 \tilde{W} \equiv \frac{N |\tilde{\Omega}_a|^2 |\tilde{\Omega}_c|^2}{\nu_c |\tilde{\Omega}_b|^2
 |\alpha_b|^2} .
 \end{equation}

Note that $\ket{\psi(t)}$ is not a coherent state unless
$\left|\alpha_b\right| \gg 1$, for which $\ket{\psi(t)} \cong
 e^{-i\, n_a n_c \phi(t)} \ket{\psi(0)}$. In \fig{coherence_15}, we have
used \eqrs{psicoh}{psi_prime} to numerically evaluate the inner
product $\left|\left\langle\psi(t)|\psi(0)\right\rangle\right|^2$
for several values of the net phase shift $\phi$, assuming that
$n_a = 1$ and $n_c = 5$. Note that for large phase shifts the
inner product differs significantly from unity even when
$\left|\alpha_b\right|^2 \approx 1000$; in fact, for $\phi = \pi$,
$\left|\left\langle\psi(t)|\psi(0)\right\rangle\right|^2 > 0.99$
only if $\left|\alpha_b\right|^2 > 2.5 \times 10^4$. Therefore,
only when the coupling field driving mode $b$ closely approximates
a classical field does EIT provide a true cross-Kerr nonlinearity.

Nevertheless, we can appreciate the magnitude of the optical Kerr
nonlinearity even at low light levels by estimating the parameters
included in \eqr{chicoh}. Let us assume that $|\tilde{\Omega}_b|
\approx |\tilde{\Omega}_c|$, and compute the phase shift induced
by a system of 1000 non-interacting atoms in the case where
$\left|\alpha_b\right|^2 = 25$. Using \eqr{Rabirule} as a guide,
we estimate $\sigma_a/\mathcal{A} \approx 20\%$, and we assume a
unit branching ratio for spontaneous emission from atomic level
$\ket{2}$ so that $\gamma_{21} \approx A_{21}$. If we let the
Fourier-limited pulse duration be $2 \pi/\Delta \omega$, then
after the pulse has interacted with the atoms we obtain a phase
shift of approximately 0.1~radians. This shift is about
\emph{seventeen orders of magnitude larger} than the corresponding
value provided by a standard Kerr cell.\cite{boyd99,kok02}

\epsf{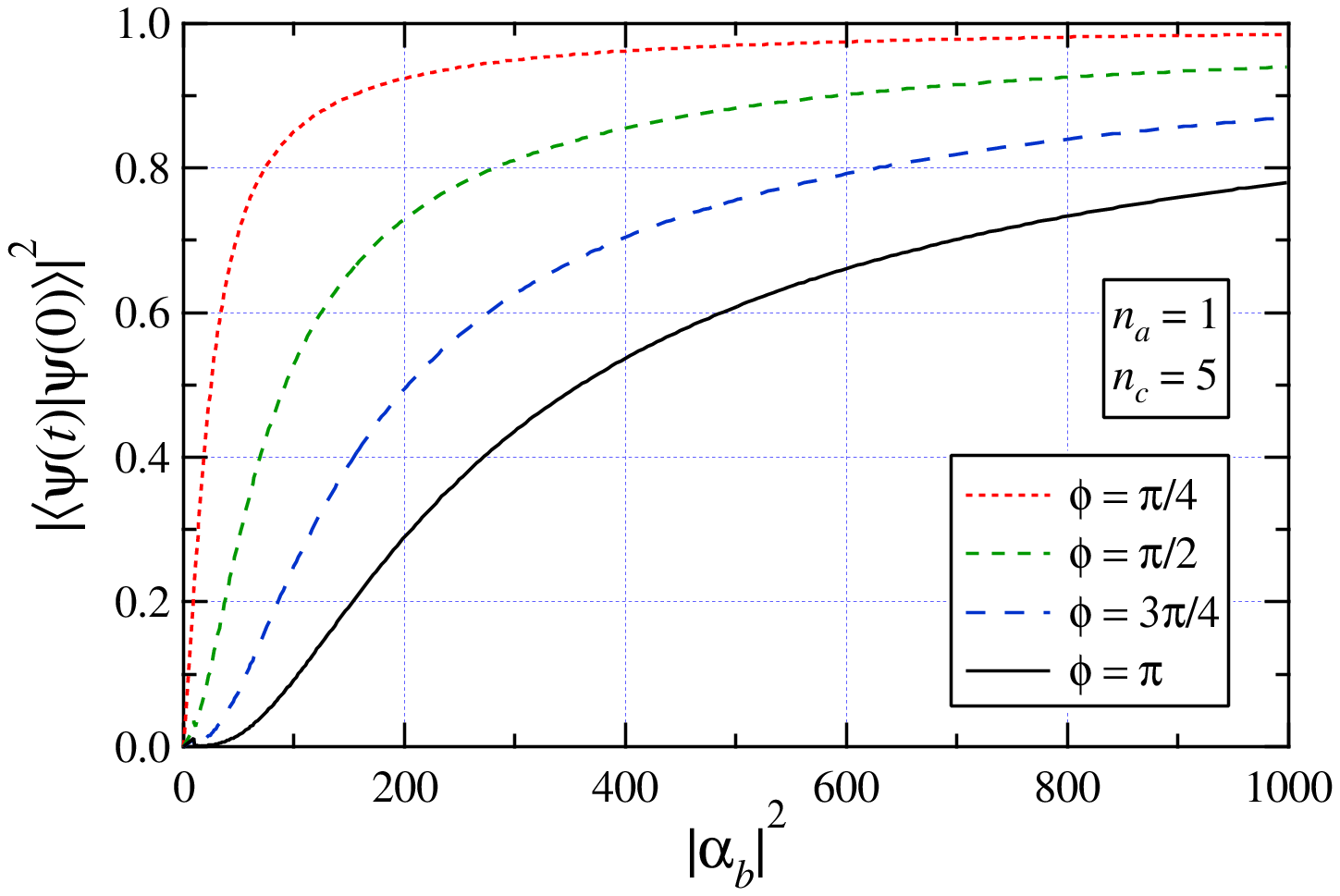}{Numerical evaluation of the
coherent-state-coupled inner product
$\left|\left\langle\psi(t)|\psi(0)\right\rangle\right|^2$ for
several values of the net phase shift $\phi$, assuming that the
input Fock state has $n_a = 1$ and $n_c = 5$. Note that for large
phase shifts the inner product differs significantly from unity
even when $\left|\alpha_b\right|^2 \approx 1000$.}

 \section{QUANTUM INFORMATION PROCESSING\labs{qip}}

We wish to assess the utility of coherent population transfer for
the creation, transmission, reception, storage, and processing of
quantum information. In particular, we must evaluate the time
dependence of coherent superpositions of discrete states of the
atom + photon field that are (at least in principle) easily
distinguished by direct detection of a photon with energy $\hbar
\omega_a$. For example, consider a system that is initially in a
pure state consisting of a superposition of two manifold states,
such as
 \begin{equation} \labe{psi0}
 \ket{\psi} = \frac{1}{\sqrt{2}} \left(\ket{1, 0} +
 \ket{1, n_a}\right)
 \end{equation}
where
 \begin{enumerate}
 \item $\ket{1, 0}$ represents the atom in the ground state and zero
 photons in the resonator of \fig{resonator}; and
 \item $\ket{1, n_a}$ represents the atom in the ground state and $n_a$
 photons in the resonator.
 \end{enumerate}
If we subsequently apply the unitary phase shift operator
$\Phi(\varphi) \equiv e^{i \varphi a^\dag a}$ to $\ket{\psi}$,
then we obtain the result
 \begin{equation} \labe{psip}
 \ket{\psi^\prime} = \Phi(\varphi) \ket{\psi} = \frac{1}{\sqrt{2}} \left(\ket{1, 0} +
 e^{i n_a \varphi} \ket{1, n_a}\right) ,
 \end{equation}
where we note that each Fock photon contributes equally to the
total accumulated phase. Similarly, if we begin with a
superposition of an empty resonator and a coherent state $\ket{1,
\alpha}$, we find
 \begin{equation} \labe{psipcoh}
 \ket{\psi^\prime} = \Phi(\varphi) \ket{\psi} = \frac{1}{\sqrt{2}} \left(\ket{1, 0} +
 \ket{1, \alpha\, e^{i \varphi}}\right) ,
 \end{equation}
in accordance with our intuition for classical fields. A simple
physical implementation of such a dual-rail coherent superposition
could be provided by the Mach-Zehnder interferometer shown in
\fig{mach_zehnder}. In one arm of the interferometer, the single
four-level atom represented by \fig{four_level_atom} is prepared
using $|\Omega_c| > 0$ to provide a phase shift at the probe
frequency $\omega_a$ while remaining largely transparent and
dispersive. In the second arm, $|\Omega_c| = 0$, and the system is
tuned to match the absorption and dispersion provided by the atom
in the first arm, allowing the interferometer to remain
time-synchronous. In principle, we can simply use the real and
complex parts of the susceptibilities given by \eqr{chi3} and
\eqr{chi4_3} to determine the classical absorption and group
velocity reduction provided by each system, as is done in standard
treatments.\cite{lo98,bouw00,niel00} In practice, however, we must
be careful to demonstrate that the interaction of either arm with
a photon at the probe frequency that has entered the
interferometer at the input port will entangle the quantum
mechanical paths of that photon with each other but \emph{not}
with either of the atoms.

\epsf{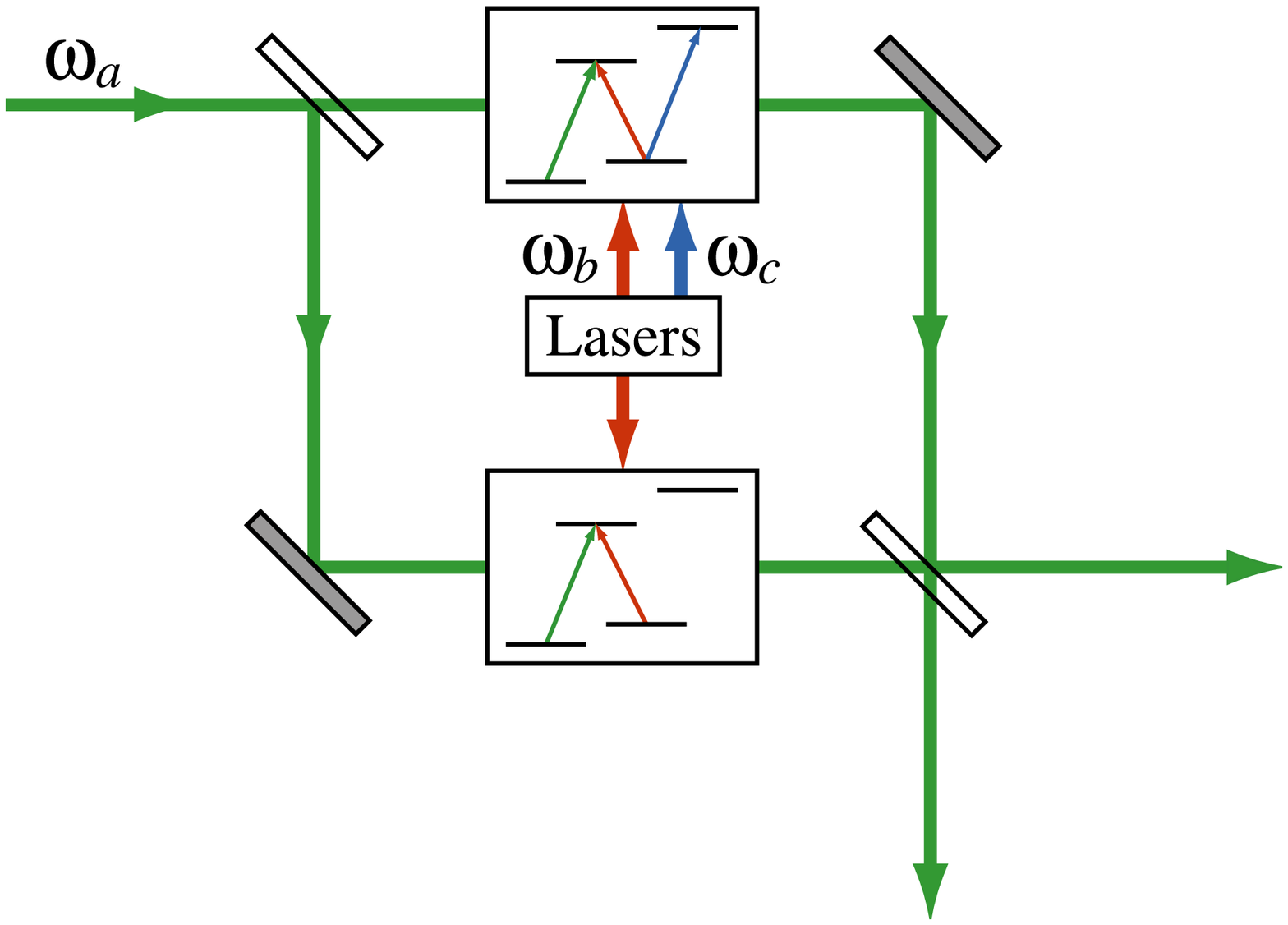}{A model Mach-Zehnder interferometer
illustrating an architecture for a quantum phase-shifter using the
four-level atom described in \sct{qip4}. }

In this section, we solve the density matrix equation of motion
given by \eqr{rhodot} for product states that include the
additional kets enumerated above, and we seek expressions for
density matrix elements that allow us to directly read out the
phase $\varphi$ in \eqr{psip} and \eqr{psipcoh} in terms of
experimentally determined parameters such as Rabi frequencies and
laser detunings. We will discover that constraints must be placed
on possible values of these parameters in systems suffering from
decoherence because of the necessity of maintaining either high
fidelity (or low entropy) in systems without active quantum error
correction, or (equivalently) high data rates in corrected
systems.

 \subsection{The Two-Level Atom\labs{qip2}}

Following the example of the dual-rail state introduced in
\eqr{psi0}, we wish to further extend the basis of \eqr{fock_2e}
to include the possibility that all $n_a$ photons have followed
another quantum trajectory, and are not found in the interaction
region containing the two-level atom(s). We add one element, and
one state vector, to include the atomic variables and the
environment in \eqr{psi0}:
 \begin{equation} \labe{fock_2e0}
 \begin{split}
 \left\{ \right. & \ket{1, n_a, 0, 0} , \\
 & \ket{2, n_a - 1, 0, 0} , \\
 & \ket{1, n_a - 1, 1, 0} , \\
 & \ket{1, 0, 0, n_a} \left. \right\}
 \end{split}
 \end{equation}
Now we can rewrite the two-level density matrix given by
\eqr{rho2} in this basis as
 \begin{equation} \labe{rho2e0}
 \rho = \begin{bmatrix}
 \rho_{1 1} & \rho_{1 2} & \rho_{1 e} & \rho_{1 0} \\
 \rho_{2 1} & \rho_{2 2} & \rho_{2 e} & \rho_{2 0} \\
 \rho_{e 1} & \rho_{e 2} & \rho_{e e} & \rho_{e 0} \\
 \rho_{0 1} & \rho_{0 2} & \rho_{0 e} & \rho_{0 0}
 \end{bmatrix} .
 \end{equation}
We can then use the corresponding total Hamiltonian
 \begin{equation} \labe{H2e0}
 H = -\hbar\, \begin{bmatrix}
 0 & \Omega^*_a & 0 & 0 \\
 \Omega_a & \nu_a & 0 & 0 \\
 0 & 0 & 0 & 0 \\
  0 & 0 & 0 & 0
 \end{bmatrix}
 \end{equation}
to solve for the density matrix element $\rho_{10}(t)$. We include
in our model the same phenomenological decoherence mechanisms
introduced in \sct{eit2}, and we do not introduce new dephasing
processes between the superposed loaded and unloaded resonators.

At $t = 0$, we assume that the system is in the pure state
superposition
 \begin{equation}
 \ket{\psi(0)} = \frac{1}{\sqrt{2}} \left(\ket{1, 0, 0, n_a} + \ket{1, n_a, 0,
 0} \right) ,
 \end{equation}
and we wish to identify a later time $t = t_1$ (if possible) where
the system state vector has evolved to the pure state
 \begin{equation}
 \ket{\psi(t_1)} = \frac{1}{\sqrt{2}} \left(\ket{1, 0, 0, n_a}
 + e^{i \varphi(t_1)} \ket{1, n_a, 0, 0} \right) ,
 \end{equation}
i.e., a state where both the atomic state and the environment can
be factored out of the Hilbert space, leaving a nonzero relative
phase difference between the remaining basis vectors. At each of
these times, the density matrix will have the form
 \begin{equation} \labe{rho2p0}
 \rho(t) = \begin{bmatrix}
 \half & 0 & 0 & \half e^{i \varphi(t)}\\
 0 & 0 & 0 & 0 \\
 0 & 0 & 0 & 0 \\
 \half e^{-i \varphi(t)} & 0 & 0 & \half
 \end{bmatrix} ,
 \end{equation}
where $\varphi(0) = 0$ establishes the initial condition.

In the absence of decoherence, the nonzero density matrix elements
are quickly found to be
 \begin{subequations} \labe{rho2e0ud}
 \begin{eqnarray}
 \rho_{11}(t) &=& \half \left[ 1 - \frac{\left|\Omega_a\right|^2}{\Omega_R^2} \sin^2\left(\Omega_R t\right) \right] , \\
 \rho_{22}(t) &=& \frac{\left|\Omega_a\right|^2}{2 \Omega_R^2} \sin^2\left(\Omega_R
 t\right), \\
 \rho_{21}(t) &=& \rho_{12}^*(t) = i \frac{\Omega_a}{2 \Omega_R} \sin\left(\Omega_R t\right) \nonumber \\
 &\times& \left[\cos\left(\Omega_R t\right) + i \frac{\nu_a}{2 \Omega_R} \sin\left(\Omega_R t\right)  \right]
 ,\\
 \rho_{10}(t) &=& \half  e^{\frac{i}{2} \nu_a t} \nonumber \\
 &\times& \left[\cos\left(\Omega_R t\right) - i \frac{\nu_a}{2 \Omega_R} \sin\left(\Omega_R t\right)
 \right], \\
 \rho_{20}(t) &=& i \frac{\Omega_a}{2 \Omega_R} e^{\frac{i}{2} \nu_a t} \sin\left(\Omega_R
 t\right),
 \end{eqnarray}
 \end{subequations}
where $\Omega_R$ is given by \eqr{rabi2def}. It is clear that
$\rho(t)$ has the form given by \eqr{rho2p0} at the times $t_q = q
\pi/\Omega_R$, where $q$ is a nonnegative integer. At these times,
the argument of $\rho_{10}(t)$ is given by
 \begin{equation}
 \varphi_q = -\left( 1 - \frac{\nu_a}{2 \Omega_R} \right) q \pi ,
 \end{equation}
where we have chosen the sign of $\varphi_q$ to be consistent with
that of the argument of $\rho_{10}(t)$ when $t$ is small. When the
system is undamped, a field with the detuning
 \begin{equation}
 \nu_a = \frac{2 (q - 1)}{\sqrt{2 q - 1}} \Omega_a
 \end{equation}
acquires a relative phase of $-\pi$ at the time
 \begin{equation}
 t_q = \frac{\sqrt{2 q - 1} \pi}{\Omega_a} .
 \end{equation}
Therefore, for a given value of $\left|\Omega_a\right|$ applied to
an undamped atomic system, a resonantly tuned field with $q = 1$
and $\nu_a = 0$ obtains a $-\pi$ phase shift \emph{earliest}.

In the presence of decoherence, we seek the quasi-steady-state
solution to the density matrix equation of motion defined by
\eqr{rhodot}, now emphasizing the single matrix element
$\rho_{10}(t)$. We build the Lindblad decoherence matrix operator
by applying \eqr{Lindblad} to the new density matrix \eqr{rho2e0},
and we extract the system of coupled linear differential equations
 \begin{subequations} \labe{rhodot0_2}
 \begin{eqnarray}
 \dot{\rho}_{10}(t) &=& -\gamma_{10} \rho_{10}(t) + i \Omega_a^*
 \rho_{20}(t) , \nd \\
 \dot{\rho}_{20}(t) &=& i (\nu_a + i \gamma_{20}) \rho_{20}(t) + i \Omega_a
 \rho_{10}(t) ,
 \end{eqnarray}
 \end{subequations}
where the decoherence constants are
 \begin{subequations} \labe{gamma20def}
 \begin{eqnarray}
 \gamma_{10} &\equiv& \gamma^{\prime \prime}_{1} , \nd \\
 \gamma_{20} &\equiv& \half \gamma^\prime_{2} + \gamma^{\prime \prime}_{2} .
 \end{eqnarray}
 \end{subequations}
Using the bootstrap method described in \sct{eit2}, we solve
\eqs{rhodot0_2} for the element $\rho_{10}(t)$ with the initial
conditions $\rho_{10}(0) = \half$ and $\rho_{20}(0) = 0$. If we
again assume that the interaction is unsaturated (i.e.,
$\left|\Omega_a\right|/\gamma_{20} \ll 1$) so that $|\rho_{10}(t)|
\gg |\rho_{20}(t)|$ for all $t$, and that
--- to zeroth order in $\left|\Omega_a\right|$ --- $|\rho_{10}(t)|$
varies slowly compared to $\sqrt{\nu_a^2 + \gamma_{20}^2}$, then
\eqs{rhodot0_2} yields the approximate solution
 \begin{equation} \labe{rho10ss}
 \begin{split}
 \rho_{10}(t) \cong \half \exp& \Bigg[\left(-\gamma_{10} + i
 W_{10}\right)t \\
 &- \frac{1 - e^{i(\nu_a + i \gamma_{20})t}}{\left(\nu_a + i
 \gamma_{20}\right)^2} \left|\Omega_a\right|^2\Bigg] ,
 \end{split}
 \end{equation}
where
 \begin{equation} \labe{W_10_2}
 W_{10} \equiv -\frac{\left|\Omega_a\right|^2}{\nu_a + i
 \gamma_{20}} = -\frac{\nu_a - i \gamma_{20}}{\nu_a^2 +
 \gamma_{20}^2}\, \left|\Omega_a\right|^2 .
 \end{equation}
It is straightforward to extend \eqr{rho10ss} to include a
coherent unsaturated interaction with $N$ independent atoms
localized within a volume that is small compared to $\left(\pi
w_0^2\right)^2/2 \lambda$ at $\mathbf{r} = 0$. We begin by
extending both the $N$-atom density matrix and Hamiltonian given
by \eqr{rho2en} and \eqr{H2en}, respectively, to include the
coherent superposition with the empty resonator, as was done in
\eqr{rho2e0} and \eqr{H2e0}. We find that \eqr{rho10ss} remains
unchanged if we make the substitution $\left|\Omega_a\right|^2
\rightarrow N \left|\Omega_a\right|^2$ and $\gamma_{10}
\rightarrow N \gamma_{21}^\prime/4$, so that---in the low-dephasing limit, after the
transient terms in \eqr{rho10ss} have decayed---the effect of the
placement of $N$ atoms in the interaction region is to replace the
time $t$ with $N t$. In this limit, \eqr{rho10ss} and \eqr{W_10_2}
are entirely consistent with \eqr{Easc} and \eqr{Wadef}, except
for the appearance of the photon number $n_a$ in the denominator
of \eqr{Wadef}. Since \eqr{W_10_2} was obtained using an
$n_a$-photon Fock state (rather than a coherent state, as was
implicitly used in the semiclassical estimate of $W_a$), we expect
an additional factor of $n_a$ from the analysis leading to
\eqr{psip}.

In the unsaturated limit, our analysis of the conditions required
to obtain a particular phase shift is significantly different from
that of the undamped case. Clearly, if we wish to accumulate a
large phase shift before the system state has become significantly
mixed, the resonant detuning should satisfy the inequality
 \begin{equation} \labe{deco2}
 \left|\Omega_a\right| \ll \gamma_{20} \ll \nu_a <
 \sqrt{\frac{N \gamma_{20}}{\gamma_{10}}} \left|\Omega_a\right|
 .
 \end{equation}
This constraint can only be met in the weak-field case if the
dephasing rate between the upper and lower atomic energy levels is
small enough that $\gamma^{\prime \prime}_{2} \ll
\gamma^\prime_{2}$. In this limit, dephasing can be neglected, and
the argument of the density matrix element $\rho_{10}(t)$ is given
approximately by the undamped value
 \begin{equation} \labe{linphase}
 \varphi(t) \approx -\frac{\left|\Omega_a\right|^2}{\nu_a}\, t ,
 \end{equation}
or, at time $t_q = q \pi/\Omega_R \cong 2 q \pi/\nu_a$,
 \begin{equation}
 \varphi(t_q) \approx -\frac{\left|\Omega_a\right|^2}{\nu_a^2}\, 2\, q \pi
 .
 \end{equation}
At time $t = t_q$, \eqr{rho10ss} gives for the magnitude of
$\rho_{10}\left(t_q\right)$
 \begin{equation}
 \left|\rho_{10}(t_q)\right| \approx \half \exp\left(-\frac{\gamma_{20}
 \left|\Omega_a\right|^2}{\nu_a^3}\, 2\, q
 \pi\right)
 .
 \end{equation}
Now, in order to achieve a phase shift of $-\pi$, we must choose a
long delay time such that $q \approx \nu_a^2/2
\left|\Omega_a\right|^2$, giving $t_q \approx \pi
\nu_a/\left|\Omega_a\right|^2$ and
 \begin{equation}
 \rho_{10}(t_q) \approx -\half \exp\left(-\frac{\gamma_{20}}{\nu_a}\,
 \pi\right) .
 \end{equation}

It is clear that we must detune the laser field such that $\nu_a
\gg \gamma_{20}$ so that we can minimize the effects of
decoherence, but it is not obvious how to choose a specific value
of $\nu_a$. First, we can define the fidelity (a measure of
distance between quantum states) of two density matrices $\rho_1$
and $\rho_2$ as\cite{niel00}
 \begin{equation} \labe{Fdef}
 F\left(\rho_1, \rho_2\right) \equiv \trace \sqrt{\rho_1^{1/2} \rho_2
 \rho_1^{1/2}} ,
 \end{equation}
or, in the case of a pure state $\ket{\psi}$ and an arbitrary
state $\rho$,
 \begin{equation} \labe{Fdefpure}
 F\left(\psi, \rho\right) = \sqrt{\bra{\psi} \rho \ket{\psi}} .
 \end{equation}
Applying \eqr{Fdefpure} to the density matrix given by
\eqr{rho2e0} and the pure state
 \begin{equation*}
 \ket{\psi\left(t_q\right)} \equiv
\frac{1}{\sqrt{2}} \left( \ket{1, 0, 0, n_a} - \ket{1, n_a, 0, 0}
\right) ,
 \end{equation*}
we obtain
 \begin{equation} \labe{Fax}
 \begin{split}
 F &= \frac{1}{\sqrt{2}} \left[ \rho_{00}\left(t_q\right) + \rho_{11}\left(t_q\right) -
 \rho_{10}\left(t_q\right) - \rho_{01}\left(t_q\right)\right]^{\half} \\
 &= \half \left[ 1 + \exp\left(-\frac{\gamma_{20}}{\nu_a} \pi\right) \right]
 \cong 1 - \frac{\gamma_{20}}{\nu_a} \frac{\pi}{2} .
 \end{split}
 \end{equation}
Second, we can compute the entropy of $\rho\left(t_q\right)$ using
the definition\cite{niel00}
 \begin{equation} \labe{Sdef}
 S \equiv -\trace\left[ \rho \log_2(\rho) \right] = -\sum_j \lambda_j
 \log_2\left(\lambda_j\right) ,
 \end{equation}
where the sum is carried over the nonzero eigenvalues of $\rho$.
Now, the density matrix elements $\rho_{2 1}\left(t_q\right)$ and
$\rho_{2 0}\left(t_q\right)$ are proportional to
$\left|\Omega_a\right|/\nu_a$, while $\rho_{2 2}\left(t_q\right)$
is proportional to $\left|\Omega_a\right|^2/\nu_a^2$. Therefore,
by \eqr{deco2}, we ignore these terms in the density matrix given
by \eqr{rho2e0}, and we apply \eqr{Sdef} in the limit
$\gamma_{20}/\nu_a \ll 1$ to obtain
 \begin{equation} \labe{Sax}
 S\left(t_q\right) \approx \frac{\gamma_{20}}{\nu_a} \pi \left[ 1 -
 \log\left( \frac{\gamma_{20}}{\nu_a} \pi \right) \right]
 \log_2(e) .
 \end{equation}
In principle, we can choose the value of $\nu_a$ to obtain
particular values of the entropy and fidelity, and then --- in an
$N$-atom system --- allow the system to evolve until time $t =
t_q/N$ to accumulate a $-\pi$ phase shift. In practice, in many
cases the probe field frequency cannot be modified \emph{post
hoc}, particularly in applications where a phase shift other than
$-\pi$ is required and/or more than one type of atom or molecule
is placed in consecutive interaction regions.

It is already clear from \eqr{Sax} and \eqr{Fax} that we must have
$\nu_a/\gamma_{20} > 30$ if we wish to hold $F > 0.95$ and $S <
0.5$ (base 2) for quantum information purposes. If, as an example,
we also have $\left|\Omega_a\right|/\gamma_{20} = 0.2$, then after
a time $N \gamma_{20} t_q = 775 \pi$ we will obtain a linear phase
shift of $-\pi$. However, if we require $F > 0.9995$ and $S <
0.01$, then we must have $\nu_a/\gamma_{20} > 4000$ and wait until
a time $N \gamma_{20} t_q = 100,000 \pi$ to obtain a linear phase
shift of $-\pi$. In other words, even though we are using a large
detuning, requirements of small entropy and high fidelity imply
that we will need either extraordinarily long interaction times or
many identical noninteracting atoms to achieve nontrivial phase
shifts.

 \subsection{The Four-Level Atom\labs{qip4}}

A calculation of the real and imaginary parts of $\rho_{10}(t)$ in
the case of the four-level atom proceeds in essentially the same
fashion as the corresponding calculation for the two-level case
described in the previous section. Again we further extend the
product state basis given by \eqr{fock_4e} to include a ``second
rail'' as an alternative quantum path for the $n_a$ probe photons,
corresponding to a $6 \times 6$ density matrix and Hamiltonian
(extended from \eqr{H4} as \eqr{H2e0} was from \eqr{H2e}). The
matrix elements $\rho_{10}(t)$, $\rho_{20}(t)$, $\rho_{30}(t)$,
and $\rho_{40}(t)$ are mutually coupled, and we seek an
approximate solution using the bootstrap method used in previous
sections. We neglect transient (homogeneous) solutions to the
coupled ODEs, and we assume that $\rho_{10}(t) \cong \half$ is
much larger than the magnitudes of the other three elements. Under
these conditions, in the unsaturated limit
$\left|\Omega_a\right|^2/\gamma_{20}^2 \ll 1$ we obtain the
quasi-steady-state solution
 \begin{equation} \labe{rho10ss4}
 \rho_{10}(t) \cong \half e^{\left(-\gamma_{10} + i
 W_{10}\right)t} ,
 \end{equation}
where
 \begin{widetext}
 \begin{equation} \labe{W_10_4}
 W_{10} \equiv -\frac{\left[(\nu_a - \nu_b + i \gamma_{30})(\nu_a - \nu_b +
 \nu_c + i \gamma_{40}) - \left|\Omega_c\right|^2\right]\, \left|\Omega_a\right|^2}
 {(\nu_a + i \gamma_{20})\left[(\nu_a - \nu_b + i \gamma_{30})(\nu_a - \nu_b +
 \nu_c + i \gamma_{40}) - \left|\Omega_c\right|^2\right] -
 (\nu_a - \nu_b + \nu_c + i \gamma_{40}) \left|\Omega_b\right|^2}
 ,
 \end{equation}
 \end{widetext}
$\gamma_{10}$ and $\gamma_{20}$ are given by \eqr{gamma20def}, and
 \begin{subequations} \labe{gamma40def}
 \begin{eqnarray}
 \gamma_{30} &\equiv& \gamma^{\prime \prime}_{3} , \nd \\
 \gamma_{40} &\equiv& \half \gamma^\prime_{4} + \gamma^{\prime \prime}_{4} .
 \end{eqnarray}
 \end{subequations}

If we set $\nu_a = \nu_b = 0$ to minimize absorption, and if the
dephasing rates are much smaller than the depopulation rates of
the upper atomic levels, then \eqr{W_10_4} becomes
 \begin{subequations} \labe{W_10_4s}
 \begin{eqnarray}
 W_{10} &=& -\frac{\left|\Omega_a\right|^2 \left|\Omega_c\right|^2}{\nu_c \left|\Omega_b\right|^2 + i \left( \gamma_{4}
\left|\Omega_b\right|^2 + \gamma_2 \left|\Omega_c\right|^2
\right)}\quad \labe{W_10_4p} \\
&=&- \frac{\tilde{\nu}_c - i \tilde{\gamma}_{20}}
 {\tilde{\nu}_c^2 + \tilde{\gamma}_{20}^2}\,
 \left|\Omega_a\right|^2, \labe{W_10_4d}
 \end{eqnarray}
 \end{subequations}
 where
 \begin{subequations}
 \begin{eqnarray}
 \tilde{\nu}_c &\equiv&
 \frac{\left|\Omega_b\right|^2}{\left|\Omega_c\right|^2}\, \nu_c , \nd \labe{nuctilde}\\
 \tilde{\gamma}_{20} &\equiv&  \gamma_{20} +
 \frac{\left|\Omega_b\right|^2}{\left|\Omega_c\right|^2}\,
 \gamma_{40}. \labe{gamma20tilde}
 \end{eqnarray}
 \end{subequations}
Note that \eqr{W_10_4d} has precisely the same form as the
corresponding two-level result given by \eqr{W_10_2}. Hence, the
nonlinear phase shift derived from \eqr{W_10_4d} can be as large
as the corresponding linear phase shift obtained from
\eqr{W_10_2}, indicating the presence of an enormous third-order
nonlinearity that couples the three
fields.\cite{schm96,niel00,vita00,wang01} In principle, this
nonlinearity requires only modest detunings to provide both a high
differential phase shift and a low absorption rate in the
semiclassical realm.

However, in the quantum information processing applications
described above, we must also check that the photon-atom system is
effectively disentangled when the dipole interaction is switched
off. In the limit $\tilde{\nu}_c \gg \tilde{\gamma}_{20}$, the
earliest elapsed time $t_1$ required to obtain a phase shift of
$-\pi$ and the corresponding fidelity and entropy are respectively
given by
 \begin{eqnarray}
 t_1 &\cong& \frac{\pi\, \tilde{\nu}_c}{\left|\Omega_a\right|^2}
 , \labe{phitilde} \\
 F &\cong& 1 - \frac{\tilde{\gamma}_{20}}{\tilde{\nu}_c} \frac{\pi}{2} , \nd \labe{Ftilde} \\
 S &\cong& \frac{\tilde{\gamma}_{20}}{\tilde{\nu}_c} \pi \left[ 1 -
 \log\left( \frac{\tilde{\gamma}_{20}}{\tilde{\nu}_c} \pi \right) \right]
 \log_2(e) . \labe{Stilde}
 \end{eqnarray}

Given a sufficiently long interaction time, it is clear from
\eqr{nuctilde} that a large phase shift can be accumulated using a
relatively small net detuning $\nu_c$ even when $\Omega_c$
represents the Hamiltonian matrix element describing the dipole
interaction of a single photon with a single atom.\cite{shim02}
However, the value of $\nu_c$ needed to maintain a high fidelity
and a low entropy depends on other system parameters. For example,
in the limit\cite{schm96}
 \begin{equation} \labe{schmlim}
  \frac{\left|\Omega_b\right|^2}{\gamma_{20}} \frac{\nu_c}{\gamma_{40}} \gg
\frac{\left|\Omega_b\right|^2}{\gamma_{20}} \gg
\frac{\left|\Omega_c\right|^2}{\gamma_{40}} ,
 \end{equation}
we note that we must have $\tilde{\nu}_c/\tilde{\gamma}_{20}
\approx \nu_c/\gamma_{40} \gg 1$ to achieve $F \longrightarrow 1$
and $S \longrightarrow 0$. Hence, relatively \emph{large}
detunings are still required when the four-level system is used
for quantum information processing applications. However, in the
limit where the spontaneous emission rate of atomic state
$\ket{4}$  in \fig{four_level_atom} has been strongly suppressed,
 \begin{equation} \labe{suppress}
 \frac{\left|\Omega_b\right|^2}{\gamma_{20}} \frac{\nu_c}{\gamma_{40}} \gg
 \frac{\left|\Omega_c\right|^2}{\gamma_{40}} \gg
 \frac{\left|\Omega_b\right|^2}{\gamma_{20}},
\end{equation}
we find from \eqr{Ftilde} and \eqr{Stilde} that
$\tilde{\nu}_c/\tilde{\gamma}_{20} \approx
\tilde{\nu}_c/\gamma_{20}$, a constraint that can be satisfied
easily for small detunings whenever
$\left|\Omega_b\right|^2/\left|\Omega_c\right|^2 \gg 1$.
Suppression of spontaneous emission from level $\ket{4}$ can be
achieved in at least two different ways. First, the interaction
region can be placed within a photonic bandgap crystal (PBC)
designed so that photons with frequency $\omega_c$ must be
injected through a defect in the crystal structure. Second, a
different atomic system could be chosen with an energy level
structure similar to that shown in \fig{lambda_4b}, where the
final dipole transition in \fig{lambda_4} has been replaced by a
two-photon transition to the metastable atomic level $\ket{4}$.
Although the details of the calculations leading to \eqr{W_10_4d}
will certainly change for this system, the relative weakness of
the two-photon transition amplitude can be effectively offset by a
suitably smaller choice of the value of the detuning frequency
$\nu_c$.

\epsf{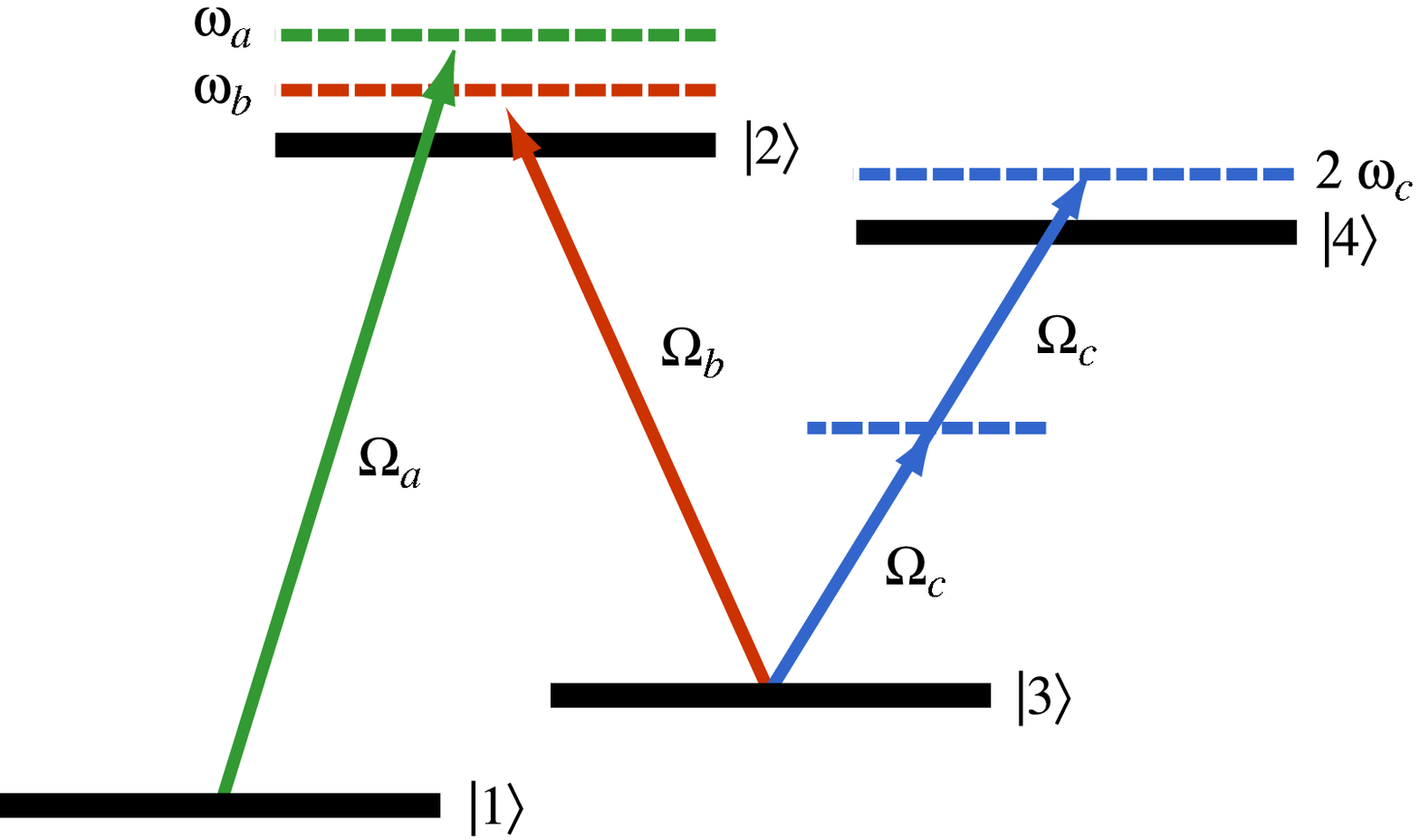}{Modified interaction between a four-level
$\mathcal{N}$ atom and a three-frequency electromagnetic field.
The final dipole transition in \fig{lambda_4} has been replaced by
a two-photon transition to the metastable atomic level $\ket{4}$.}

If the controlled coupling transition $\ket{2} \longleftrightarrow
\ket{3}$ is driven by a Fock state, then the four-level atom +
field system provides a large cross-Kerr nonlinearity of the form
$H_\text{Kerr} = \hbar \chi\, \hat n_a\, \hat n_c$ when the
constraint given by \eqr{suppress} is satisfied. Neglecting the
imaginary part of \eqs{W_10_4s}, we find that the evolution
corresponds to that of the Kerr Hamiltonian given by
\eqr{Kerrevol}, with the nonlinear coefficient given by
\eqr{chifock}. In this case, the entanglement of the atoms and
fields is negligible, and the fidelity of the final state is high.
However, if the coupling transition is driven by a coherent state,
then --- as shown in \fig{coherence_15} --- the electromagnetic
intensity of that state (i.e., the magnitude of the Poynting
vector) must be quite large to ensure that the atoms and fields
are completely disentangled at the conclusion of the gate
operation. This condition requires that $\left|\alpha_b\right|^2
\gg 1$, reducing the magnitude of the effective cross-Kerr
nonlinearity given by \eqr{chicoh}.

We have demonstrated so far that it is possible to apply an
arbitrary phase shift to an initial state $c_0 \ket{0} + c_1
\ket{1}$ of a photonic qubit tuned to the $\ket{1}
\longleftrightarrow \ket{2}$ transition in \fig{four_level_atom},
resulting in the final state $c_0 \ket{0} + c_1 e^{i \phi}
\ket{1}$. Clearly it is straightforward to perform Hadamard gates
on qubits encoded in photons in this manner, through use of simple
linear optics (beamsplitters). With the tuneable EIT phase shift
gate, the required range of single qubit gates needed for
universal quantum information processing is therefore covered. The
other necessary ingredient for universal quantum processing
\cite{divi95,lloy95,bare95,niel00} is a two qubit entangling gate.
For photonic qubits
which generally interact very weakly with each other, this is the
more difficult gate to realize. One solution is to use measurement
and feedback to create an effective strong non-linearity between
photonic qubits.\cite{knil01} However, it is clearly of real
significance for photonic quantum information processing to
consider the possibility of a direct non-linear coupling between
photonic qubits using an EIT system, to realize, for example, a
conditional two-qubit phase gate.

Consider a case (depicted schematically in
\fig{two-qubit-phase-shift}) with two photon number encoded qubits
(target and control), where the target qubit is tuned to the
$\ket{1} \longleftrightarrow \ket{2}$ transition frequency
$\omega_a$ and the control qubit is tuned to the $\ket{3}
\longleftrightarrow \ket{4}$ transition frequency $\omega_c$ of
\fig{four_level_atom}, so this mode is now a quantum rather than
classical control field. As shown above, if no photon is present
in the $\ket{3} \longleftrightarrow \ket{4}$ transition, then the
target qubit $c_0 \ket{0} + c_1 \ket{1}$ acquires no phase shift.
However, if a photon with frequency $\omega_c$ is present in the
$\ket{3} \longleftrightarrow \ket{4}$ transition, then the target
qubit evolves to $c_0 \ket{0} + c_1 e^{i \phi} \ket{1}$. Hence,
this system implements a conditional phase shift and is extremely
useful for quantum information processing. In fact, for a
conditional phase shift of $-\pi$ the operation provides a
universal two-qubit gate capable of maximally entangling two
initially unentangled photonic qubits. The input product state
$\frac{1}{2}(\ket{00} + \ket{01} + \ket{10} + \ket{11})$ can be
transformed to the maximally entangled state $\frac{1}{2}(\ket{00}
+ \ket{01} + \ket{10} - \ket{11})$, as shown schematically in
\fig{two-qubit-phase-shift}. Therefore, in principle, universal
optical quantum information processing can be performed with such
EIT systems.

\epsf{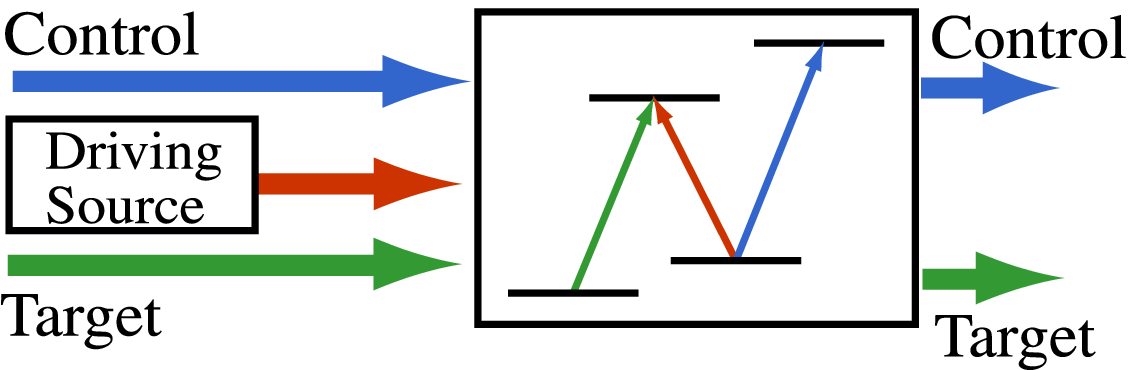}{Schematic diagram of a two qubit
conditional phase shift which transforms the state $c_{00}
\ket{00} + c_{01} \ket{01} +c_{10} \ket{10} + c_{11} \ket{11}$ to
$c_{00} \ket{00} + c_{01} \ket{01} +c_{10} \ket{10} + c_{11} e^{i
\chi t} \ket{11}$. When $\chi t = \pi$ the two qubit phase gate is
implemented.}

To illustrate this we consider the error probability in generating
the conditional phase shift on the $\ket{11}$ amplitude.
\fig{fidelity_11} shows the error probability in generating the
phase-shifted amplitude for single photons in the $a$ and $c$
modes of the four-level system, as a function of the average
photon number in the coherent drive applied to mode $b$, for
various values of the chosen phase shift. Clearly as long as the
applied drive in mode $b$ is pushed towards being a classical
field and so the EIT system provides an accurate cross-Kerr
nonlinearity, the $\ket{11}$ amplitude can in principle receive a
large and accurate phase shift. The dependence of the nonlinearity
on the various parameters is given in \eqr{chicoh}. From
\fig{fidelity_11} we see that the gate requires a drive field with
a large $|\alpha_b|^2$, but from \eqr{chicoh} this reduces the
strength of the nonlinearity, which increases the value of $N t$
required for the chosen phase shift (for a given detuning). So, as
discussed in \sct{qip2} and \sct{qip4}, it is crucial that the
effects of decoherence are kept small, in particular the
spontaneous emission from level $\ket{4}$, in order to perform a
two-qubit gate with small error. Detailed analysis of the effects
of decoherence on the EIT phase gates will be addressed in future
work.

\epsf{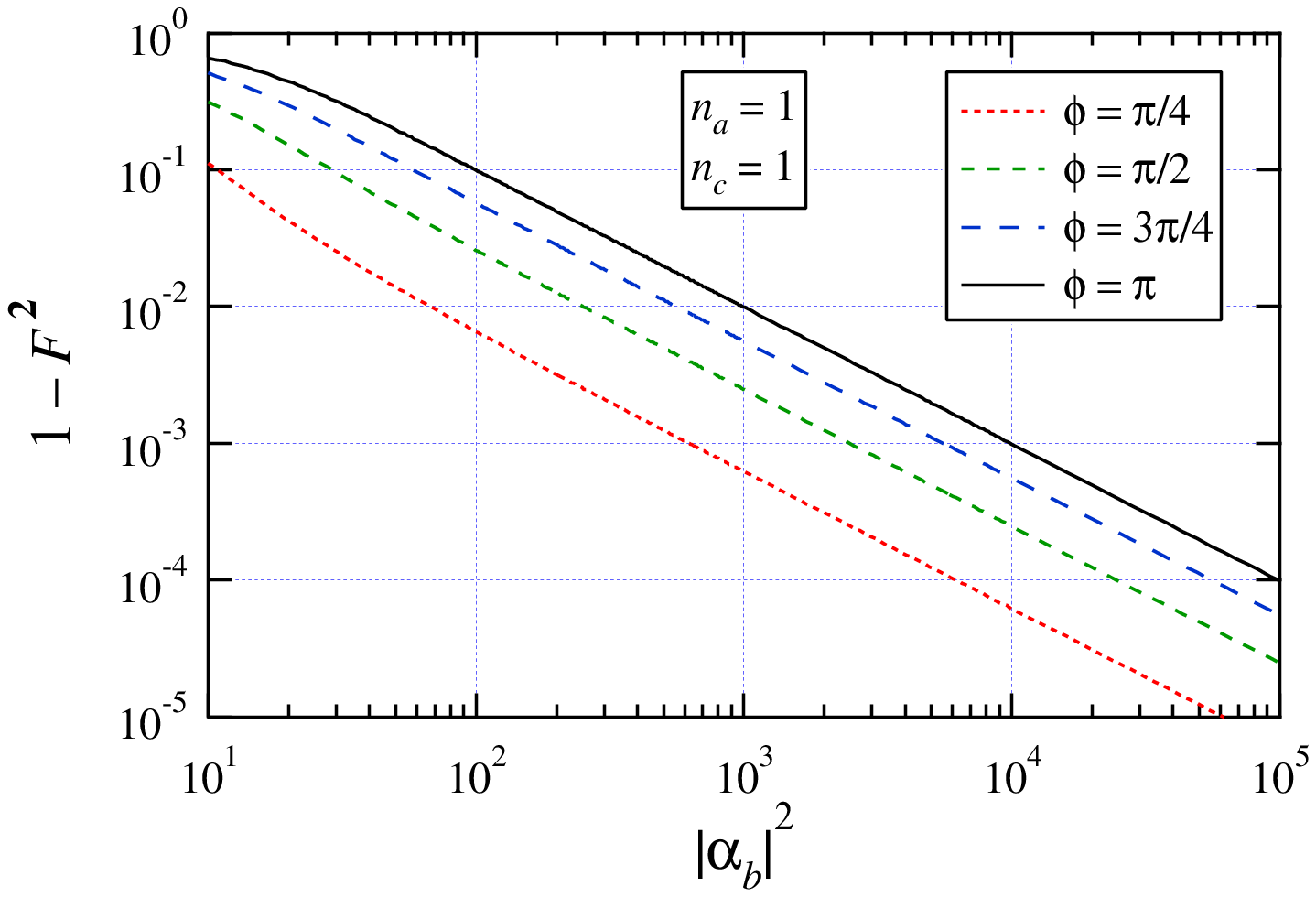}{Numerical evaluation of the error probability
$1 - F^2$ for several values of the net phase shift $\phi$,
assuming that the input Fock state has $n_a = 1$ and $n_c = 1$. We
have used \eqr{Fdefpure} to compare an ideal phase-shifted state
\eqr{psicoh} to the actual final state \eqr{psi_prime}.}

Clearly the conditional phase shift can be put to good use in
other quantum processing applications. One extremely useful device
is a high efficiency non-demolition detector for photons. If
instead of a single photon state on the $\ket{3}
\longleftrightarrow \ket{4}$ transition, a weak coherent state is
input, a measurable (by standard techniques) phase shift arises
conditional on the single-photon Fock amplitude of a qubit in the
$\ket{1} \longleftrightarrow \ket{2}$ transition. In effect a
projective measurement in the computational basis can be performed
on the photonic qubit, with the qubit being available for re-use
afterwards rather than being absorbed. This can be achieved with
high efficiency ($>99\%$) and with just a few hundred atoms in the
EIT system. Details of this detector are reported
elsewhere.\cite{munr03}

 \section{CONCLUSIONS}

In \sct{eit4}, we demonstrated that it should be possible to use
the four-level atomic system of \fig{four_level_atom} as the
foundation for both a (so-called) quantum switch and a
phase-shifter. For example, the quantum switch is, in principle,
simple to implement: given the availability of a photon with
frequency $\omega_b = \omega_{32}$, if a photon with frequency
$\omega_c = \omega_{43}$ is present (i.e., $\Omega_c \ne 0$), then
a probe photon with frequency $\omega_a = \omega_{21}$ will be
absorbed; otherwise, if $\Omega_c = 0$, then the probe photon will
be transmitted. All this shows that conventional classical
information processing operations are possible on optical
data---single bit phase shifts and conditional (two-bit)
switching. These effects have significant potential for
applications to conventional optical communications.

In \sct{qip}, we demonstrated that it should be possible to
operate the four-level atomic system as a ``dual rail'' photon
qubit phase shifter, provided that the spontaneous emission from
the atomic level $\ket{4}$ can be suppressed. The size of the
phase shift and the fidelity of the gate are quantified in terms
of the system parameters in our model, so the trade-off between
the size of the phase shift, the accuracy of the gate, the time of
operation and the atomic and control parameters can be analyzed in
detail. A single photon phase gate, coupled with others that can
be performed using linear optical elements, enables the
performance of arbitrary single qubit operations. Universal
quantum information processing requires the addition of a suitable
entangling two-qubit gate.\cite{divi95,lloy95,bare95,niel00} We
have also demonstrated that in principle, the phase shifter
arrangement can be turned into such a gate---a two-qubit
conditional phase gate---by using a qubit input also on the
control field at frequency $\omega_c$. Therefore such EIT systems
can in principle be used to enable universal quantum information
processing with ``dual rail'' photon qubits. Coupled with ideas
such as quantum memory for photons
\cite{flei02,juze02,mewe02,flei01} and non-absorbing  photon
detectors,\cite{munr03} it is clear that EIT systems present a
very promising route forward for few-qubit quantum information
processing.

Indeed, given the experimental progress with EIT phenomena
over the last few years, it seems likely that these QIP
applications can tested and developed over the next few years.
Of course, more detailed research still needs to be done.
For example, further refinements to our model to
include coherent wavepacket representations of the fields are
needed to realistically assess in detail the performance of the
two-qubit gate. This will be addressed in future work.

\acknowledgments

We thank Kae Nemoto and Pieter Kok for helpful conversations, and
Adrian Kent and Sandu Popescu for numerous detailed suggestions
after carefully reading the manuscript.

\bibliography{cpt_qip}

\end{document}